\documentclass{aa}
\usepackage{graphicx}
\usepackage{amssymb}
\usepackage{natbib}

\usepackage{supertabular}
\usepackage{lscape}
\usepackage{longtable}
\usepackage{rotating}%

\def\ssim{\setbox0=\hbox{$\propto$}%
\setbox1=\hbox{$<$}\dimen0=\ht1%
\advance\dimen0by-1.2pt\,\lower.6\dimen0%
\copy0\kern-\wd0\raise.4\dimen0\copy1 \,}

\def\gsim{\setbox0=\hbox{$\propto$}%
\setbox1=\hbox{$>$}\dimen0=\ht1%
\advance\dimen0by-1.2pt\,\lower.6\dimen0%
\copy0\kern-\wd0\raise.4\dimen0\copy1\,}

\def\lambdab{\lambda\mkern-9mu\lower1.2pt\hbox{$\mathchar'26$}}%

\begin{document}
   \title{Very low metallicity massive star models:}
\subtitle{Pre--SN evolution and primary nitrogen production.}
   \titlerunning{Pre--SN evolution and primary nitrogen @ very low $Z$.}


 \author{Raphael Hirschi}
   \institute{Dept. of physics and Astronomy, University of Basel,           
	      Klingelbergstr. 82, CH-4056 Basel
             }
\offprints{R. Hirschi \email{Raphael.Hirschi@unibas.ch}}

   \date{Received  / Accepted }

\abstract
{Precise measurements
of surface abundances of extremely low metallicity stars have recently 
been obtained and provide new constraints for the
stellar evolution models.}
{Compute stellar evolution models in order to explain the surface abundances
observed, in particular of nitrogen.}
{
Two series of models were computed.
The  first series consists of 20 $M_\odot$ models with varying initial
metallicity ($Z=0.02$ down to $Z=10^{-8}$) and rotation
($\rm \upsilon_{ini}=0-600$\,km\,s$^{-1}$). 
The second one consists of
models with an initial metallicity of $Z=10^{-8}$, 
masses between 9 and
85 $M_\odot$ and fast initial rotation velocities 
($\rm \upsilon_{ini}=600-800$\,km\,s$^{-1}$).
}
{
The most interesting models are the models 
with $Z=10^{-8}$ ([Fe/H]$\sim-6.6$).
In the course of helium burning, carbon and oxygen are mixed into the
hydrogen burning shell. This boosts the importance of the shell and
causes a reduction of the CO core mass. Later in the evolution,
the hydrogen shell deepens and produces large amount of primary
nitrogen. For the most massive models ($M\gtrsim 60$\,$M_\odot$),
significant mass
loss occurs during the red supergiant stage. This mass loss is due to
the surface enrichment in CNO elements via rotational and convective
mixing. The 85 $M_\odot$ model ends up as a WO type
wolf-Rayet star. Therefore the models predict SNe of type Ic and 
possibly long and soft GRBs at very low metallicities.

The rotating 20 $M_\odot$ models can 
best reproduce the observed CNO abundances at the surface 
of extremely 
metal poor (EMP) stars and the metallicity trends when their angular
momentum content is the same as at solar metallicity (and therefore have
an increasing surface velocity with decreasing metallicity).
The wind of the massive star models can also reproduce the CNO 
abundances of the most metal-poor carbon--rich star known to
date, HE1327-2326. 
}
{}

\keywords{Stars: abundances, evolution,
rotation, mass loss, Wolf--Rayet, supernova, Gamma rays: theory, bursts}

\maketitle
%

\section{Introduction}
Precise measurements
of  surface abundances of extremely metal poor (EMP) stars have recently 
been obtained by \citet{FS5}, \citet{FS6} and \citet{IER04}. 
These provide new constraints for the
stellar evolution models \citep[see][]{CMB05,Fr04,Pr05}.
The most striking constraint is the need for primary $^{14}$N production in very low
metallicity massive stars. Other constraints are an upturn of the 
[C/O] ratio with a [C/Fe] ratio 
 \citep[see for example Fig. 14 for C/O and Fig. 13 for C/Fe in][]{FS6} 
about constant or slightly decreasing (with increasing metallicity) at very low metallicities, 
which requires an increase (with increasing metallicity) of oxygen yields
below [Fe/H]$\sim$ -3. 
About one quarter of EMP stars are carbon rich (C-rich EMP, CEMP stars).
\citet{RANB05} and \citet{BC05} propose a classification for these CEMP stars. They find two
categories: about three quarter are main s--process enriched (Ba-rich) 
CEMP stars and one quarter are enriched with a weak component of
s--process (Ba-normal). The two most metal poor stars known to date, 
HE1327-2326 \citep{Fr05,Ao06} and HE 0107-5240 \citep{Ch04} are both
CEMP stars. These stars are believed to have been enriched by only one
to several stars and the yields of the models can therefore be compared to
their observed abundances without the filter of a galactic chemical
evolution model (GCE).

The evolution of low metallicity or metal free stars is not a new
subject \citep[see for example][]{CC83,EE83,CBA84,A96}. The observations
cited above have however greatly increased the interest in very metal
poor stars. There are many recent works studying the evolution of metal
free (or almost) massive \citep{HW02,LC05,UN05,MEM06}, intermediate mass
\citep{SLL02,H04,SAMFI04,GS05} and low
mass \citep{PCL04,W04} stars 
in an attempt to explain the origin of the surface abundances observed.
In this work pre-supernova
evolution models of rotating single stars were computed 
with metallicities ranging 
from solar metallicity down to $Z=10^{-8}$ to study the impact of
rotation and to see which initial rotation velocities can lead to the
chemical enrichments observed.
In Sect. 2, the model
physical ingredients and the calculations are presented. 
In Sect. 3, the evolution of the models is described. 
In Sect. 4, the stellar yields of light elements are presented and compared to
observations. In Sect. 5, the conclusions are given. 
\section{Description of the stellar models}\label{phys}
The stellar evolution model used to calculate the stellar models
 is described in detail in \citet{psn04a}.
Convective stability is determined by the 
Schwarzschild criterion. 
Convection is treated no longer as an instantaneous mixing but as a 
diffusive process from oxygen burning onwards.
The overshooting parameter is 0.1 H$_{\rm{P}}$ 
for H and He--burning cores 
and 0 otherwise.  
The reaction rates are taken from the
NACRE \citep{NACRE} compilation for the experimental rates
and from the NACRE website (http://pntpm.ulb.ac.be/nacre.htm) for the
theoretical ones. 

\subsection{Initial composition}
The initial composition of the models is given in Table \ref{inic}.
The solar metallicity composition is described in \citet{ywr05} and only 
low metallicity compositions are presented here.
For a given metallicity $Z$ (in mass fraction), 
the initial helium mass fraction
$Y$ is given by the relation $Y= Y_p + \Delta Y/\Delta Z \cdot Z$, 
where $Y_p$ is the primordial
helium abundance and $\Delta Y/\Delta Z$ the slope of 
the helium--to--metal enrichment law. 
$Y_p$ = 0.24 and $\Delta Y/\Delta Z$ = 2.5 were used according to recent
determinations \citep[see][ for example]{IT04}.
For the mixture of the heavy elements, 
the same mixture as the one
used to compute the opacity tables for Weiss 95's alpha--enriched 
composition \citep{IR96} was adopted. 

\begin{table}
\caption{Initial abundance (in mass fraction) of the chemical elements
followed in the calculations.}
\begin{tabular}{l r r r r}
\hline
\hline
Element   & $Z=10^{-3}$ & $Z=10^{-5}$ & $Z=10^{-8}$ \\
\hline \\
 $^{1}$H   & 7.5650D-01 & 0.759965   & 0.759999965\\
 $^{3}$He  & 2.5702D-05 & 2.5440D-05 & 2.5437D-05 \\
 $^{4}$He  & 2.4247D-01 & 0.23999956 & 0.239974588\\
 $^{12}$C  & 7.5542D-05 & 7.5542D-07 & 7.5542D-10 \\
 $^{13}$C  & 9.0930D-07 & 9.0930D-09 & 9.0930D-12 \\ 
 $^{14}$N  & 2.3358D-05 & 2.3358D-07 & 2.3358D-10 \\ 
 $^{15}$N  & 9.2242D-08 & 9.2242D-10 & 9.2242D-13 \\ 
 $^{16}$O  & 6.7105D-04 & 6.7105D-06 & 6.7105D-09 \\ 
 $^{17}$O  & 2.7196D-07 & 2.7196D-09 & 2.7196D-12 \\ 
 $^{18}$O  & 1.5162D-06 & 1.5162D-08 & 1.5162D-11 \\ 
 $^{20}$Ne & 7.8366D-05 & 7.8366D-07 & 7.8366D-10 \\ 
 $^{22}$Ne & 6.3035D-06 & 6.3035D-08 & 6.3035D-11 \\
 $^{24}$Mg & 3.2475D-05 & 3.2475D-07 & 3.2475D-10 \\
 $^{25}$Mg & 4.2685D-06 & 4.2685D-08 & 4.2685D-11 \\
 $^{26}$Mg & 4.8956D-06 & 4.8956D-08 & 4.8956D-11 \\
 $^{28}$Si & 3.2769D-05 & 3.2769D-07 & 3.2769D-10 \\
 $^{32}$S  & 1.8897D-05 & 1.8897D-07 & 1.8897D-10 \\
 $^{36}$Ar & 1.9797D-06 & 1.9797D-08 & 1.9797D-11 \\
 $^{40}$Ca & 5.1728D-06 & 5.1728D-08 & 5.1728D-11 \\
 $^{44}$Ti & 0          & 0          & 0          \\
 $^{48}$Cr & 0          & 0          & 0          \\
 $^{52}$Fe & 0          & 0          & 0          \\
 $^{56}$Ni & 0          & 0          & 0          \\
\hline
\end{tabular}
\label{inic}
\end{table}

\subsection{Mass loss}\label{mdot}
Since mass loss rates are a key ingredient for the evolution of massive 
stars, 
the prescriptions used are summarised here.
The changes of the mass loss rates, $\dot{M}$, with  
rotation are taken into account as explained in \citet{ROTVI}.
As reference mass loss rates, 
the adopted mass loss rates are the ones of \citet{Vink00,VKL01},
who account for the occurrence of bi--stability
limits, which change the wind properties and mass loss rates.
For the domain not covered by these authors
the empirical law devised by \citet{Ja88} was used.
Note that this empirical law, which presents
a discontinuity in the mass flux near the Humphreys--Davidson limit,
implicitly accounts for the mass loss rates of LBV stars. 
For the non--rotating
models, since the empirical values
for the mass loss rates are based on 
stars covering the whole range of rotational velocities, 
one must apply a reduction factor to the empirical rates to make
them correspond to the non--rotating case. The same reduction factor 
was used as in \citet{ROTVII}.
During the Wolf--Rayet phase the mass loss rates by \citet{NuLa00}
were used. 
The mass loss rates depend 
on metallicity as $\dot{M} \sim (Z/Z_{\odot})^{0.5}$, where
$Z$ is the mass fraction of heavy elements at the surface
of the star. 

The mass loss rates (and opacity) are rather well 
determined for chemical compositions which are similar to solar 
composition or similar to a fraction of the
solar composition (or of the
alpha--enriched mixing). However, very little was known about the mass
loss of very low metallicity stars with a strong enrichment in CNO
elements until recently. \citet{VdK05} study the case of WR stars 
and find a clear dependence with iron group mass fractions. 
For red supergiant stars (RSG), recent studies 
\citep[see][ and references therein]{VL05} show that dust--driven 
winds at cool temperature show no metallicity dependence for
$Z>0.1\,Z_\odot$. 
As we shall see later, due to rotational and
convective mixing, the surface of the star is strongly enriched in CNO
elements during the RSG stage. It is implicitly assumed 
in this work \citep[as in][]{MEM06} 
that CNO elements have a significant contribution to
opacity and mass loss rates.   
The mass loss rates used depend 
on metallicity as $\dot{M} \sim (Z/Z_{\odot})^{0.5}$, where
$Z$ is the mass fraction of heavy elements at the surface
of the star, also when the iron group elements content is much lower
than
the CNO elements content. The largest mass losses in the present calculations
occur during the RSG stage. This means that if the independence of the
mass loss rates on the metallicity in the RSG stage
\citep{VL05} is confirmed at very low metallicities, the mass loss rate used
in this work possibly underestimate the real mass loss rate. This point
surely deserves to be studied in more detail in the future.

A specific treatment for mass loss was applied at break-up 
\citep[see][]{MEM06}.
At break-up, the mass loss rate adjusts itself in such a way that an
equilibrium is reached between the two following opposite effects. 
1) The radial inflation due to evolution, combined with the growth 
of the surface velocity due to the
internal coupling by meridional circulation, brings the star to 
break-up, and thus some
amount of mass at the surface is no longer bound to the star. 
2) By removing the most external layers,
mass loss brings the stellar surface down to a level in the star that
is no longer critical. Thus, at break-up, one should adapt 
the mass loss rates, in order
to maintain the surface layers at the break-up limit.
In practise, however, since the critical limit contains mathematical
singularities, it was considered that during the break-up phase, 
the mass loss rates should be such
that the model stays near a constant fraction (around 0.95) 
of the limit.
Note that wind anisotropy \citep[described in][]{ROTVI} was not taken 
into account for the present work.

\subsection{Chemical elements and angular momentum
transport}\label{mixel}
The instabilities induced by rotation
taken into account in this work are meridional circulation 
and secular and dynamical shears. Meridional circulation is an advective
process and shears are diffusive processes. 
The equations for the transport of angular momentum and chemical
elements are given in Sect. 2.3 of \citet{MM00}. For more details on
the equation for the transport of angular momentum, the reader can 
refer to \citet{ROTIII}.
The equation for the change of the mass fraction of chemical 
species $i$ is the following:
\begin{equation}
\left(\frac{dX_i}{dt} \right)_{M_r} =
\left(\frac{\partial  }{\partial M_r} \right)_t
\left[ (4\pi r^2 \rho)^2 D \left( \frac{\partial X_i}
{\partial M_r}\right)_t
\right] + \left(\frac{d X_i}{dt} \right)_{\rm n}
\end{equation} 

\noindent The second term on the right accounts for
 composition changes due
to nuclear reactions. Note that the nuclear burning term and the
diffusive term are treated in a serial 
way in the simulations. 
The coefficient $D$ is the 
sum of the different diffusion coefficients:
$D = D_{\rm conv} + D_{\rm eff}+D_{\rm shear}$,
where $D_{\rm conv}$ is the convective diffusion coefficient,
$D_{\rm eff}$ accounts
for the combined effect of advection and horizontal 
turbulence
and $D_{\rm shear}$ represents the sum of secular \citep[see Eqs. 5.32 in][]{ROTII} and 
dynamical \citep[see Eq. 5 in][]{psn04a} shear coefficients.
Although meridional circulation is an advective process, its effect on
the change in chemical abundances can be approximated by a diffusion
process.  
The coefficient $D_{\rm eff}$ (accounting
for the combined effect of advection and horizontal 
turbulence) is calculated in this work using the
following formula:

\begin{equation}
D_{\rm eff} = \frac{\mid rU(r) \mid^2}{30 D_h} \; ,
\end{equation} 

\noindent where $D_h$ is the coefficient of horizontal
turbulence, for which the estimate is $D_h = |r U (r)|$ \citep{ZA92}.
This equation expresses that the vertical
advection of chemical elements is severely inhibited by the
strong horizontal turbulence characterised by $D_h$. 

Even though there is no free parameter in the prescriptions above to 
increase or decrease the
importance of the coefficients, different authors use different
prescriptions for the various processes \citep[see for example][]{HLW00}.
The coefficient of horizontal turbulence was also recently revised
\citep{Mh03}. The new coefficient was used in \citet{MEM06} but not in this
work. The impact of the new coefficient is discussed briefly in Sect.
\ref{mdotm}.

\subsection{Initial rotation}\label{vini}
The value of 300 km\,s$^{-1}$ as the initial rotation velocity at solar 
metallicity
corresponds to an average velocity of about 220\,km\,s$^{-1}$ on the Main
Sequence (MS) which is
very close to the average observed value \citep[see for instance][]{FU82}. 
It is unfortunately not possible to measure the rotational velocity of very low
metallicity massive stars since they all died a long time ago.
Higher observed ratio of Be to B stars in
the Magellanic clouds compared to our Galaxy \citep{MGM99} could point
out to the fact the stars rotate faster at lower metallicities. 
Also a low metallicity star containing the same angular momentum
as a solar metallicity star has a higher surface rotation velocity due to
its smaller radius (one quarter of $Z_\odot$ radius for 20 $M_\odot$
stars). Since there is however not yet firm evidence for fast surface
rotation velocities at low metallicities, we explore in this work with 20 $M_\odot$ models 
different velocities ranging between no rotation and surface velocities
corresponding to the same total angular momentum as in solar
metallicity stars.

In order to compare the
models at different metallicities and with different initial masses
with another quantity than the surface velocity, 
the ratio $\upsilon_{\rm ini}/ \upsilon_{\rm crit}$ is used (see Table
\ref{table1}).
The critical velocity is reached when the gravitational acceleration is
balanced by radiative and centrifugal forces.
The critical velocity is given by the following formula if the star 
is far from its Eddington limit ($\Gamma_{\rm max}<0.64$):

\begin{equation}
v_\mathrm{crit, 1} = \Omega \; R_\mathrm{eb} = 
\left( \frac{2}{3} \frac{GM}{R_\mathrm{pb}} \right)^{\frac{1}{2}} \; .
\end{equation}

\noindent
$R_\mathrm{eb}$  and $R_\mathrm{pb}$ are respectively the 
equatorial  and polar radius at  the break--up velocity.
If the star gets closer to the Eddington limit, then the following
critical velocity has to be used:

\begin{eqnarray}
v_\mathrm{crit, 2}^2 =
\frac{9}{4} \;v_\mathrm{crit, 1}^2 \; 
\frac{1-\Gamma_{\mathrm{max}}}{V^{\prime}(\omega)} \; \frac{
R^{2}_\mathrm{e}(\omega)}{R^2_{\mathrm{pb}}} \; ,
\end{eqnarray}

\noindent
where R$_\mathrm{e}(\omega)$ is the equatorial radius for a given 
value of the rotation parameter $\omega$.
More details can be found in \citet{ROTVI}.
Towards lower metallicities, $\upsilon_{\rm ini}/ \upsilon_{\rm crit}$ 
 increases only as $r^{-1/2}$ for
models with the same angular momentum ($J$), whereas the
surface rotational velocity increases as $r^{-1}$ 
($J\sim \upsilon r$).
The angular momentum varies significantly
for models with different initial masses.
Finally,
$\upsilon_{\rm ini}/ \upsilon_{\rm crit}$ is a good indicator for the
impact of rotation on mass loss.
  
In the first series of models, the aim is to scan the parameter space of
rotation and metallicity with 20 $M_\odot$ models since a 20 $M_\odot$
star is not far from the average massive star when a \citet{Sa55} like 
IMF is used. 
For this series, on top of non-rotating models,
two initial rotational velocities were 
used at very low metallicities. 
The first velocity is the same as at solar metallicity, {\it i. e.}
300\,km\,s$^{-1}$.
The ratio $\upsilon_{\rm ini}/ \upsilon_{\rm crit}$ decreases with
metallicity (see Table \ref{table1}) 
for the initial velocity of 300\,km\,s$^{-1}$.
The second $\upsilon_{\rm ini}$ is 
500\,km\,s$^{-1}$ at Z=10$^{-5}$ ([Fe/H]$\sim $-3.6) and 600\,km\,s$^{-1}$
at Z=10$^{-8}$ ([Fe/H]$\sim$-6.6). These values have
ratios of the initial velocity to the break--up velocity, 
$\upsilon_{\rm ini}/\upsilon_{\rm crit}$ around 0.55, which is only slightly
larger than the solar metallicity value (0.44). 
The 20 $M_\odot$ model at Z=10$^{-8}$
 and with 600\,km\,s$^{-1}$ has a total initial angular momentum $J_{\rm
tot}=3.3\,10^{52}$\,erg\,s, which is the same as for 
the solar metallicity 20 $M_\odot$ model with 300\,km\,s$^{-1}$ ($J_{\rm
tot}=3.6\,10^{52}$\,erg\,s).
So even though a star at Z=10$^{-8}$ with a velocity of 
600\,km\,s$^{-1}$ at first glance seems to be an extremely fast rotator, 
it is in fact similar to a solar metallicity star in terms of
angular momentum and ratio $\upsilon_{\rm ini}/ \upsilon_{\rm crit}$.
In the second series of models, following the work of
\citet{MEM06}, models were computed at Z=10$^{-8}$ with initial masses of
40, 60 and 85 $M_\odot$ and initial rotational velocities of 700, 800 and
800\,km\,s$^{-1}$ respectively. Note that, for these models as well, the
initial total angular momentum is similar to the one contained in solar 
metallicity models with rotational velocities equal to 300\,km\,s$^{-1}$.

\begin{table*}
\caption{Initial parameters of the models (columns 1--5): 
mass, metallicity, rotation velocity [km\,s$^{-1}$], 
total angular momentum [$10^{53}$\,erg\,s] and $\upsilon_{\rm ini}/ \upsilon_{\rm crit}$.
Lifetimes [Myr] (6--8): total and core hydrogen and helium burnings.
Various masses [$M_\odot$] (9--14): final mass, masses of the helium, carbon--oxygen,
silicon and iron cores and the remnant mass.}
\begin{tabular}{r r r r r r r r r r r r r r}
\hline \hline 
$M_{\rm{ini}}$ & $Z_{\rm{ini}}$ & $\upsilon_{\rm{ini}}$ & $J_{\rm{tot}}^{\rm{ini}}$ & $\upsilon_{\rm ini}/ \upsilon_{\rm crit}$
& $\tau_{\rm{life}}$ & $\tau_{\rm{H}}$ & $\tau_{\rm{He}}$ 
& $M_{\rm{final}}$ & $M_{\alpha}$ & $M_{\rm{CO}}$ & $M_{\rm{Si}}$ &
$M_{\rm{Fe}}$ & $M_{\rm{rem}}^a$  \\ 
\hline
20 & 2e-2 & 300 & 0.36 & 0.44 & 11.0 & 10.1 & 0.798 &  8.7626 & 8.66 & 6.59 & 2.25 & 1.27 & 2.57 \\
20 & 1e-3 & 000 &  --  & 0.00 & 10.0 & 9.02 & 0.875 & 19.5567 & 6.58 & 4.39 & 1.75 & 1.16 & 2.01 \\
20 & 1e-3 & 300 & 0.34 & 0.39 & 11.5 & 10.6 & 0.813 & 17.1900 & 8.32 & 6.24 & 2.27 & 1.33 & 2.48 \\
\hline
20 & 1e-5 & 000 &  --  & 0.00 & 9.80 & 8.86 & 0.829 & 19.9795 & 6.24 & 4.28 & 1.67 & 1.18 & 1.98 \\
20 & 1e-5 & 300 & 0.27 & 0.34 & 11.1 & 10.2 & 0.806 & 19.9297 & 7.90 & 5.68 & 1.99 & 1.30 & 2.34 \\
20 & 1e-5 & 500 & 0.42 & 0.57 & 11.6 & 10.6 & 0.812 & 19.5749 & 7.85 & 5.91 & 2.18 & 1.35 & 2.39 \\
\hline
20 & 1e-8 & 000 &  --  & 0.00 & 8.96 & 8.24 & 0.598 & 19.9994 & 4.43 & 4.05 & 1.91 & 1.05 & 1.92 \\
20 & 1e-8 & 300 & 0.18 & 0.28 & 9.98 & 9.20 & 0.610 & 19.9992 & 6.17 & 5.18 & 1.96 & 1.29 & 2.21 \\
20 & 1e-8 & 600 & 0.33 & 0.55 & 10.6 & 9.71 & 0.703 & 19.9521 & 4.83 & 4.36 & 2.01 & 1.29 & 2.00 \\
\hline
09 & 1e-8 & 500 & 0.80 & 0.08 & 30.5 & 26.8 & 3.24  &  8.9995 & 1.90 & 1.34 &  --  &  --  & 1.21 \\
40 & 1e-8 & 700 & 1.15 & 0.55 & 5.77 & 5.31 & 0.402 & 35.7954 & 13.5 & 12.8 & 2.56 & 1.49 & 4.04 \\
60 & 1e-8 & 800 & 2.41 & 0.57 & 4.55 & 4.19 & 0.321 & 48.9747 & 25.6 & 24.0 &  --  &  --  & 7.38 \\
85 & 1e-8 & 800 & 4.15 & 0.53 & 3.86 & 3.50 & 0.322 & 19.8677 & 19.9 & 18.8 & 3.19 & 1.84 & 5.79 \\
\hline 
\end{tabular}
\\ $^a$ estimated from the CO core mass.
\label{table1}
\end{table*}

\section{Evolution}\label{evo}

The evolution of the models was in general followed until core Si--burning. 
The 60 $M_\odot$ model was evolved until neon burning and the 9 $M_\odot$
model until carbon burning. The main characteristics of the models are presented 
in Table \ref{table1}.

\begin{figure*}[!tbp]
\centering
\includegraphics[width=6cm]{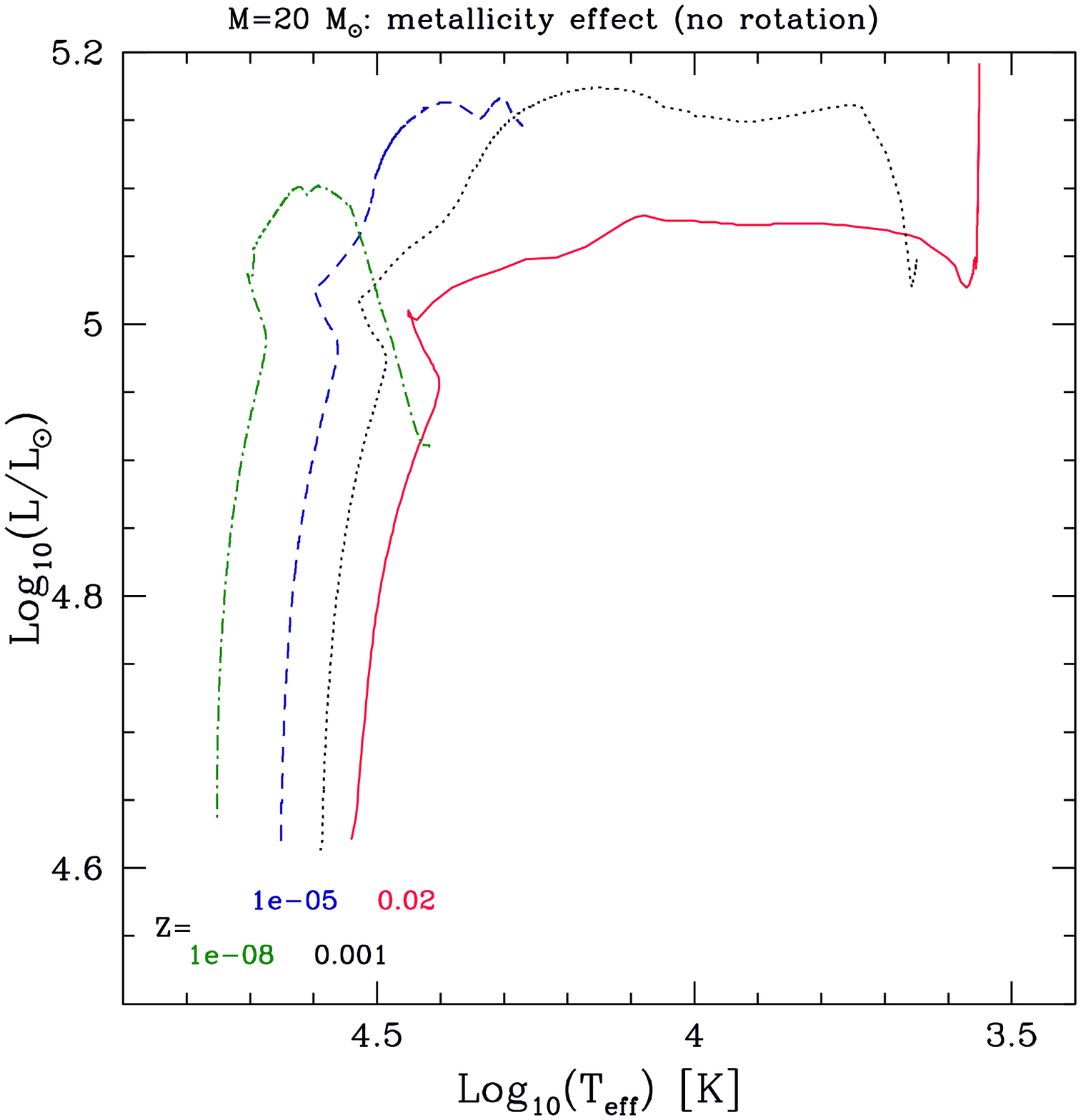}\includegraphics[width=6cm]{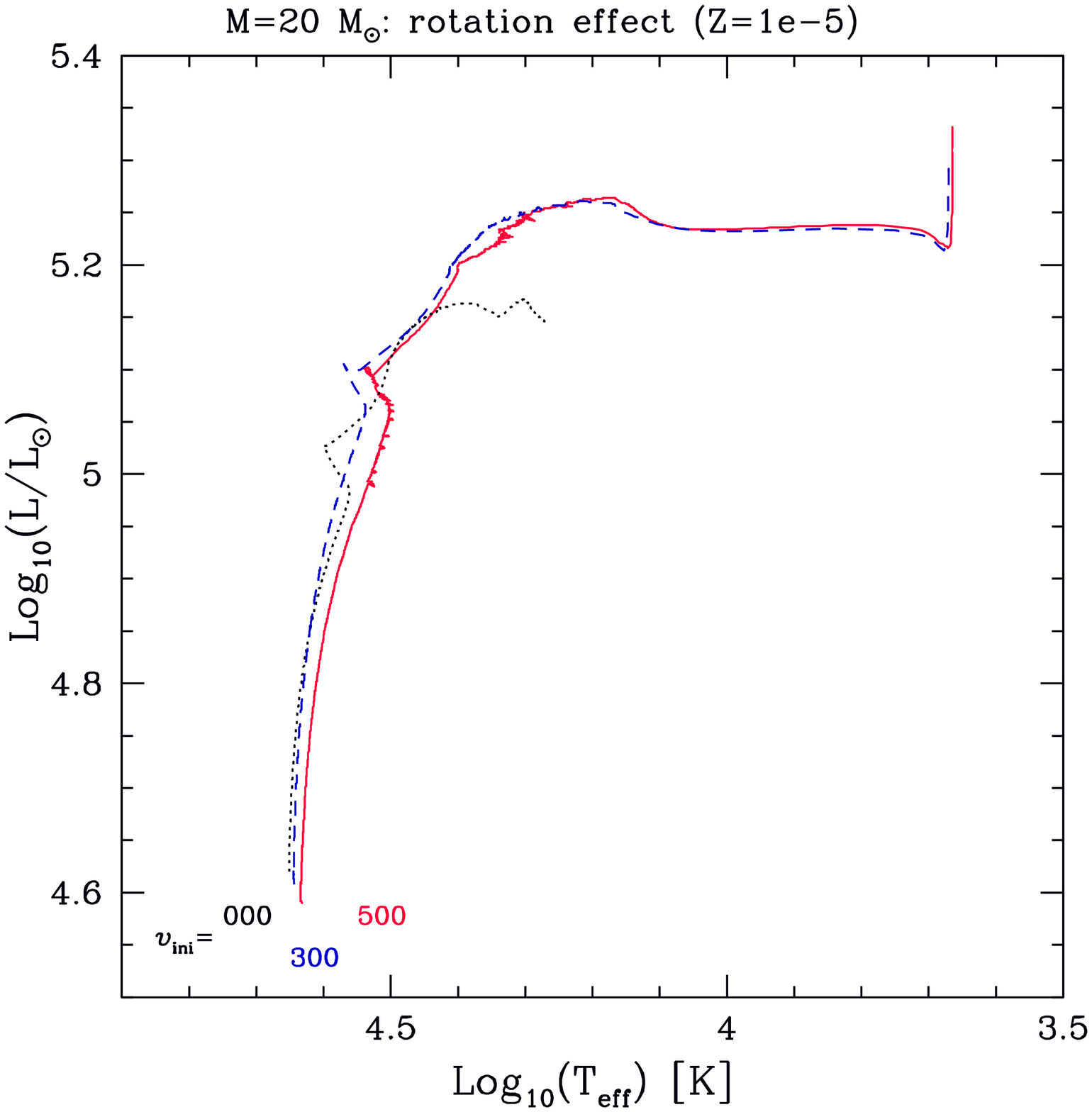}\includegraphics[width=6cm]{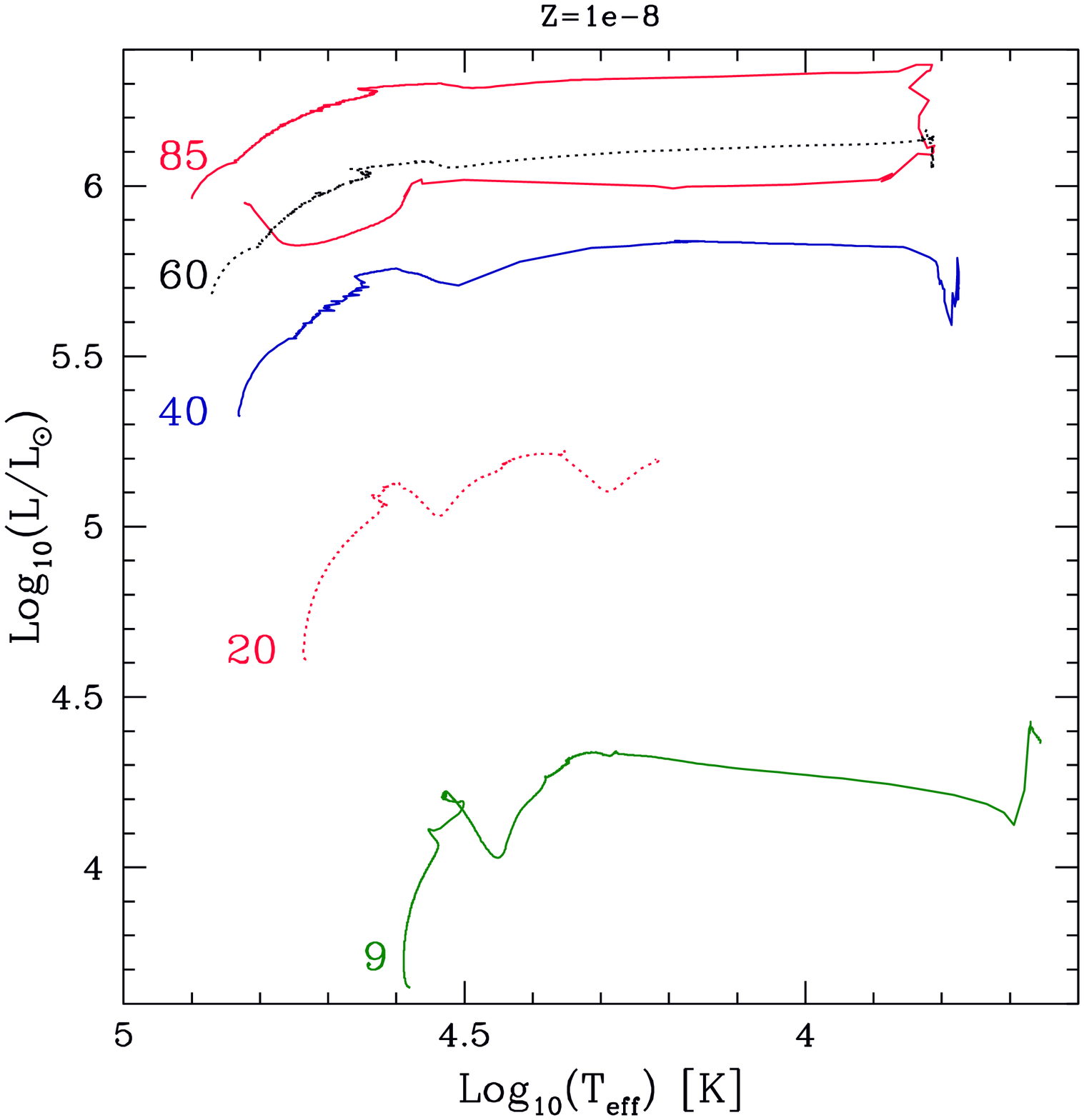}
\caption{HR diagrams: ({\it left}) non--rotating 20 $M_\odot$
models, ({\it middle}) 20 $M_\odot$ models with $Z=10^{-5}$ 
and ({\it right}) $Z=10^{-8}$ models with average rotation.}
\label{hrd}
\end{figure*}
\subsection{Metallicity effects}
The effects of metallicity on stellar evolution have already been
discussed in the literature 
\citep[see for example][]{HFWLH03,CL04,MM94}.
But before looking at the very low metallicity models and the impact of
rotation, it is useful to summarise the effects low metallicity has on the
evolution of massive stars. 
A lower metallicity implies a lower luminosity which leads to slightly smaller
convective cores. This can be seen in Table \ref{table1} by comparing 
the core masses of 
non--rotating the 20 $M_\odot$ models at different metallicities. 
A lower metallicity implies lower opacity and lower
mass losses (as long as the chemical composition has not been changed by
burning or mixing in the part of the star one considers). So at the start
of the evolution lower metallicity stars are more compact. This can be
seen in the Herzsprung--Russell (HR) diagram (Fig. \ref{hrd} {\it left})
where the lower metallicity models have bluer tracks during the main
sequence.
They also lose less 
mass as can be seen by looking at the final masses in Table \ref{table1}. 
The lower metallicity models also have a harder time reaching
the red supergiant (RSG) stage 
\citep[see][ for a detailed discussion]{ROTVII}. The non--rotating model 
at $Z=10^{-3}$
becomes a RSG only during shell He--burning (see Fig. \ref{kip}) and the
lower metallicity non--rotating models never reach the RSG stage. As long as the
metallicity is above about $Z=10^{-10}$, no significant differences have
been found in non--rotating models. Below this metallicity and for metal
free stars, the CNO cycle cannot operate at the start of H--burning. At
the end of its formation, the
star therefore contracts until it starts He-burning because the
pp--chains cannot balance the effect of the gravitational force. Once
enough carbon and oxygen are produced, the CNO cycle can operate and the
star behaves like stars with $Z>10^{-10}$ for the rest of the main
sequence. Shell H--burning still differs between $Z>10^{-10}$ and metal
free stars. Metal free stellar models are presented in \citet{CL04},
\citet{HW02} and \citet{UN05}.

\subsection{Rotation effects}
\begin{figure*}[!tbp]
\centering
\includegraphics[width=5.5cm]{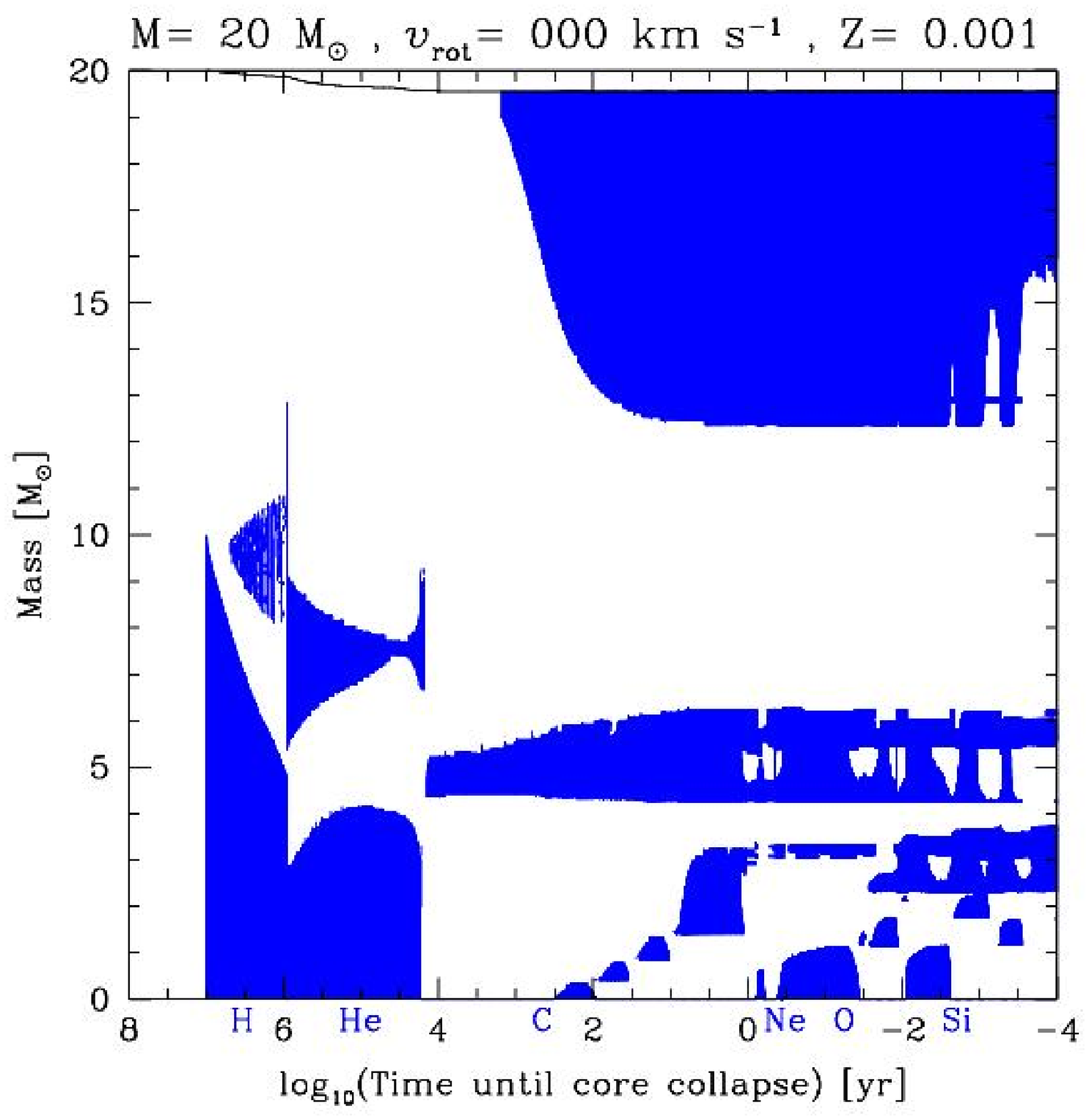}\includegraphics[width=5.5cm]{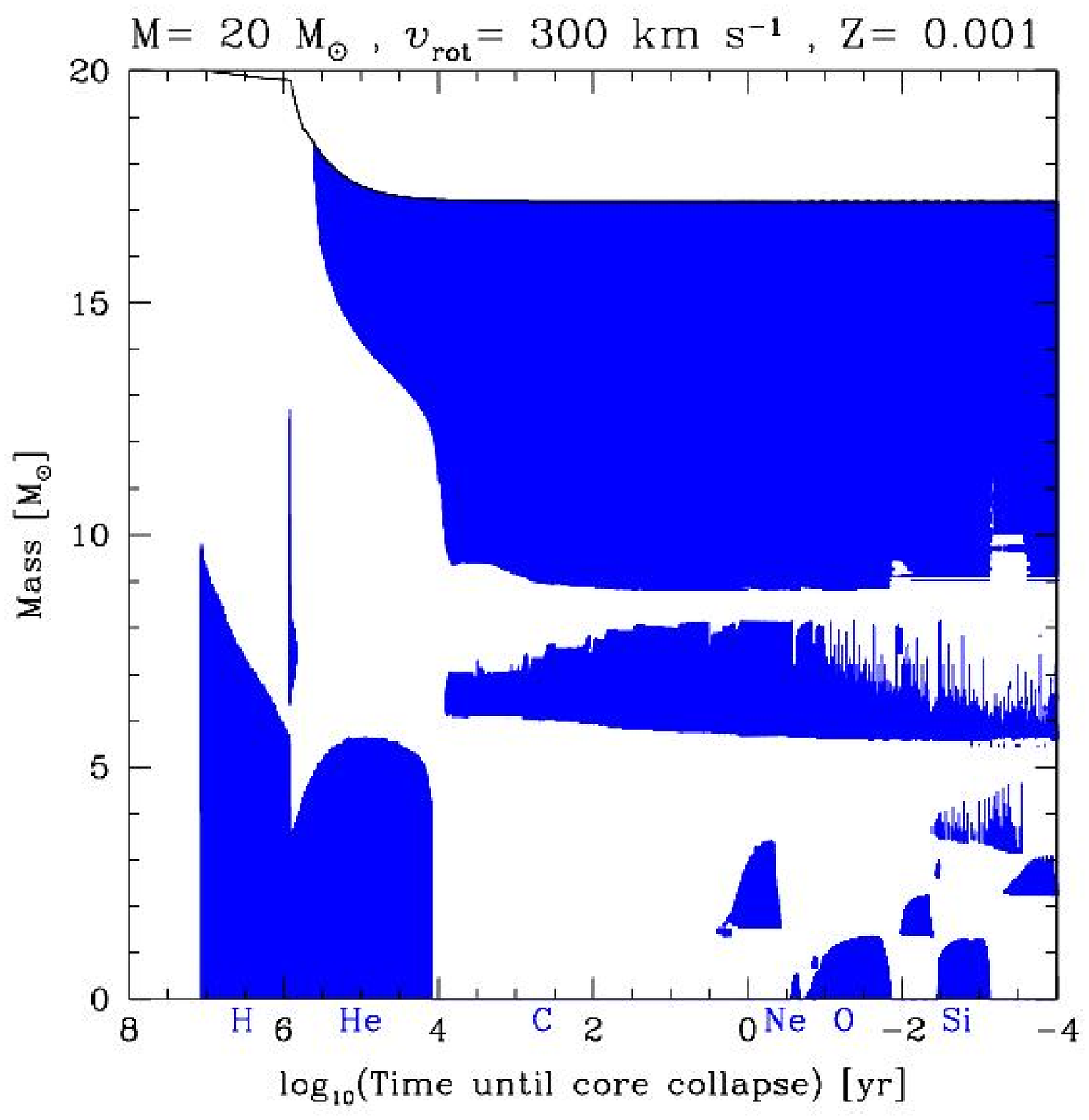}\includegraphics[width=5.5cm]{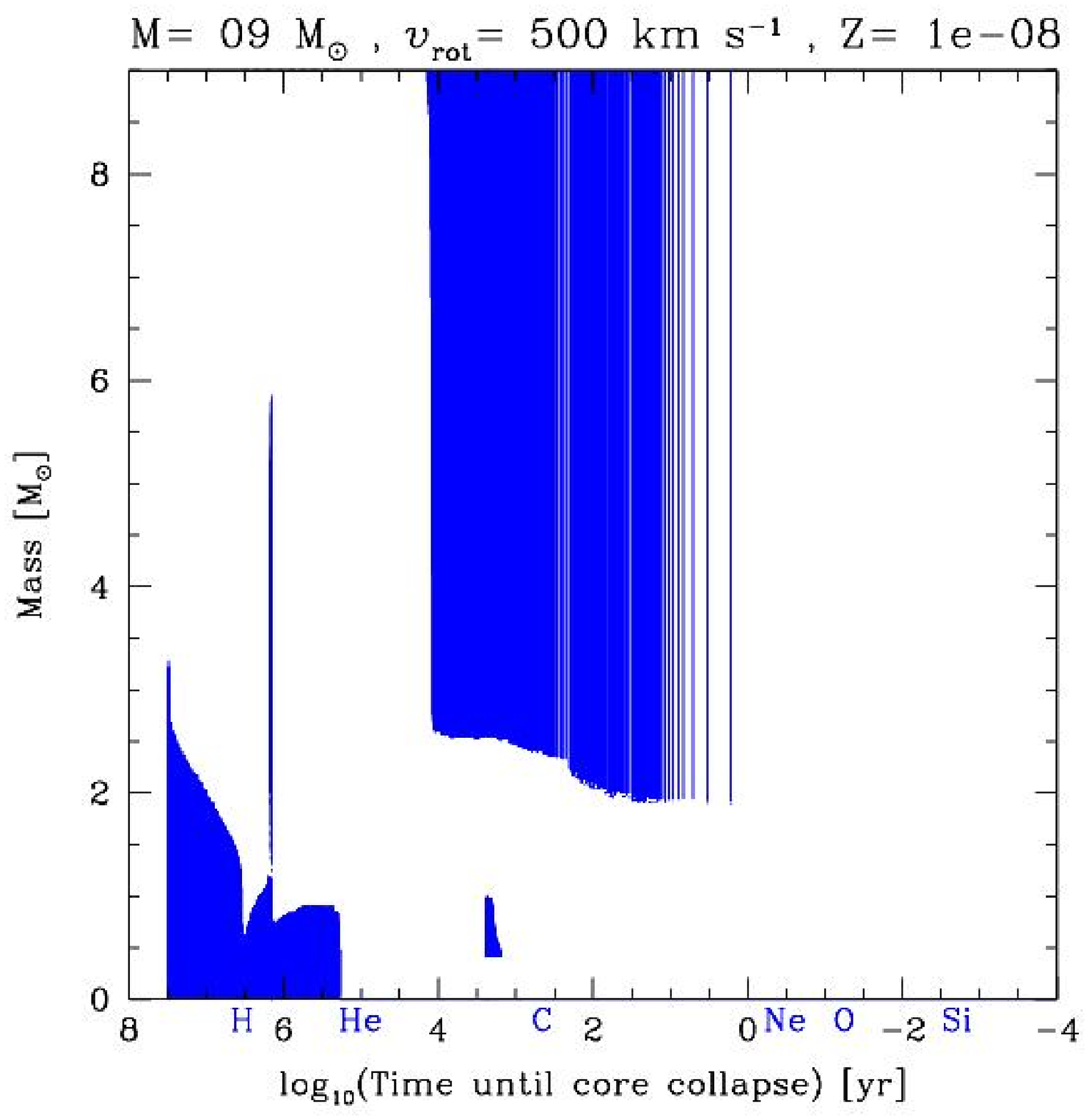}
\includegraphics[width=5.5cm]{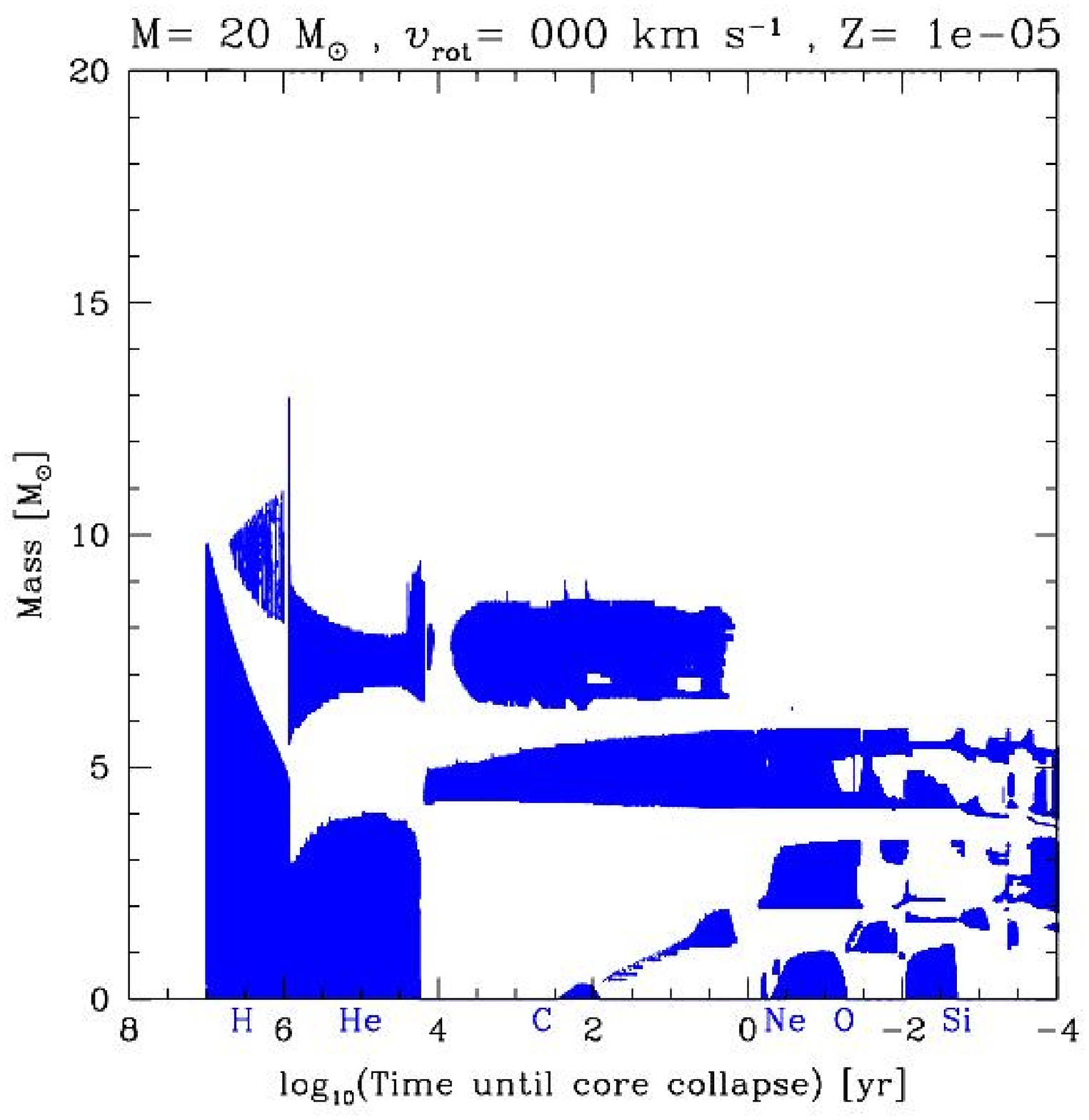}\includegraphics[width=5.5cm]{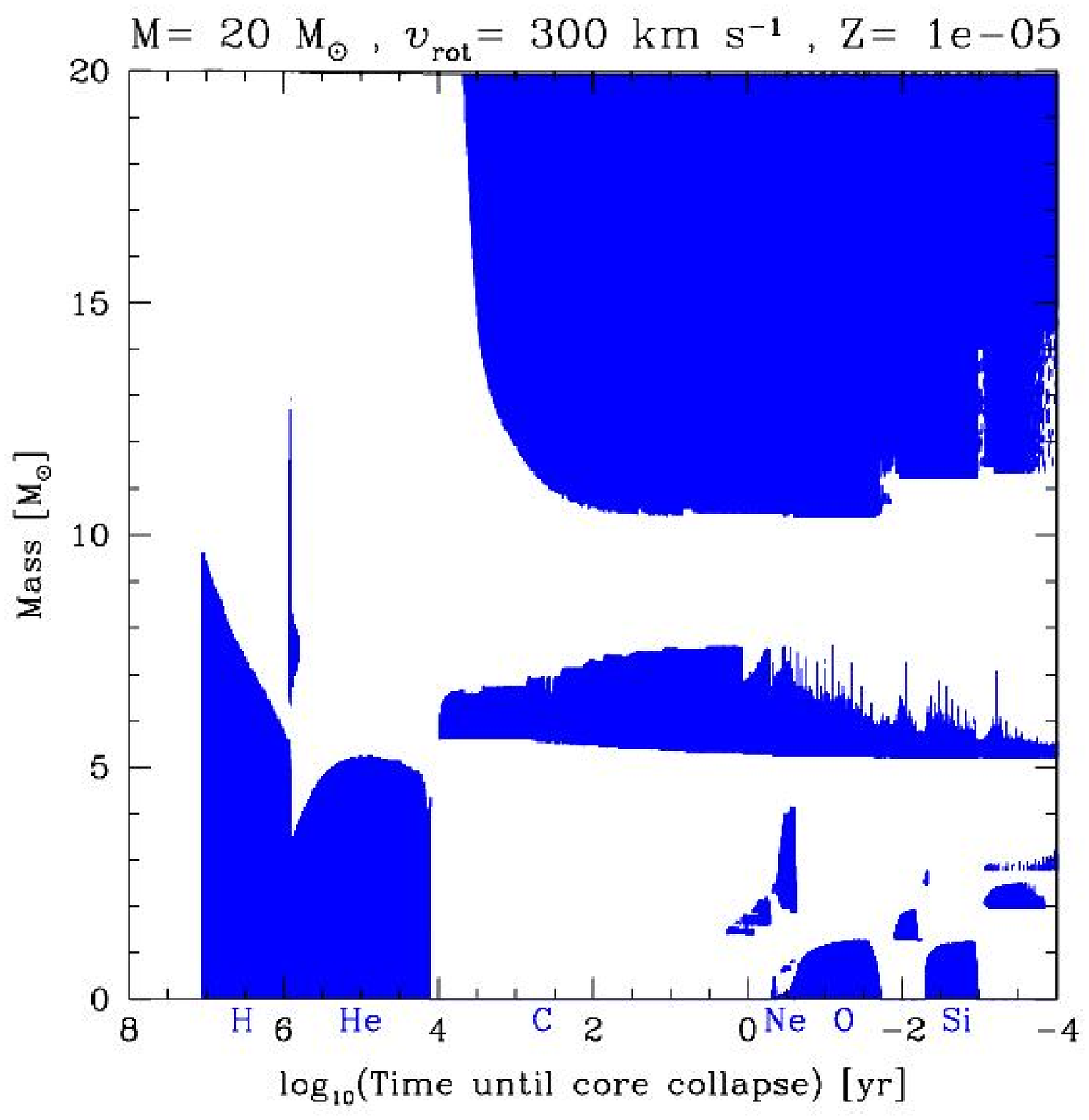}\includegraphics[width=5.5cm]{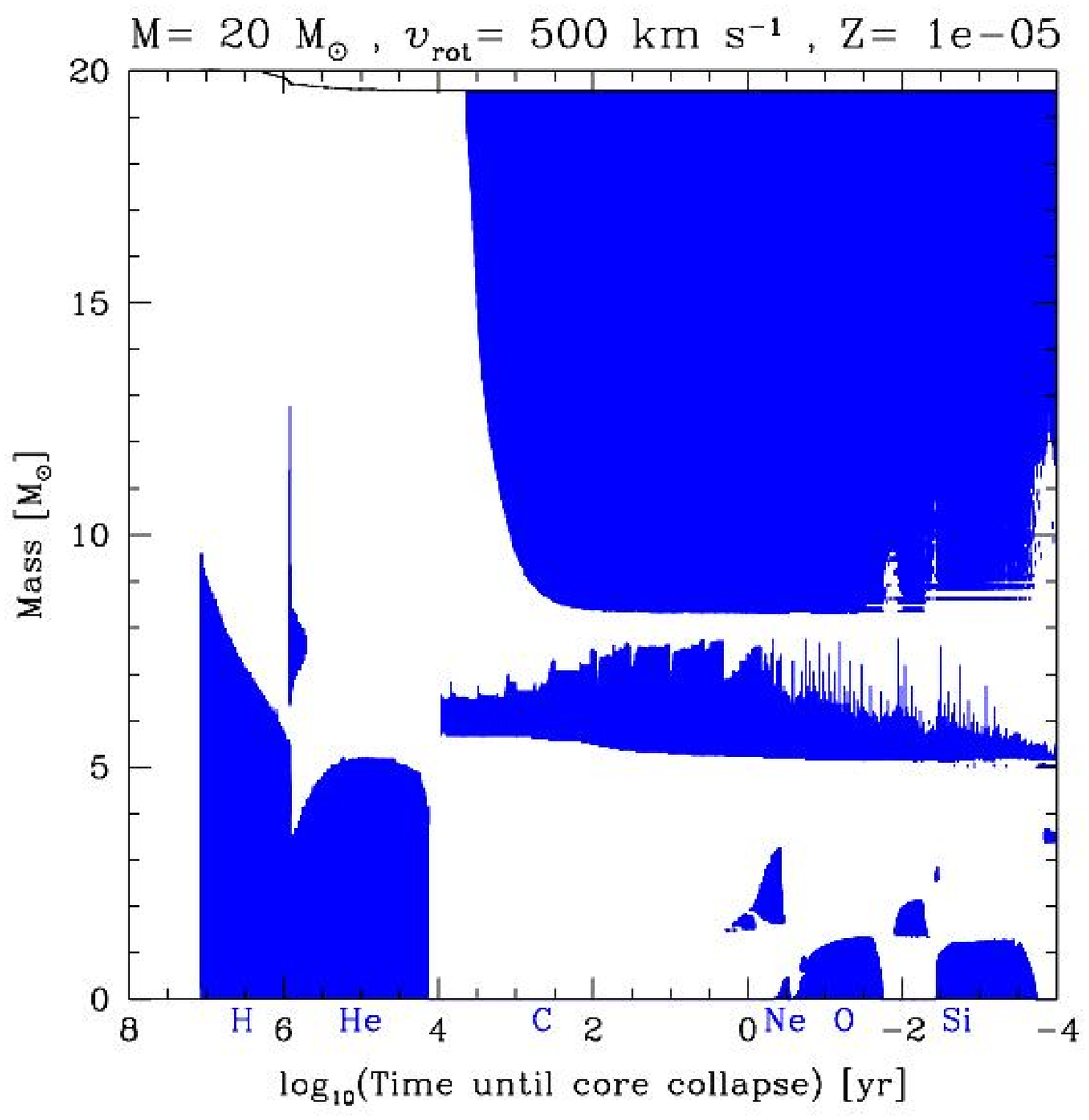}
\includegraphics[width=5.5cm]{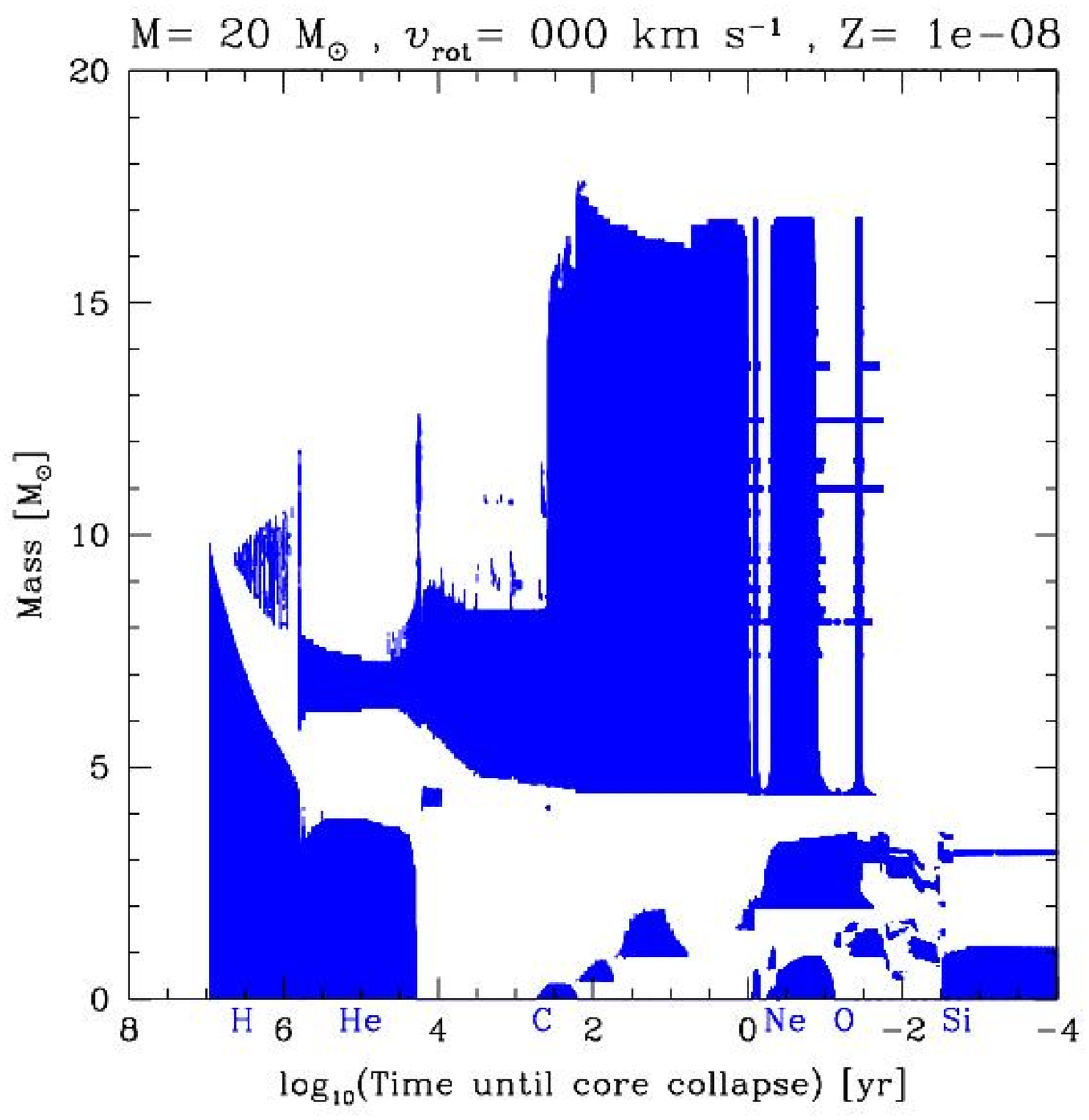}\includegraphics[width=5.5cm]{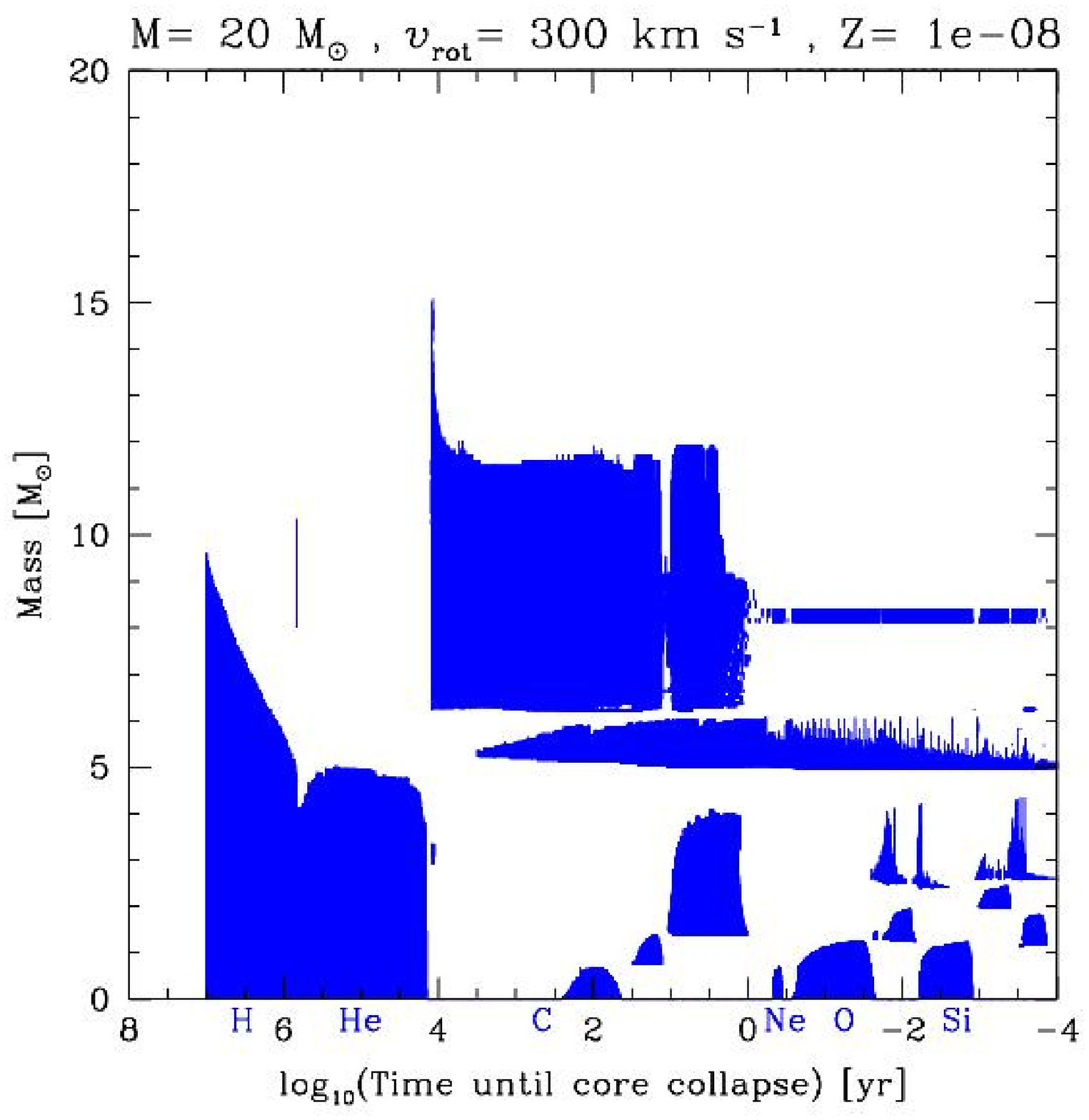}\includegraphics[width=5.5cm]{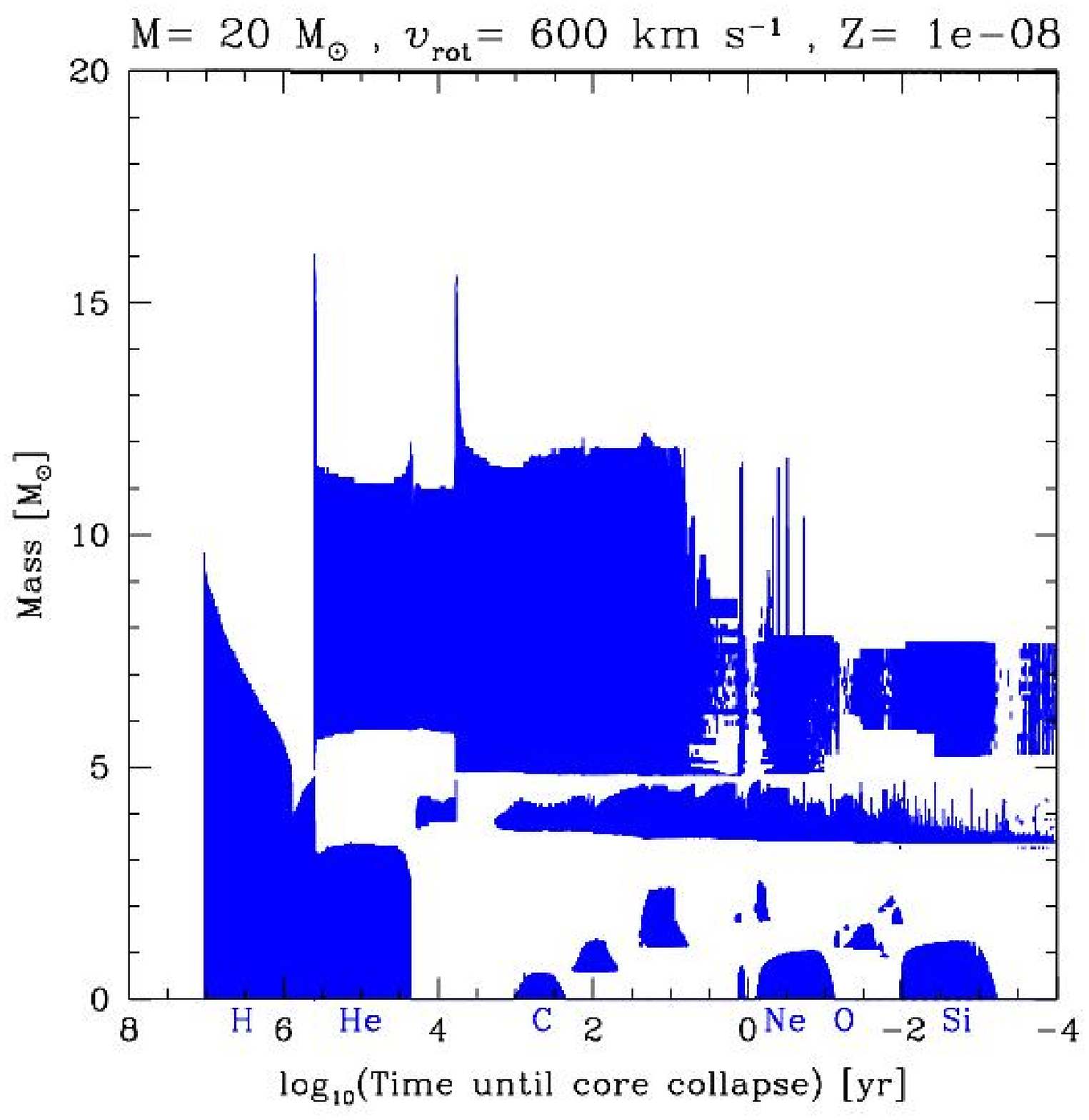}
\includegraphics[width=5.5cm]{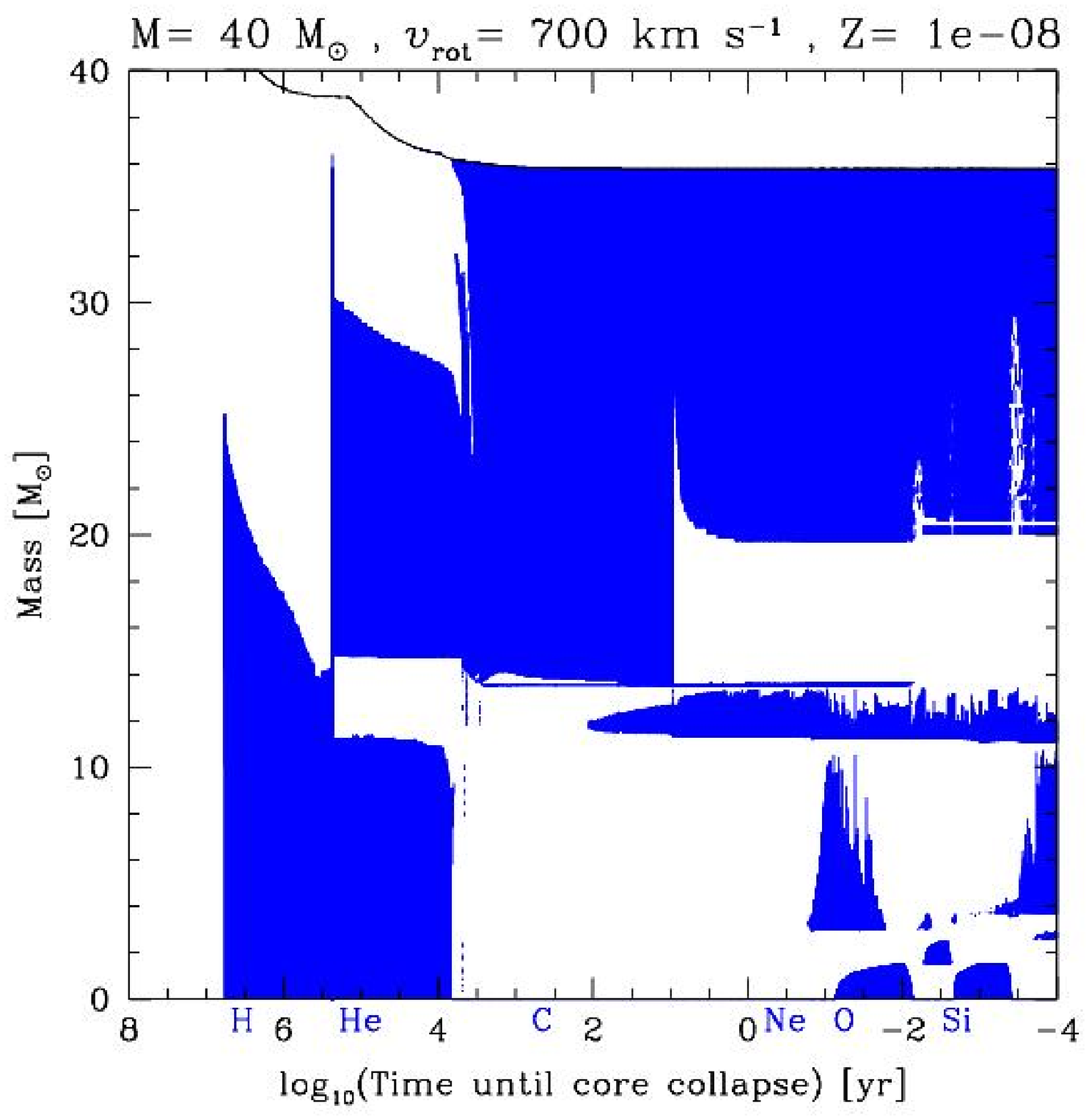}\includegraphics[width=5.5cm]{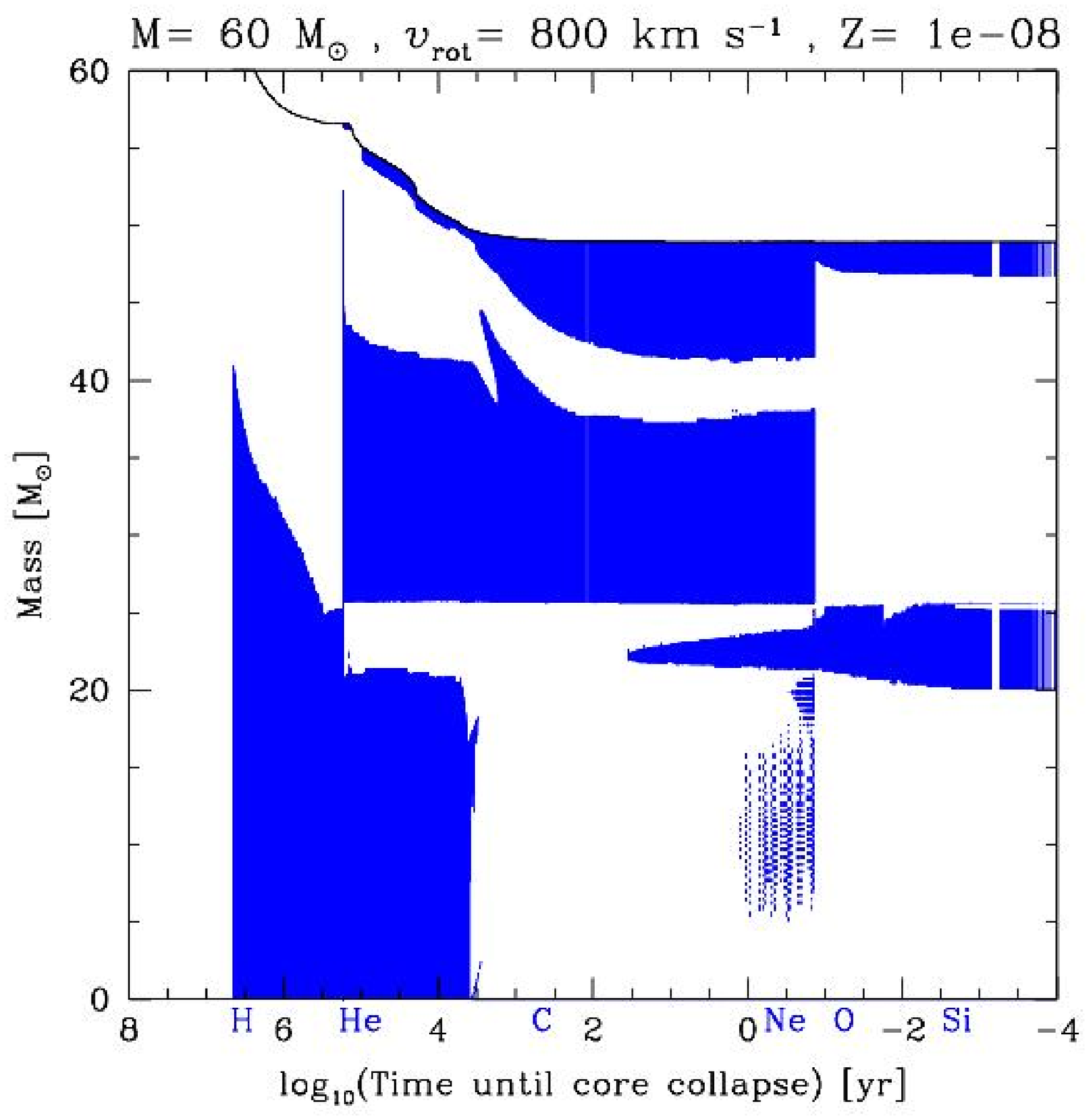}\includegraphics[width=5.5cm]{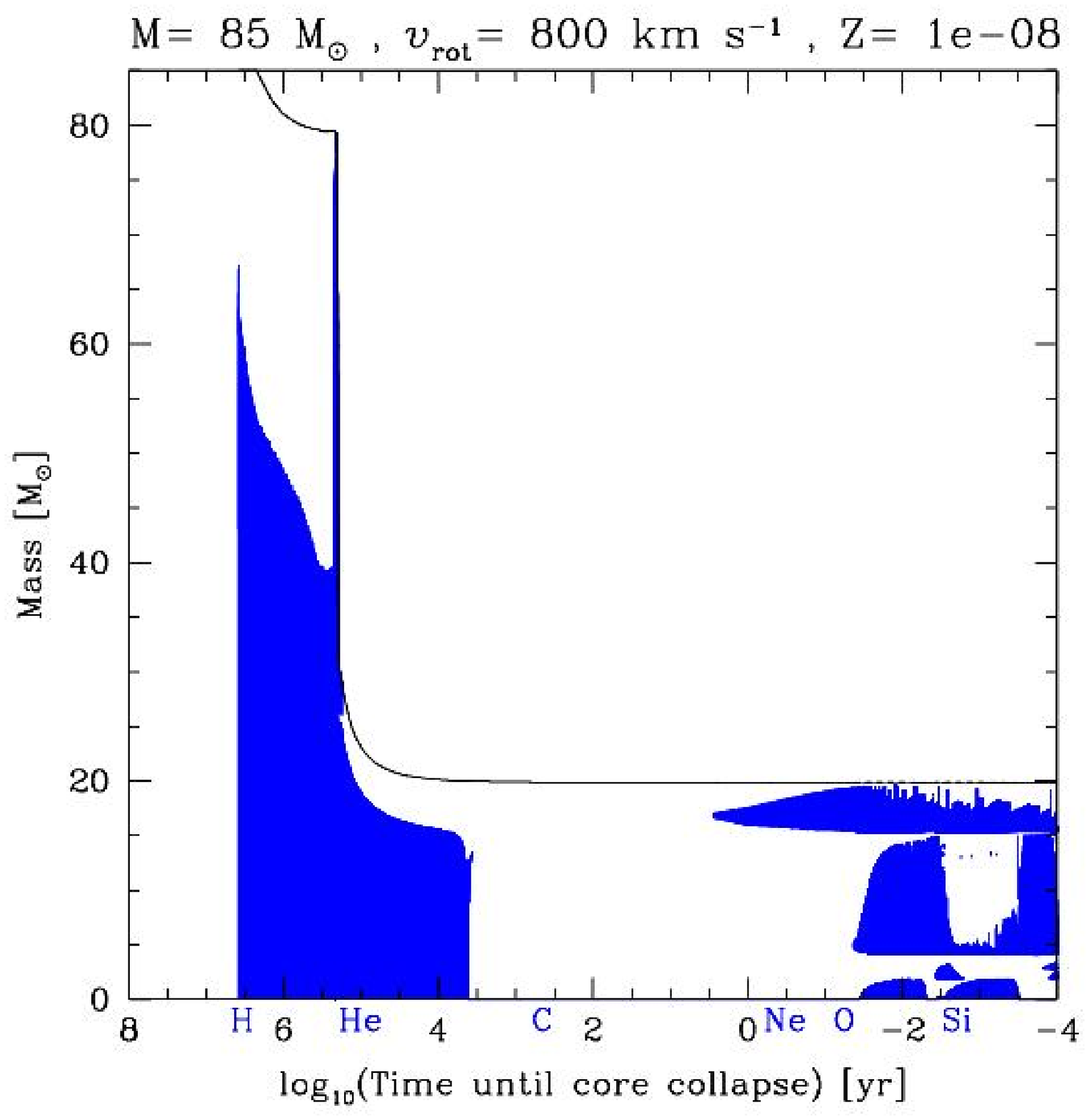}
\caption{Stellar structure (Kippenhahn) diagrams, which show the
evolution of the structure as a function of the time left until the core
collapse. 
The initial parameters of the models are given on top of each plot.
The coloured zone correspond to the convective zones and the symbols of
the burning stages are given below the time axis.}
\label{kip}
\end{figure*}

How does rotation change this picture? At all metallicities, rotation 
usually increases the core sizes, the lifetimes, the
luminosity and the mass loss.
\citet{ROTVII} and \citet{ROTVIII} have
already studied the impact of rotation down to a metallicity of
$Z=10^{-5}$. They find in \citet{ROTVII} that rotation 
favours a redward evolution and that rotating models can reproduce the
observed ratio of blue to red supergiants (B/R) in the small Magellanic cloud.
We can see in Fig. \ref{hrd} ({\it middle}) that the rotating models
at $Z=10^{-5}$ become RSGs during shell He--burning (See Fig. \ref{kip}). 
This does not
change the ratio B/R but changes the structure of the star when the SN
explodes.
At $Z=10^{-8}$ (Fig. \ref{hrd} {\it right}), the 20 $M_\odot$ models do not become RSG. However other
mass models do reach the RSG stage and the 85 $M_\odot$ model even
becomes a WR star of type WO (see below).
\citet{ROTVII} also find that a larger fraction of stars reach break-up velocities
during the evolution. This will be further discussed in Sect. \ref{mdotm}).

In \citet{ROTVIII}, they show that low metallicity ($Z=10^{-5}$)
models have strong internal $\Omega$--gradients, which favours an
important mixing. This mixing leads to primary nitrogen production during
He--burning by rotational diffusion of carbon and oxygen into the
H-burning shell. Their results already point out that the primary
nitrogen yields strongly depend on the initial rotation velocity. 
This dependence is further studied in Sect. \ref{yields}.

\citet{MEM06}  present the evolution of 60 $M_\odot$ models at
$Z=10^{-8}$ and confirm the effects that were predicted in their
previous papers. The fast rotating model, with an initial rotational velocity
of 800\,km\,s$^{-1}$ reaches break-up, becomes a red supergiant and 
produces significant amount of primary
nitrogen. The model becomes a Wolf--Rayet
(WR) star due to large mass losses during the RSG stage. These effects
are further discussed below and models at $Z=10^{-8}$ with different
initial masses are presented.

\subsection{Strong impact of mixing at very low metallicities}\label{mix}
\begin{figure}[!tbp]
\centering
\includegraphics[width=8cm]{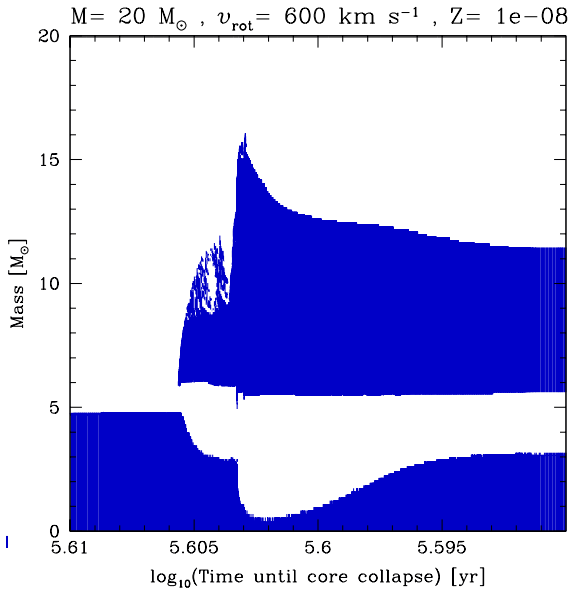}
\caption{Stellar structure (Kippenhahn) diagram of the 20 $M_\odot$ with 
$\upsilon_{\rm ini}=$600\,km\,s$^{-1}$ at $Z=10^{-8}$. The central
convective zone is created by core He--burning and the intermediate
convective zone is created by shell H--burning.
This is a zoom around the time when the shell H--burning is boosted due
to strong mixing of carbon and oxygen from the core.
The graph shows the apparition of the convective zone induced by shell
H--burning and that the mass of the helium burning core is smaller after
than before the H--burning shell boost.}
\label{kipz20}
\end{figure}

At solar metallicity and metallicities higher than about
$Z=10^{-5}$, rotational mixing increases the helium and CO core masses
(see Table \ref{table1}). In particular, the oxygen yield is increased.
The impact of mixing on models at $Z=10^{-8}$ 
\citep[and at $Z=0$ see][]{EMM06} 
is however different for fast rotation
($\upsilon_{\rm ini}=$600-800\,km\,s$^{-1}$). 
The impact of mixing on the structure and convective zones is 
represented
in the Kippenhahn diagram for the $Z=10^{-8}$ models (see Fig. \ref{kip}). 
During hydrogen burning and the start of helium burning,
the impact of mixing is the same as at higher metallicity: mixing
increases the core sizes and mixing of helium above the core suppresses
the intermediate convective zones linked to shell H--burning.
The difference from higher metallicity models occurs during He--burning.
As shown in Fig. \ref{ab20} ({\it left}) for the 20 $M_\odot$ with 
$\upsilon_{\rm ini}=$600\,km\,s$^{-1}$, primary carbon and oxygen are
mixed outside of the convective core into the H--burning shell. Once the
enrichment is strong enough, the H--burning shell is boosted (the CNO
cycle depends strongly on the carbon and oxygen mixing at such low
initial metallicities). The shell then
becomes convective, as can be seen in Fig. \ref{kipz20}, which is a zoom
of the Kippenhahn diagram. 
The boost phase, which could look like an instantaneous event in Fig.
\ref{kip} ({\it third line, right}), is fully revealed in this zoom. 
The calculations have been
repeated with different time steps to 
verify that the results did not depend on the numerical details.
The evolution of the final model was followed 
with 2000 time steps between the time axis measures of 5.607 and 5.596.

In response to the shell boost, the core
expands and the convective core mass decreases.
At the end of He--burning, the CO core is less massive than in 
the non--rotating
model (Fig. \ref{kip}, {\it third line, left and right}).
The yield of $^{16}$O being closely
correlated with the mass of the CO core, it is therefore
reduced due to the strong mixing. 
At the same time the carbon yield is slightly increased (See Table
\ref{ytot}).
It is interesting to note that the shell H--burning boost occurs in all
the different initial mass models at $Z=10^{-8}$ as can be seen in Fig.
\ref{kip} (noticeable by the appearance of a strong intermediate
convective shell due to H--burning and the corresponding sharp 
decrease in the He--burning convective core mass).
This means that the relatively "low" oxygen yields and "high" carbon
yields are produced over a large mass range at $Z=10^{-8}$. This could
be an explanation for the possible high [C/O] ratio observed in the most
metal poor halo stars \citep[ratio between the surface
abundances of carbon and oxygen relative to solar; see Fig. 14 in][]{FS6}.

Figure \ref{ab20} ({\it left}) shows the abundance profiles before the
shell H--burning boost and Fig. \ref{ab20} ({\it middle}) shows the
profiles after. These profiles show that the carbon and oxygen brought
to the shell H--burning are transformed into primary nitrogen. 
The bulk of
primary nitrogen is however produced later in the evolution. 
It happens
when the shell H--burning deepens in mass during core helium burning for the very massive models
and during shell He--burning for the 20 $M_\odot$ model. 
The deepening of the convective H--burning shell is caused
by dynamical shear instabilities 
\citep[see Sect. 2.3 in][and references therein]{psn04a}.
Dynamical shear instabilities take place just below the bottom of the 
convective zone due to the strong differential rotation between the
convective H--burning shell and the layers below (deeper in the star). 
Dynamical shear
instabilities occur on a dynamical time scale as
opposed to secular shear and they therefore induce a fast mixing just below
the convective zone. 
Note that the dynamical shear instabilities are not influenced by mean molecular 
weight gradients.
The consequences are the same as during the first boost. Carbon
and oxygen are mixed inside the H--burning shell and a new boost occurs, 
producing this time more nitrogen because the newly mixed material is richer 
in carbon and oxygen.
Figure \ref{ab20} ({\it right}) shows the abundance profiles after the
second boost. The abundance of nitrogen does not change later during the
advanced stages and the pre--SN profile for nitrogen (see Fig. \ref{ab1})
stays the same. The total production of primary nitrogen is further
discussed in Sect. \ref{yields}. Rotational mixing also influences strongly the mass loss
of very massive stars as is discussed below.

\subsection{Mass loss}\label{mdotm} 
Mass loss becomes gradually unimportant as the metallicity decreases in
the 20 $M_\odot$ models. At solar metallicity, the rotating 20 $M_\odot$
model loses more than half of its mass, at $Z=0.001$, the models lose
less than 15\% of their mass, at $Z=10^{-5}$ less than 3\% and at
$Z=10^{-8}$ less than 0.3\% (see Table \ref{table1}). 
\citet{MEM06} show that the situation can be very different for a 60 $M_\odot$
star at $Z=10^{-8}$. Indeed, their 60 $M_\odot$ model loses about half 
of its initial mass. About ten percents of the initial
mass is lost when the surface of the star reaches break--up velocities
during the main sequence. The largest mass loss occurs during the red
supergiant (RSG) stage due to the mixing of primary carbon and oxygen
from the core to the surface through convective and rotational mixing.
The large mass loss is due to the fact that the star crosses the 
Humphreys-Davidson limit.

\begin{figure}[!tbp]
\centering
\includegraphics[width=4.5cm]{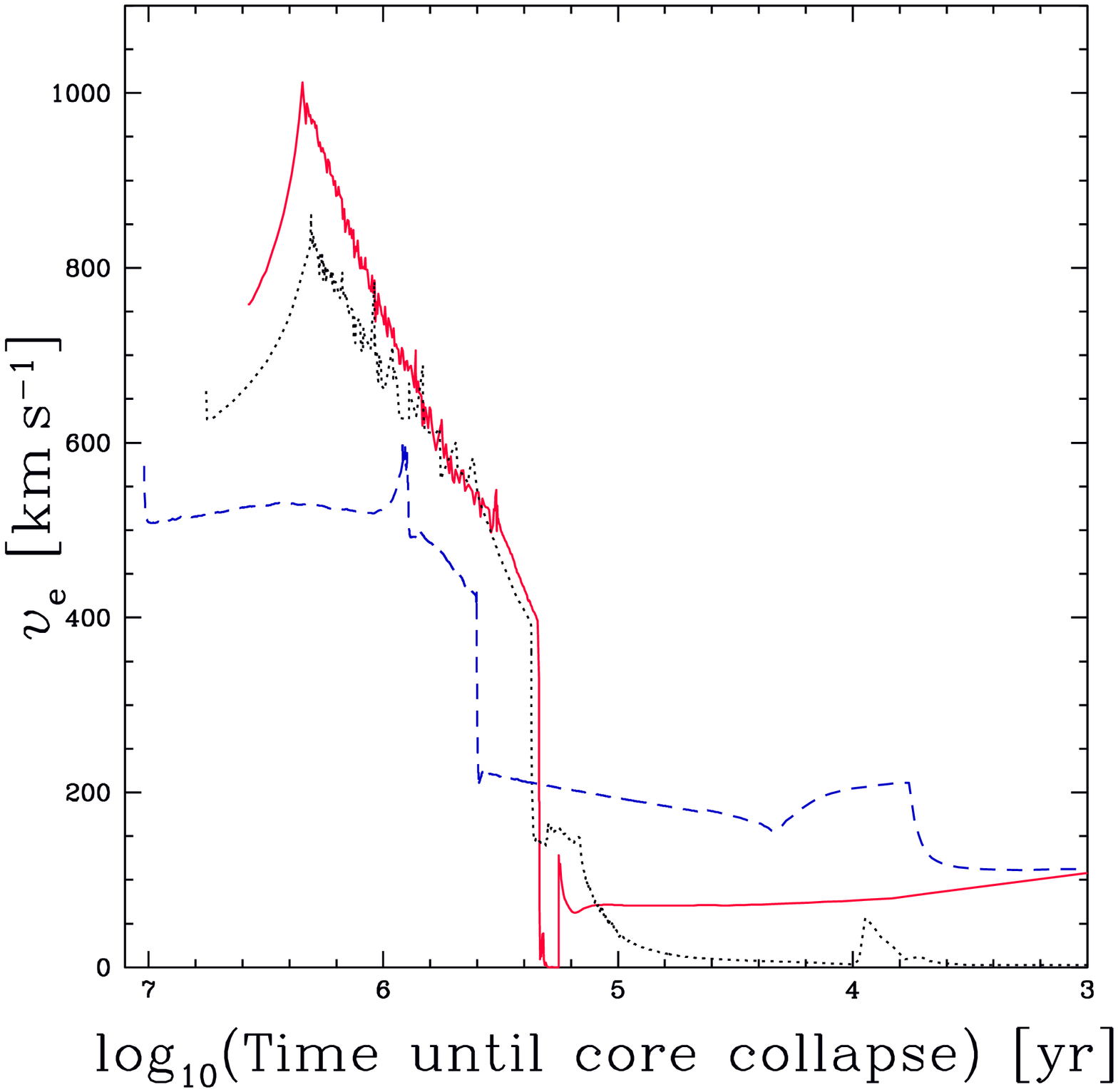}\includegraphics[width=4.5cm]{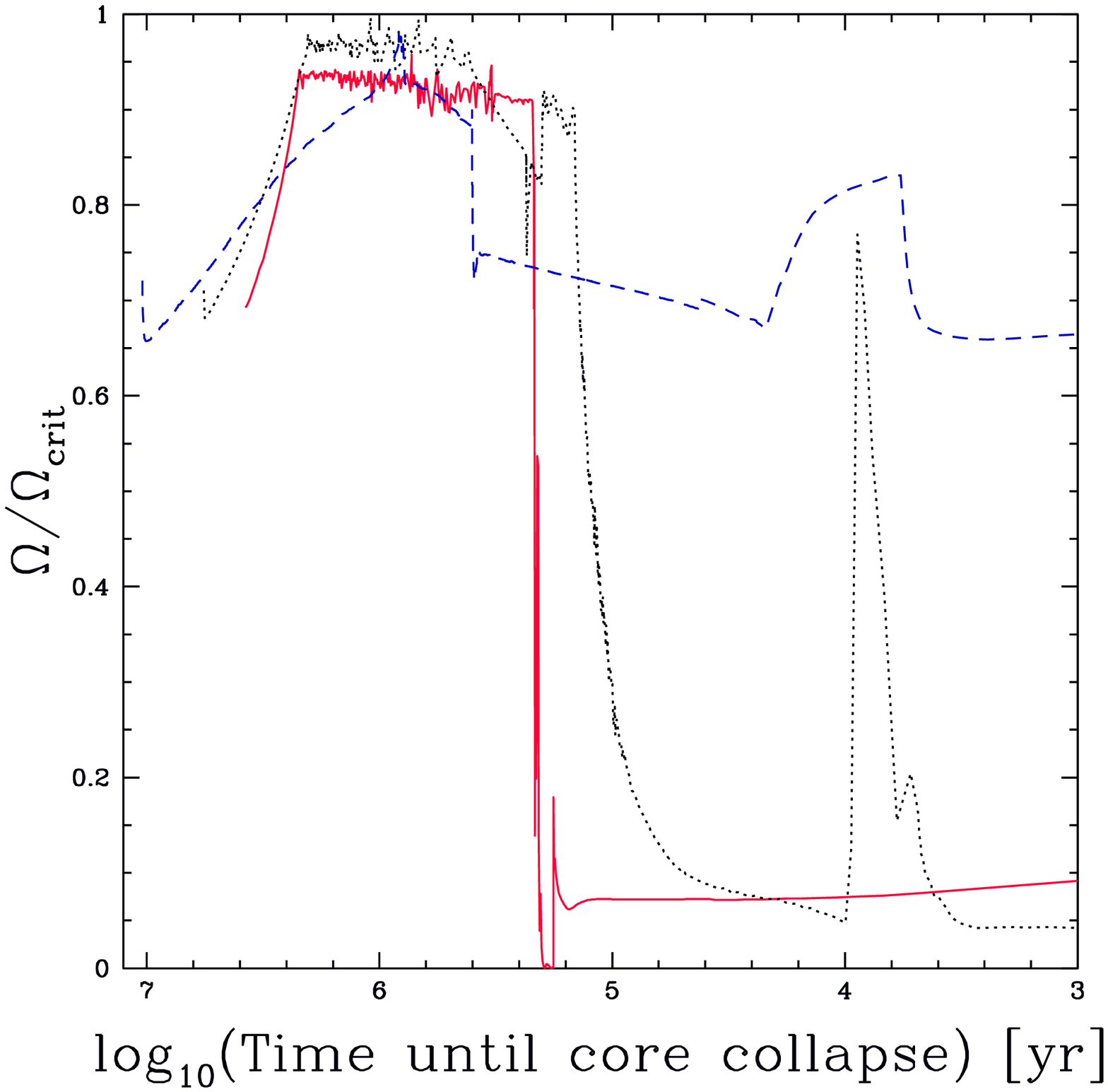}
\caption{Evolution of rotation for the 85 $M_\odot$ (red solid line), 40
$M_\odot$ (dotted black line) and 20 $M_\odot$ (dashed blue line)
models with fast rotation velocities, 
$\upsilon_{\rm ini}=$600-800\,km\,s$^{-1}$, at $Z=10^{-8}$: 
({\it left}) surface equatorial velocity, 
and ({\it right}) ratio of the surface angular velocity to the critical
angular velocity, $\Omega/\Omega_{\rm crit}$.}
\label{vev}
\end{figure}

\begin{figure}[!tbp]
\centering
\includegraphics[width=8cm]{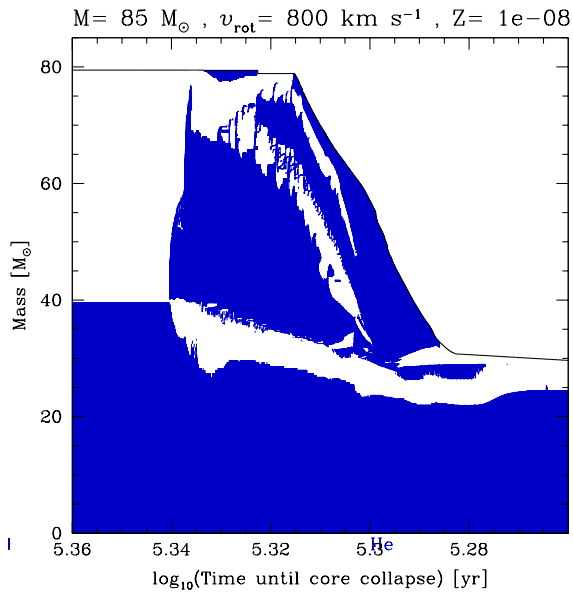}
\caption{Stellar structure (Kippenhahn) diagram of the 85 $M_\odot$ 
model at $Z=10^{-8}$. 
This is a zoom covering the period when C and O are transported from the
core to the surface and induce a strong mass loss. The central
convective zone is related to core He--burning. The main intermediate
convective zone is caused by shell H--burning and the surface convective
zone appears when the star becomes a RSG.}
\label{kipz85}
\end{figure}

\begin{figure}[!tbp]
\centering
\includegraphics[width=9cm]{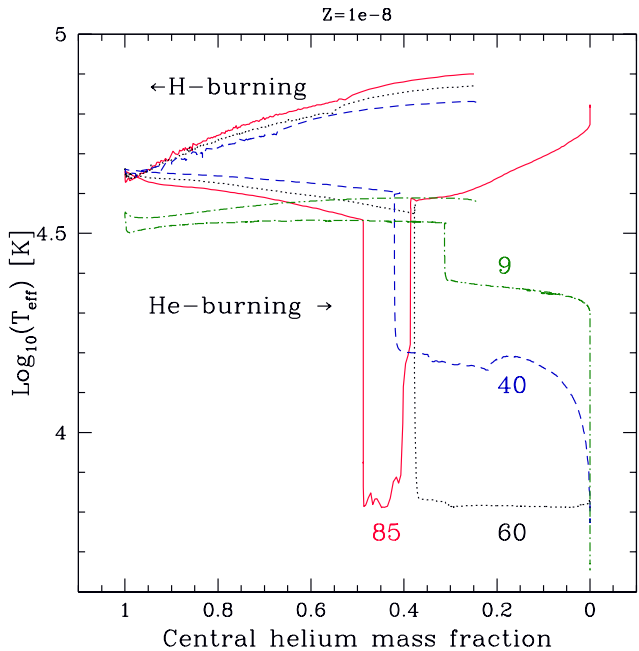}
\caption{Evolution of the effective temperature as a function of the
central helium mass fraction for the $Z=10^{-8}$ models.
Models evolve from the top right corner to the top left corner during
the MS. They
then evolve to the right during helium burning and downwards when they
become RSG. The 20 $M_\odot$ track resembles the 9 $M_\odot$ model track
except that it doesn't reach the RSG stage at the end of He--burning.}
\label{ycteff}
\end{figure}

What happens in the models calculated in this study?
First let us study the mass loss at break--up. Figure \ref{vev} presents
the evolution of the surface velocity and of the ratio of the surface angular velocity to the critical
angular velocity, $\Omega/\Omega_{\rm crit}$ for the 85 $M_\odot$ (red solid line), 40
$M_\odot$ (dotted black line) and 20 $M_\odot$ (dashed blue line)
models with fast rotation velocities, 
$\upsilon_{\rm ini}=$600-800\,km\,s$^{-1}$, at $Z=10^{-8}$.
It shows that the 20 $M_\odot$ model only reaches break--up velocities
at the end of the main sequence (MS) and therefore does not lose mass due to
this phenomenon. However, more massive models reach critical velocities
early during the MS (the earlier the more massive the model). The
evolution of rotation for the 60 $M_\odot$ model is very similar to the 40 $M_\odot$
model and is therefore not shown here for the clarity of the plot.
The mass lost due to break--up increases with the initial mass and
amounts to 1.1, 3.5 and 5.5 $M_\odot$ for the 40, 60 and 85 $M_\odot$
models respectively (see the top solid line in Fig. \ref{kip} going down
during the MS). At the end of core H--burning,
the core contracts and the envelope expands, thus decreasing the surface
velocity and $\Omega/\Omega_{\rm crit}$. The mass loss rates becomes
very low again until the star crosses the HR diagram and reaches the RSG
stage. At this point the convective envelope dredges up CNO elements to
the surface increasing its overall metallicity. As said in Sect.
\ref{mdot}, the total metallicity, $Z$, is used (including CNO elements)
for the metallicity dependence of the mass loss.
Therefore depending on how much CNO is brought up to the surface, the
mass loss can become very large again. The CNO brought to the surface
comes from primary C and O produced in He--burning. As described in the
above subsection, rotational and convective mixing brings these elements
into the H--burning shell. A large fraction of the C and O is then 
transformed into primary nitrogen via the CNO cycle. 
Additional convective and rotational
mixing is necessary to bring the primary CNO to the surface of the star.
The whole process is complex and depends on 
mixing (see Fig. \ref{kipz85}). 
Of particular importance is the surface convective zone, which appears
when the star becomes a RSG. This convective zone dredges-up the CNO
to the surface. For a very large mass loss to occur, it is necessary that the
star becomes a RSG in order to develop a convective envelope. It is also
important that the extent of the convective envelope is large enough to
reach the CNO rich layers. Finally, the star must reach the RSG stage 
early enough (before the end of core He--burning) so that there will be 
time remaining to lose mass. Figure \ref{ycteff} shows the evolution of
the effective temperature as a function of the central helium mass
fraction. This figure shows that the 9 and 40 $M_\odot$ models reach the
RSG stage only after the end of helium burning, so too late for a large
mass loss. The 60 $M_\odot$ model reaches the RSG stage during
He--burning.
It would therefore have time to lose large amounts of mass. 
However, the dredge--up is not strong enough.
The 85 $M_\odot$ model becomes a RSG during
He--burning earlier than the 60 $M_\odot$ model. 
The dredge-up is stronger for this model and the surface CNO abundance
becomes very high (see Fig. \ref{absw} {\it bottom}). The
series of models presented here constrain the minimum initial mass for
significant mass loss (more than half of the initial mass) to be between
60 and 85 $M_\odot$. 

The dependence on mixing of the lower initial mass for a large mass loss
to occur can be estimated by
comparing the 60 $M_\odot$ model calculated here and the one presented by
\citet{MEM06}. The model calculated by \citet{MEM06}, which does
not include overshooting and uses a different prescription for the 
horizontal diffusion coefficient, 
$D_{\rm h}$ \citep{Mh03}, loses a large fraction of its mass (and
becomes a WR star with high effective temperature) just before
the end of core helium burning \citep[see Fig. 4 from][]{MEM06}.  
The $D_{\rm h}$ used in \citet{MEM06}, compared to the $D_{\rm h}$ used in
the present calculations, tends to 
allow a larger enrichment of the surface 
in CNO processed elements. This different physical ingredient
explains the differences between the two 60 $M_\odot$ 
models.
The fact that, out of two 60 $M_\odot$ models, one model does not lose
much mass and the other model with a different physics just does,
could mean that the minimum initial mass for the star to
lose a large fraction of its mass is around 60 $M_\odot$.
\begin{figure}[!tbp]
\centering
\includegraphics[width=8cm]{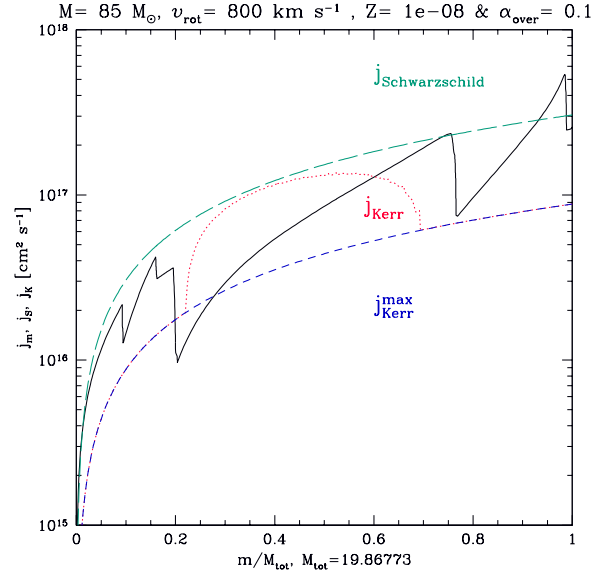}
\caption{Profile of the specific angular momentum at the pre--SN stage
for the 85 $M_\odot$ model at $Z=10^{-8}$ (solid line). 
The red dotted line shows the minimum angular momentum necessary in
order to form an accretion disk around a rotating black hole.
The blue short dashed and green long dashed lines show the minimum angular momentum
necessary for a maximally rotating and a non--rotating black hole
respectively.
}
\label{j85}
\end{figure}

\begin{figure*}[!tbp]
\centering
\includegraphics[width=15cm]{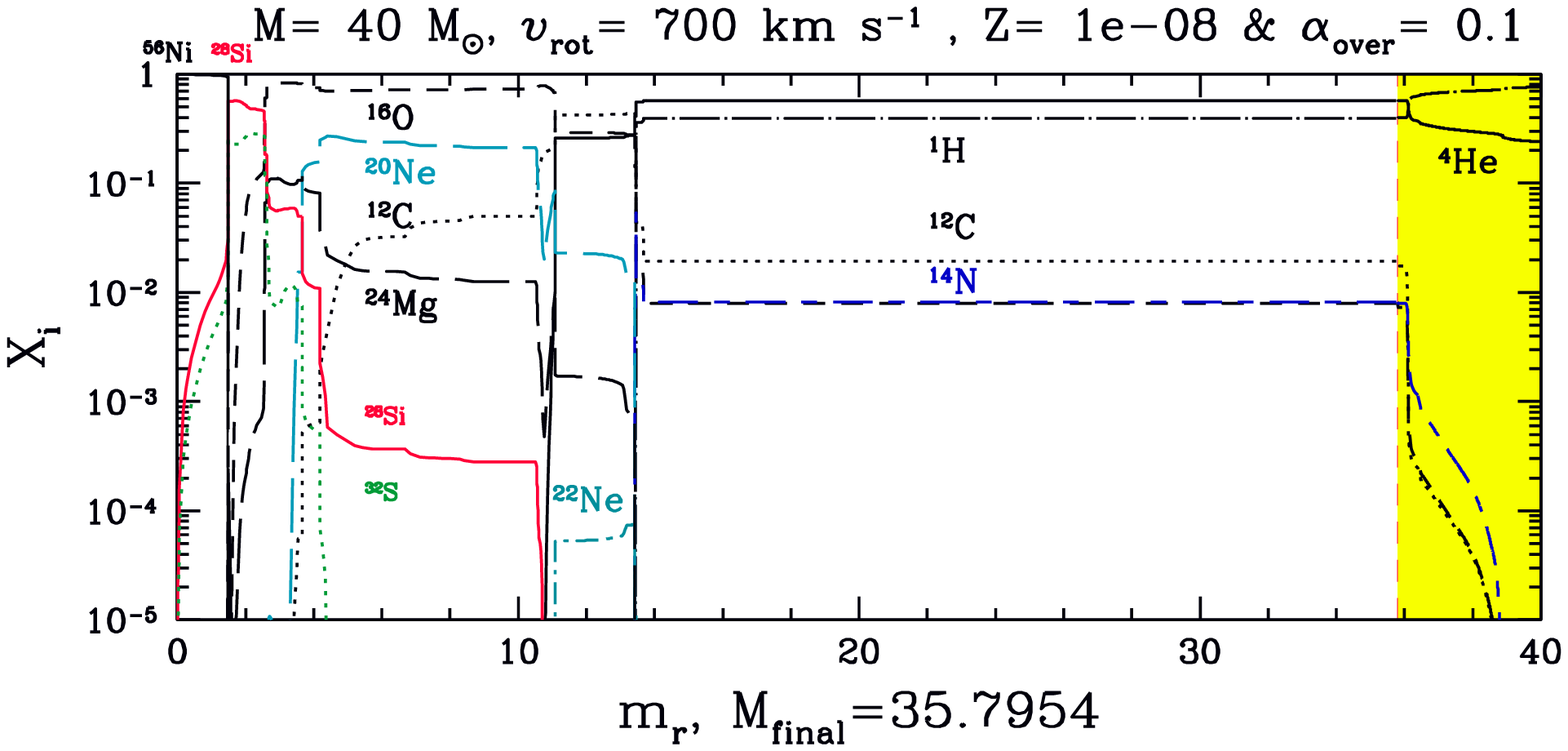}
\includegraphics[width=15cm]{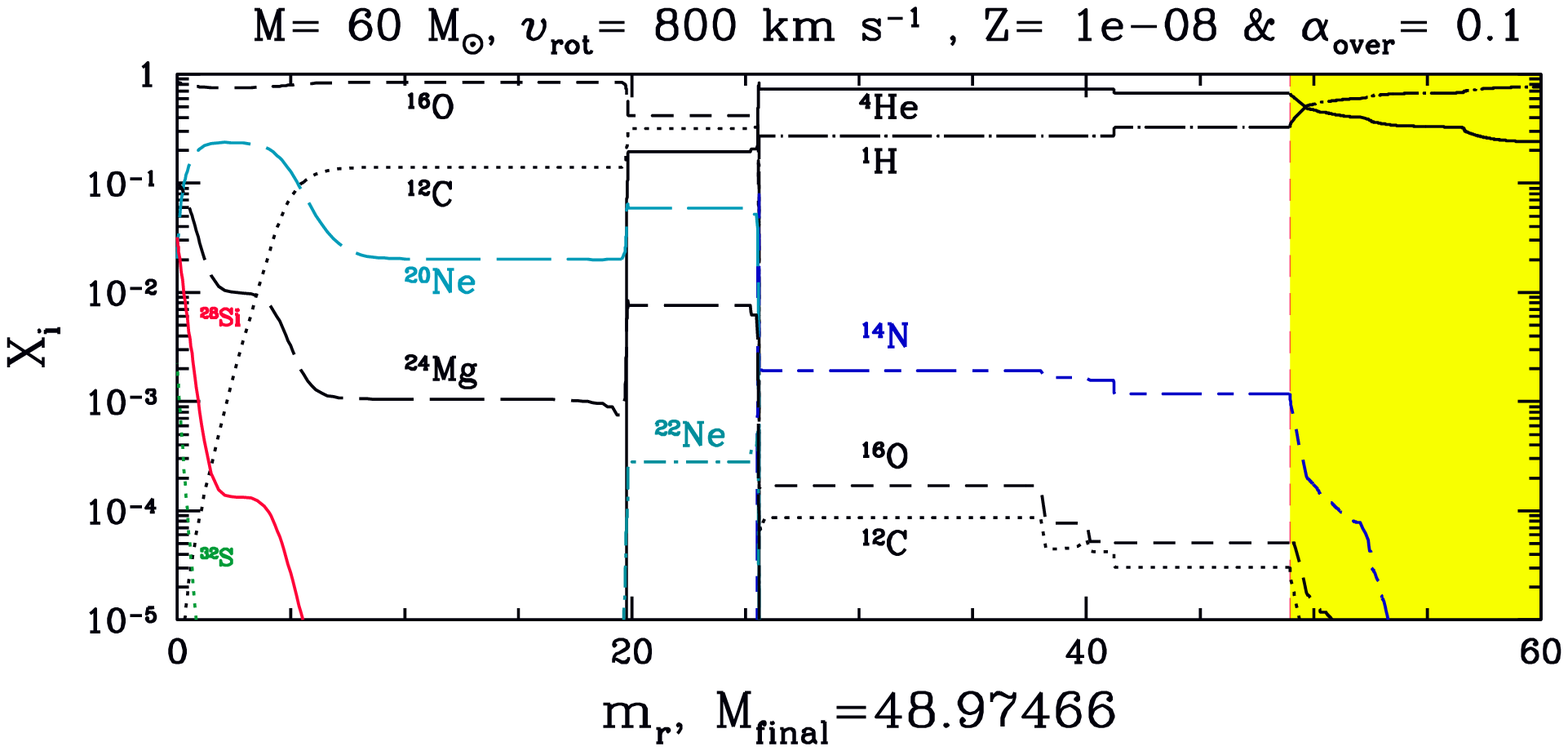}
\includegraphics[width=15cm]{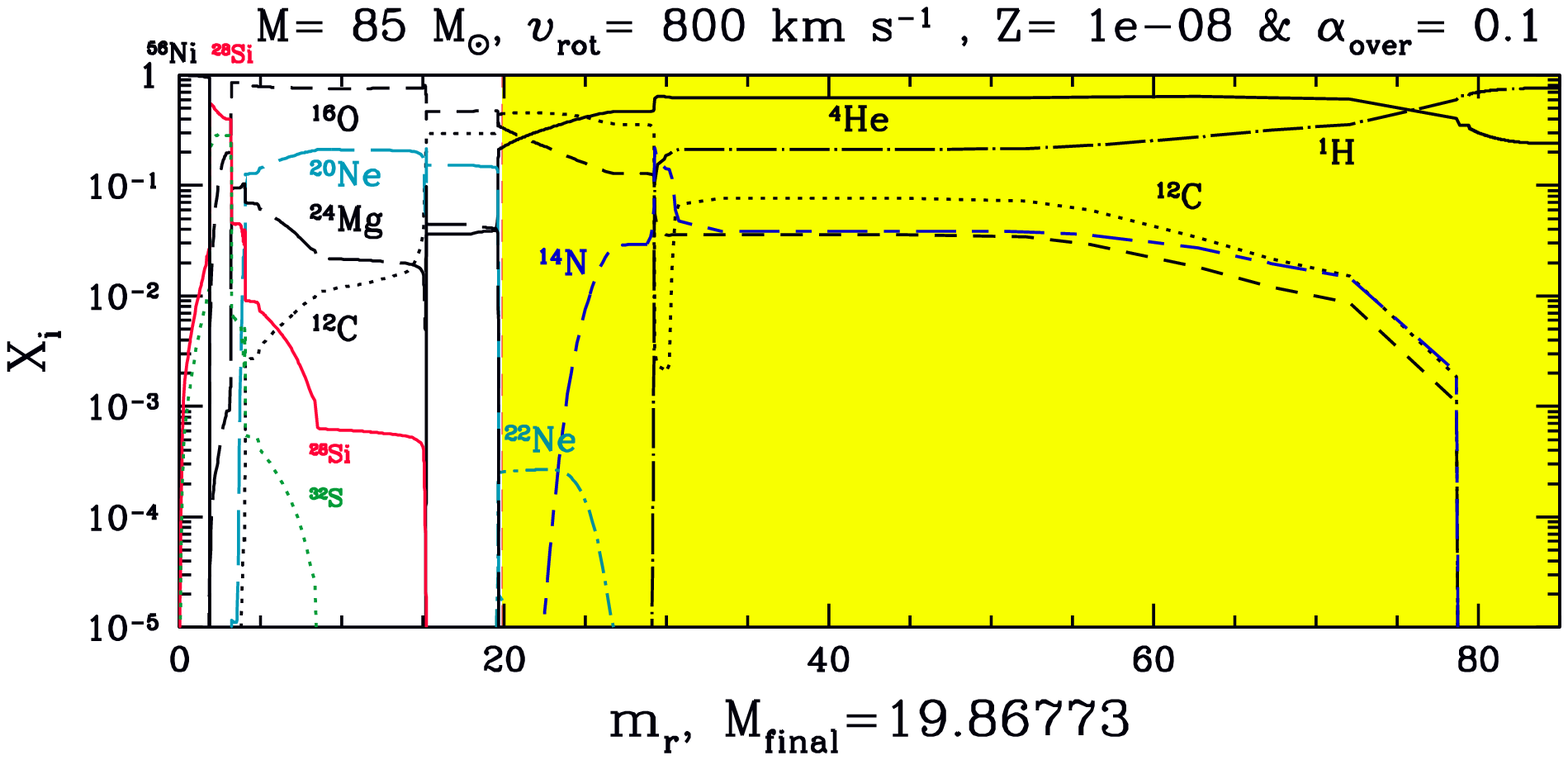}
\caption{Abundance profiles for the 40 ({\it top}), 60 ({\it middle}) and 85 ({\it
bottom}) $M_\odot$ models. The pre--SN and wind  (yellow shaded
area) chemical compositions are
separated by a red dashed line located at the pre--SN total mass
($M_{\rm final}$), given below each plot.}
\label{absw}
\end{figure*}

\begin{table*}
\caption{Initial mass, metallicity and velocity and 
{\bf total stellar yields}$^a$ ($mp^{\rm{pre-SN}}_{im} + mp^{\rm{wind}}_{im}$) 
 of the models. All masses and yields are in
solar mass units and velocities are in km\,s$^{-1}$.
For the metallicity, $Z_{\rm{ini}}$ is in log10 units 
(-3 corresponds to $Z_{\rm{ini}}=10^{-3}$) except for $Z=0.02$ ($\odot$).
Z is the total
metal content and is defined by: Z$=1-X_{\rm{^1 H}}-X_{\rm{^3 He}}- 
X_{\rm{^4 He}}$. These are the yields to be used for chemical evolution 
models using Eq. 2 from \citep{AM92}.}
\begin{tabular}{r r r r r r r r r r r r r r r}
\hline \hline
\vspace{0.1cm}
$M$ & $Z_{\rm{ini}}$ & $\upsilon_{\rm{ini}}$ 
        & $^{4}$He & $\rm \frac{^{3}He}{10^4}$ & $^{12}$C & $^{13}$C & $^{14}$N & $^{16}$O & $^{17}$O & $^{18}$O & $^{22}$Ne & $Z$  \\  
\hline 
 20 & $\odot$ & 300 & 1.62 &   -3.36 & 0.433 & 1.01e-3 & 4.33e-2 & 2.57 & -2.75e-6 & -1.96e-4 & 4.26e-2 &  3.98  \\
\hline 
 20 & -3      & 000 & 2.47 &   -2.84 & 0.373 & 2.58e-5 & 3.31e-3 & 1.46 & -5.48e-7 & -1.12e-5 & 2.48e-3 &  2.38  \\
 20 & -3      & 300 & 2.11 &   -3.31 & 0.676 & 2.84e-5 & 3.10e-3 & 2.70 &  4.83e-7 & -1.89e-5 & 1.04e-2 &  3.77  \\
\hline
 20 & -5      & 000 & 2.50 &   -3.32 & 0.370 & 1.93e-7 & 4.27e-5 & 1.50 &  3.05e-7 & -9.43e-8 & 2.11e-5 &  2.30  \\
 20 & -5      & 300 & 2.34 &   -3.91 & 0.481 & 2.42e-6 & 1.51e-4 & 2.37 &  3.40e-7 &  5.27e-7 & 2.74e-3 &  3.35  \\
 20 & -5      & 500 & 2.26 &   -4.08 & 0.648 & 1.53e-5 & 5.31e-4 & 2.59 &  4.79e-7 &  5.49e-6 & 1.07e-2 &  3.54  \\
\hline
 20 & -8      & 000 & 2.27 &   -4.17 & 0.262 & 2.27e-4 & 8.52e-3 & 1.20 &  1.94e-7 & -2.15e-0 & 1.85e-6 &  2.14  \\
 20 & -8      & 300 & 2.03 &   -4.37 & 0.381 & 1.80e-6 & 1.20e-4 & 1.96 &  1.70e-8 &  2.14e-7 & 5.48e-5 &  2.97  \\
 20 & -8      & 600 & 3.15 &   -4.50 & 0.823 & 5.55e-3 & 5.90e-2 & 1.35 &  1.73e-5 &  2.52e-7 & 7.72e-5 &  2.49  \\
\hline
 09 & -8      & 500 & 1.43 &   -1.76 & 0.082 & 1.35e-4 & 2.53e-3 & 5.85 &  5.64e-7 &  4.07e-5 & 1.86e-4 & 0.143  \\
 40 & -8      & 700 & 6.01 &   -9.10 & 1.79  & 6.31e-2 & 1.87e-1 & 5.94 &  6.31e-5 &  4.74e-7 & 1.51e-4 &  9.64  \\
 60 & -8      & 800 & 8.97 &   -13.3 & 3.58  & 5.00e-4 & 4.14e-2 & 12.8 &  6.26e-6 &  3.56e-7 & 1.66e-3 &  17.1  \\
 85 & -8      & 800 & 16.8 &   -2.00 & 7.89  & 5.60e-1 & 1.75e+0 & 12.3 &  6.66e-4 &  4.95e-5 & 1.55e-3 &  25.5  \\
\hline 
\end{tabular}
\\ $^a$ Note that the corresponding ejected masses
can be calculated by adding the initial composition given in 
Table \ref{inic} multiplied by the mass interval, the mass boundaries of
which (initial and remnant masses) are given in Table \ref{table1}. 

\label{ytot}
\end{table*}
\begin{table*}
\caption{Initial mass and velocity and 
{\bf stellar wind} yields$^a$ ($mp^{\rm{wind}}_{im}$)
of the $Z=10^{-8}$ models. 
All masses and yields are in
solar mass units and velocities are in km\,s$^{-1}$.
}
\begin{tabular}{r r r r r r r r r r r r r}
\hline \hline 
$M$ & $\upsilon_{\rm{ini}}$ 
               & $^{4}$He & $^{3}$He & $^{12}$C & $^{13}$C & $^{14}$N & $^{16}$O & $^{17}$O & $^{18}$O & $^{22}$Ne & $Z$ \\  
\hline 
09 & 500 & 2.80e-5 & -8.34e-9 &  5.86e-08 & 1.72e-09 & 2.53e-08 &  2.33e-08 & 6.19e-12 &  1.20e-11 & 9.77e-12 & 1.10e-7  \\
20 & 600 & 2.36e-4 & -1.04e-6 & -3.27e-11 & 5.06e-13 & 3.08e-10 & -2.54e-10 & 5.30e-13 & -7.22e-13 & 2.72e-16 & 7.59e-9  \\
40 & 700 & 3.29e-1 & -1.03e-4 &  5.34e-03 & 8.06e-04 & 3.63e-03 &  2.42e-03 & 1.05e-06 &  2.18e-09 & 2.35e-07 & 1.17e-2  \\
60 & 800 & 1.21e+0 & -2.72e-4 &  1.80e-05 & 4.94e-06 & 6.87e-04 &  5.48e-05 & 7.94e-08 & -9.49e-11 & 4.80e-08 & 7.25e-4  \\
85 & 800 & 2.00e+1 & -1.64e-3 &  6.34e+00 & 5.60e-01 & 1.75e+00 &  3.02e+00 & 6.66e-04 &  4.95e-05 & 1.46e-03 & 1.16e+1  \\
\hline 
\end{tabular}
\\ $^a$ Note that the corresponding ejected masses
can be calculated by adding the initial composition given in 
Table \ref{inic} multiplied by the mass interval, the mass boundaries of
which (initial and final masses) are given in Table \ref{table1}. 
For heavy elements with yields larger than 1e-6, ejected masses and
stellar yields are essentially the same here since the initial total
metallicity is 1e-8.
\label{yw}
\end{table*}
\subsection{WR stars at very low metallicities}\label{WR} 
The lower limit for WR star formation is roughly the same as for large 
mass loss to occur, {\it i.e.} probably about 60 $M_\odot$, 
since models calculated here that
lose a lot of mass become WRs. Therefore our models of single
rotating massive stars produce WR stars at very low metallicities.
The 85 $M_\odot$ model even becomes a WO type WR star. SNe of type Ib and Ic are
therefore predicted to ensue from the death of single massive stars at
very low metallicities.
The next question is whether or not these stars
can produce long and soft gamma ray bursts (GRBs) via the collapsar model
\citep{W93}. Figure \ref{j85} shows the angular momentum distribution
at the pre--SN stage. This figure shows that the core has enough
angular momentum to produce a GRB \citep[see][for more details on GRB
progenitors]{YL05,WH06,grb05}. 
The effects of magnetic fields are however not included in the
calculations. It is therefore possible that the core will not retain
enough angular momentum to form an accretion disk. 
It is interesting to
point out here that the outer part of the star also contains sufficient 
angular
momentum to form an accretion disk, which is less likely to be removed
due to the effects of magnetic fields.
In a very optimistic outcome, one could imagine that the model could
produce two jet episodes, one when the core collapses and one when the 
outer parts collapse, since for such a high initial mass
the central black hole could swallow the whole pre--SN structure.

\section{Stellar yields of light elements and comparison with observations}\label{yields}
\subsection{Stellar yields}
The stellar yields are calculated using the same formulae as in \citet{ywr05}.
The wind contribution from a star of initial mass, $m$, 
to the stellar yield of an element $i$ is:
\begin{equation}\label{wdef}  
mp^{\rm{wind}}_{im}= \int_{0}^{\tau(m)} \dot{M}(m,t)[X_{i}^S(m,t)-X_{i}^{0}]\,dt
\end{equation}where 
$\tau(m)$ is the final age of the star, 
 $\dot{M}(m,t)$ the mass loss rate when the age of the star is equal to
 $t$,
 $X_{i}^S(m,t)$ the surface abundance in mass fraction of element $i$ 
and $X_{i}^{0}$ its initial mass fraction 
(see Table \ref{inic}). 
The pre--SN contribution from a star of initial mass, $m$, 
to the stellar yield of an element $i$ is:
\begin{equation}\label{sndef}
mp^{\rm{pre-SN}}_{im}= \int_{m(\rm{rem})}^{m(\tau)} [X_{i}(m_r)-X_{i}^{0}]\,dm_r  
\end{equation}where 
$m(\rm{rem})$ is the remnant mass estimated from the CO core mass 
\citep[see][]{AM92},
$m(\tau)$ the final stellar mass,
$X_{i}^{0}$ the initial abundance in mass fraction of element $i$
and $X_{i}(m_r)$ the final abundance in mass fraction at the lagrangian
mass coordinate, $m_r$. 
The pre--SN contribution is calculated at the end of Si--burning. 
Therefore the contribution from explosive nucleosynthesis is not 
included. However, elements
lighter than neon are marginally modified by explosive
nucleosynthesis \citep{CL03,TNH96} and 
are mainly determined by the hydrostatic
evolution.  
The total stellar yield of an element $i$ from a star of initial 
mass, $m$, is then :
\begin{equation}\label{ydef}
mp^{\rm{tot}}_{im} = mp^{\rm{pre-SN}}_{im} + mp^{\rm{wind}}_{im}
\end{equation}$mp^{\rm{tot}}_{im}$ corresponds to the amount of element 
$i$ newly synthesised and ejected by a star
of initial mass $m$. One notes that, using the above
formulae, negative yields are obtained if the star destroys an
element (see yields of $^3$He). 
The total stellar yields for the chemical elements which are not significantly 
affected by the evolution beyond the calculations done in this work
are presented in Table \ref{ytot}.
The stellar wind contribution is presented for 
the $Z=10^{-8}$ models in Table \ref{yw}. 
The SN contribution to the yields (layers of the stars ejected during the
SN) is not presented but the pre--SN abundance profiles are
presented in Figs. \ref{absw} and \ref{ab1}.
The 60 $M_\odot$ model was evolved until neon burning and the 9 $M_\odot$
model until carbon burning. This means that, for
these two models only, the abundance profiles can still vary in the
central regions
before the collapse. Since the remnant mass probably contains a large
fraction of 
the CO core for these two models, the SN contribution to the yields 
should not be affected by these variations. 

The 20 $M_\odot$ models are expected to produce neutron stars and
eject most of their envelope.
On the other hand, stars more massive than about 40
$M_\odot$ on the ZAMS are expected to form black holes directly and not
to eject anything during or after their collapse \citep{HFWLH03}.
The present 40 and 60 $M_\odot$ models probably follow this scenario.
If this is the case, the wind 
contribution is the only contribution to be taken into account.
The outcome is uncertain for the 85 $M_\odot$ model because the final
mass is only about 20 $M_\odot$ but the alpha and CO core masses are
very large (see Table \ref{table1}). This model could produce a GRB, in
which case jets would be produced and some iron rich matter would be
ejected.
\begin{figure}[!tbp]
\centering
\includegraphics[width=4.5cm]{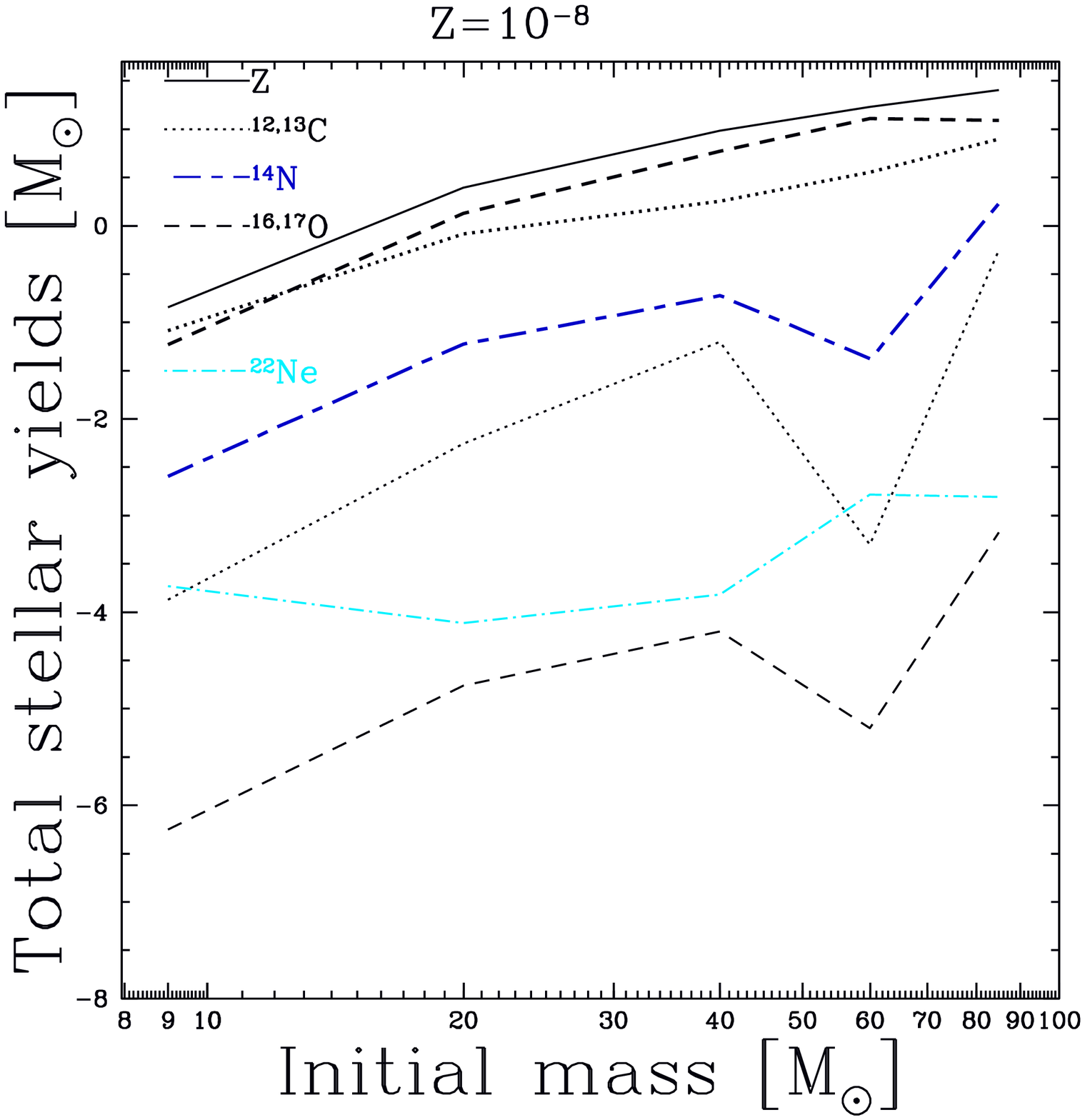}\includegraphics[width=4.5cm]{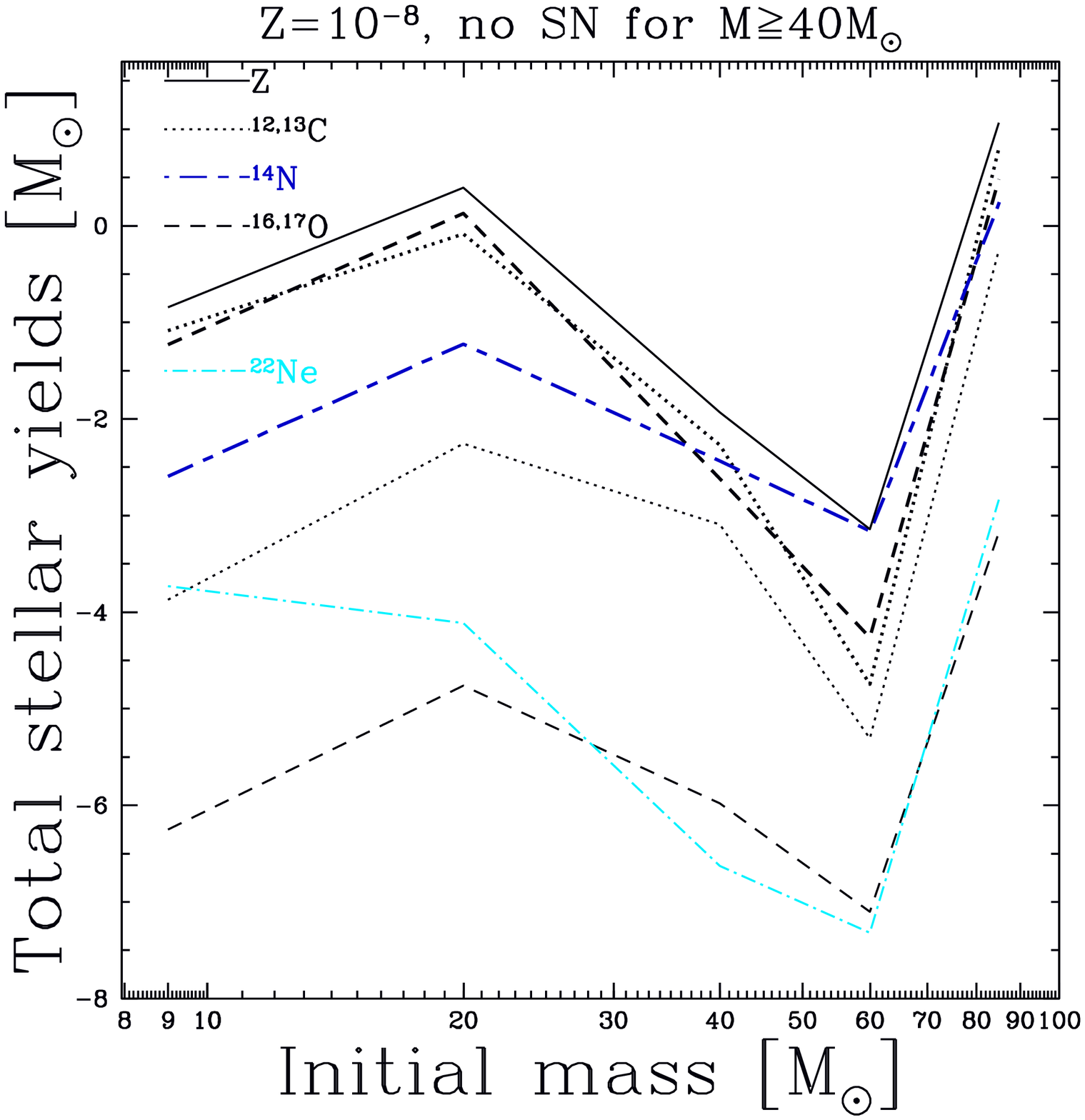}
\caption{Total stellar yields as a function of the initial mass of the
models with $Z=10^{-8}$ and 
({\it left}) assuming the SN ejects matter for $M \gtrsim 40$ $M_\odot$
and ({\it right}) assuming direct black hole formation and no matter
ejection by the SN for $M \gtrsim 40$ $M_\odot$.}
\label{zm8t}
\end{figure}
In Fig. \ref{zm8t}, the two possible outcomes for very massive
stars are compared. On the left, the stellar yields include the 
SN contribution and on the right, the stellar yields only include the
wind contribution for stars with $M \gtrsim 40$ $M_\odot$.
As can be expected, the two outcomes give very different yields for the 40
and 60 $M_\odot$ since
they do not lose much mass before the SN explosion. For stars above about
60 $M_\odot$, the difference is much smaller because the strong winds
peal off most of the CNO rich layers before the final collapse.

\subsubsection{Metallicity dependence of the CNO yields}
The metallicity dependence of the yields can be studied using the 20
$M_\odot$ models. Indeed when the yields are convolved by a \citet{Sa55} 
like initial
mass function, 20 $M_\odot$ is not far from the average massive star.
Note that it is however uncertain whether a standard IMF applies at very low
metallicities. Recent studies tend to argue that the IMF is top
heavy but very massive stars (VMS, $M\gtrsim 140 M_\odot$) are not
necessary to reproduce observations \citep[see for example][]{SF05,TVS04}.

\begin{figure*}[!tbp]
\centering
\includegraphics[width=6cm]{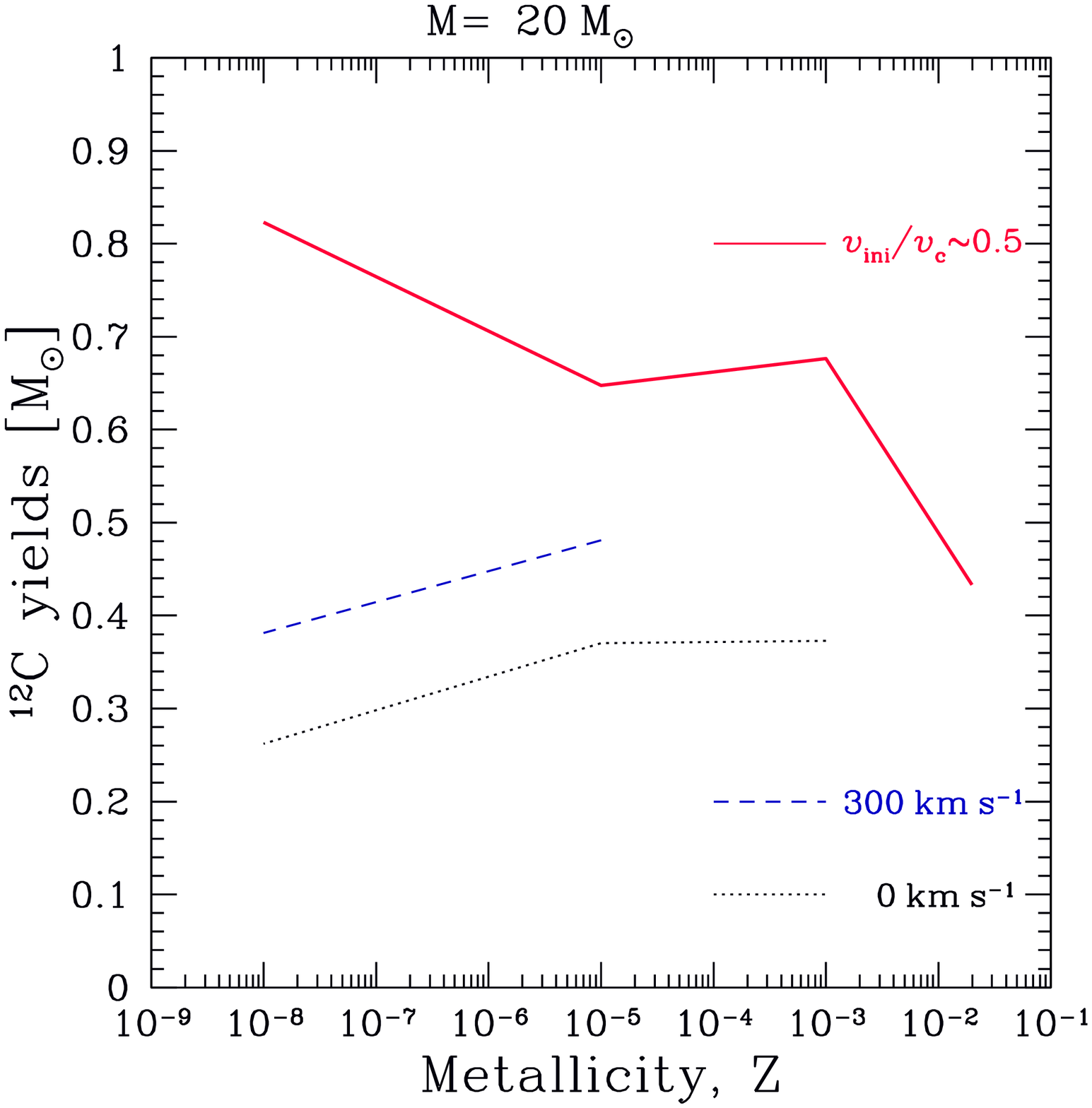}\includegraphics[width=6cm]{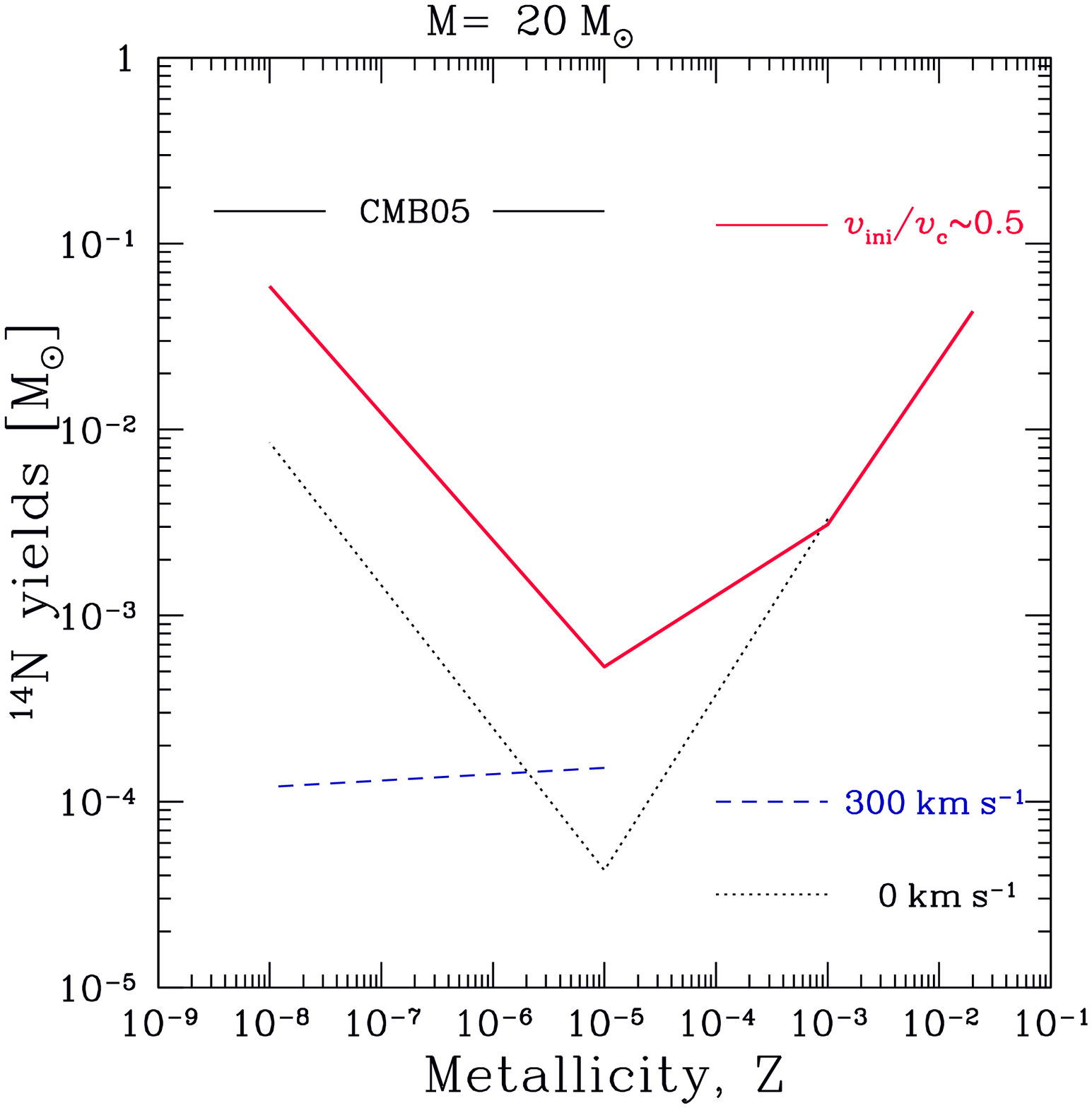}\includegraphics[width=6cm]{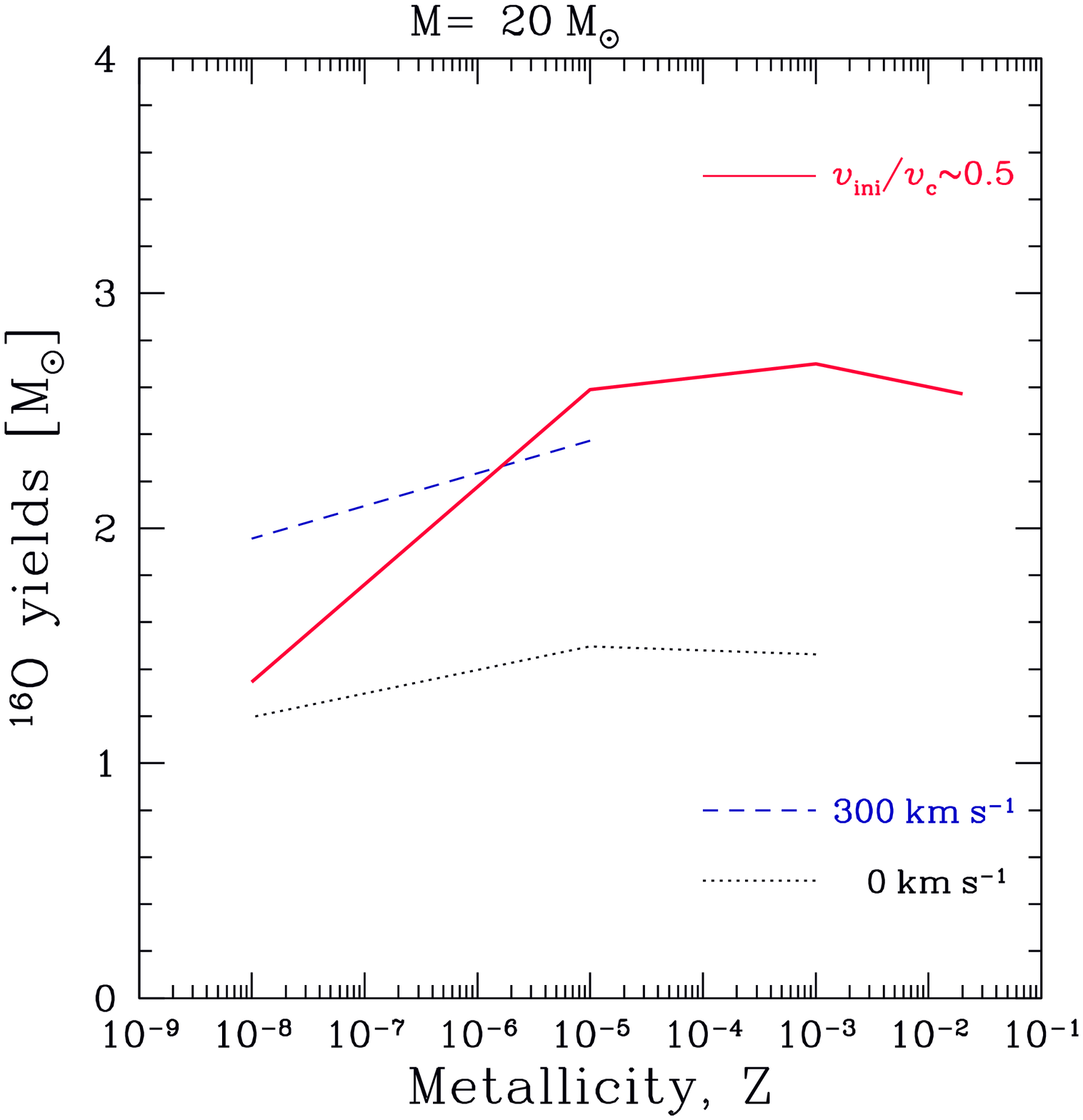}
\caption{$^{12}$C ({\it left}), $^{14}$N ({\it centre}) and $^{16}$O 
({\it right}) total stellar yields as a function of the initial 
metallicity of the 20 $M_\odot$ models.
The red solid curves corresponds to the models with fast rotation, the
blue dashed curve to the very low metallicity models with 
$\upsilon_{\rm ini}$=300\,km\,s$^{-1}$ and the dotted black curve to the
non--rotating models. In the graph of the $^{14}$N yields, the yields
used in the chemical evolution models of \citet{CMB05} in order to
reproduce observations is marked with the symbol CMB05.
For the fast rotating models, one sees that the carbon yield increases
slightly between $Z=10^{-5}$ and $Z=10^{-8}$. At the same time, the
oxygen yield decreases towards lower $Z$ by a factor about two. }
\label{ycno}
\end{figure*}

The yields for $^{12}$C, $^{14}$N and $^{16}$O are presented in Fig.
\ref{ycno}.
The most stringent observational constraint at very low Z is a very
high primary $^{14}$N production.
This requires extremely high primary $^{14}$N production in massive
stars, of the order of 0.1 $M_\odot$ per star \citep[$\sim$0.15 $M_\odot$ used 
in the heuristic model
of][]{CMB05}.
In Fig. \ref{ycno}, we can see that only the model at $Z=10^{-8}$ and 
with a fast rotation ($\upsilon_{\rm ini}=$600\,km\,s$^{-1}$) 
gets close to such high values.
The bulk of $^{14}$N is produced in the convective
zone created by shell hydrogen burning (see Sect. \ref{mix}). If this
convective zone deepens enough
to engulf carbon (and oxygen) rich layers, then significant amounts of
primary $^{14}$N can be produced (0.01$\sim$0.1$\,M_\odot$).
This occurs in both the non--rotating model
and the fast rotating model but for different reasons.
In the non--rotating model, it occurs due to
structure rearrangements similar to the third dredge--up at the end of
carbon burning \citep[see also][for a similar process in Z=0 models]{CL04}. In the model with $\upsilon_{\rm ini}=$600\,km\,s$^{-1}$
it occurs during shell helium burning
because of the strong mixing of carbon and oxygen into the
hydrogen shell burning zone.
Another interesting feature is the possible upturn of the [C/O] ratio 
observed at very low metallicities \citep[ratio between the surface
abundances of carbon and oxygen relative to solar; see Fig. 14][]{FS6}. 
Indeed, the models at $Z=10^{-8}$ with a fast rotation
have high C and low O yields compared to the $Z=10^{-5}$ models and could
reproduce such an upturn. As explained in Sect. \ref{mix}, the
reason is the reduction of the He--burning core mass after the shell
H--burning boost induced by strong mixing. As can be seen in Fig.
\ref{kip}, this occurs for all initial masses calculated in the
$Z=10^{-8}$ series.
The stellar yields of the fast rotating models calculated here were 
used in a galactic 
chemical evolution model and successfully reproduce the early 
evolution of CNO elements \citep{CH06}. This is a good argument in
favour of fast rotation at very low metallicities.
Note that in \citet{CH06}, the SN contribution
was included in the stellar yields for all masses. 
The impact of including only the wind contribution for masses above 40
$M_\odot$ will be studied in the future. 
The impact of the important primary nitrogen production and of the other
yields on the initial composition and therefore on the
evolution of the next 
stellar generations and their yields is an interesting aspect that will
also be studied in the future.

\subsubsection{$Z=10^{-8}$ models}
The stellar yields of the fast rotating $Z=10^{-8}$ models are presented in
Fig. \ref{zm8t}. It shows that the significant (above 0.01 $M_\odot$) 
production of primary nitrogen occurs for the entire mass range. 
For massive stars with $40<M<60\,M_\odot$, the yields depend on
whether or not the SN contributes to the total yields as discussed
earlier. The production of nitrogen is accompanied by a production of
$^{13}$C and $^{17}$O (and to a lesser extent $^{18}$O). The ratio
$^{12}$C/$^{13}$C is very different between the wind and the SN
contributions. It is around 5 for the wind and more than 100 for the SN
and SN+wind contributions. This, with the N/C and N/O ratios 
are good tests to differentiate between
the two contributions. Part of the primary nitrogen also captures 
two $\alpha$--particles and becomes $^{22}$Ne. The primary $^{22}$Ne
yields are of the order of $10^{-4}\,M_\odot$. $^{22}$Ne is one of the
main neutron sources for the weak s--process in massive stars. With a low
initial iron content due to the low initial overall metallicity, the
s--process could occur with a high neutron to seed ratio and produce
surprising results. This will be the subject of a future study.

\subsection{Carbon rich EMP stars}\label{crumps}
The zoo of the extremely metal poor stars has been classified by \citet{BC05}. 
Carbon rich extremely metal poor stars (CEMPs also called CRUMPs at a recent
meeting at Tegernsee, http://www.mpa-garching.mpg.de/$\sim$crumps05) are 
different from normal EMP star 
and are rarer \citep{RANB05}. About three quarter of the CEMPs show a
standard s--process enrichment pointing to the fact that they have
accreted matter from an AGB companion in a binary system \citep{SAMFI04}.
The other CEMP stars show a weak
s--process enhancement. Their peculiar abundances are therefore thought 
to originate from the previous generation of stars.
The two most metal poor stars known to date, 
HE1327-2326 \citep{Fr05,Ao06,Fr06} and HE 0107-5240 \citep{Ch04,BCG04} are both
CEMP stars with a weak s--process enrichment. 
\citet{CL02} and \citet{NTU05} have studied the enrichment due to PopIII
SNe. By using one or a few SNe and using a very large mass cut, they are
able to reproduce the abundance of most elements \citep{LCB03,IUTNM05}. 
However they are not
able to reproduce the nitrogen surface abundance of
HE1327-2326 without rotational mixing or the oxygen surface abundance of 
HE 0107-5240 without mixing and fall--back mimicking an aspherical
explosion. In this work, the impact of rotation is explored.
HE1327-2326 is characterised by very high N, C and O abundances,
high Na, Mg and Al abundances, a weak s--process enrichment and depleted
lithium. The star is not evolved so has not had time to bring
self--produced CNO elements to its surface and is most likely a subgiant 
(Korn, presentation at the CRUMPS meeting).
A lot of the features of this star are similar to the properties of the
stellar winds of very metal poor rotating stars \citep{MEM06}. 
\begin{figure}[!tbp]
\centering
\includegraphics[width=4.5cm]{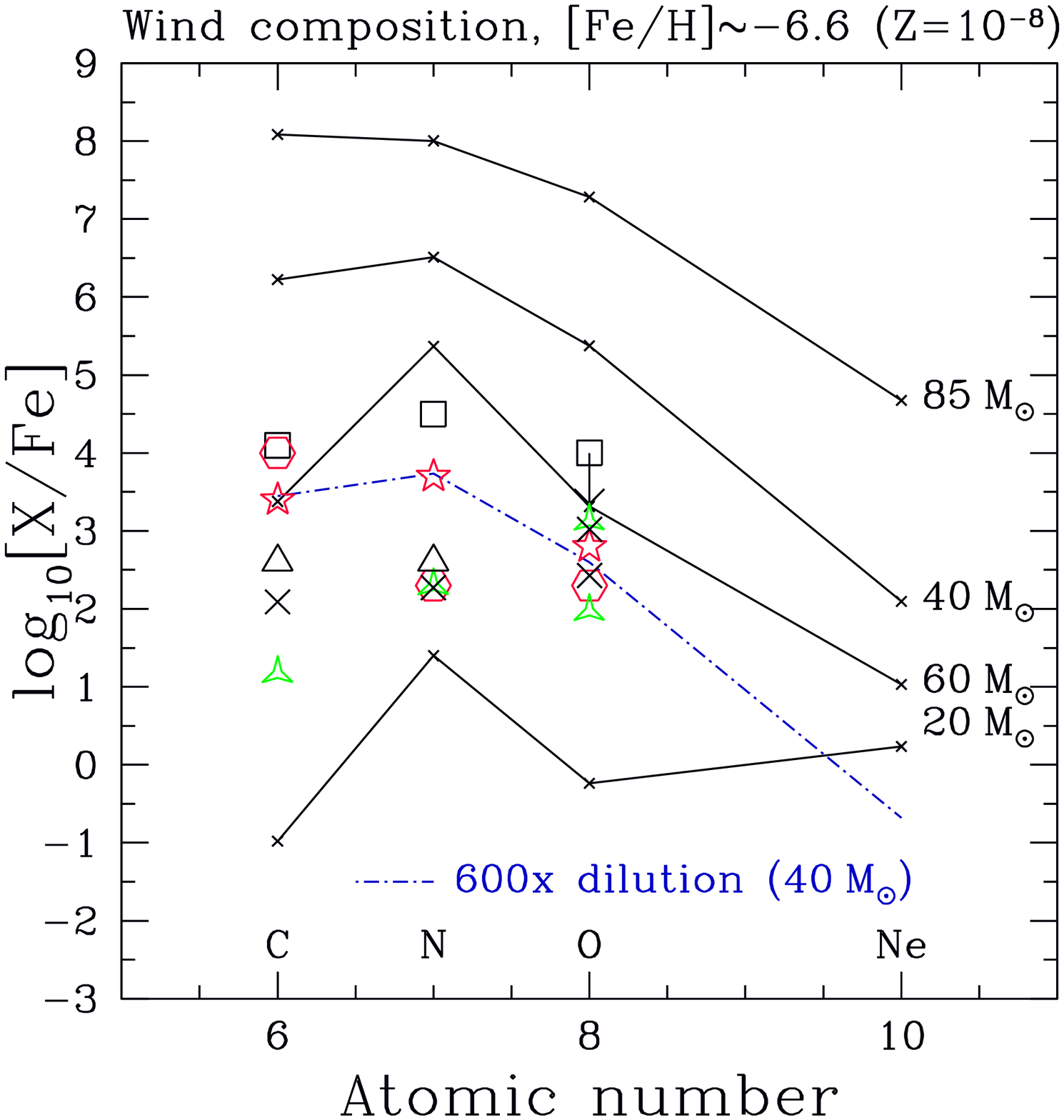}\includegraphics[width=4.5cm]{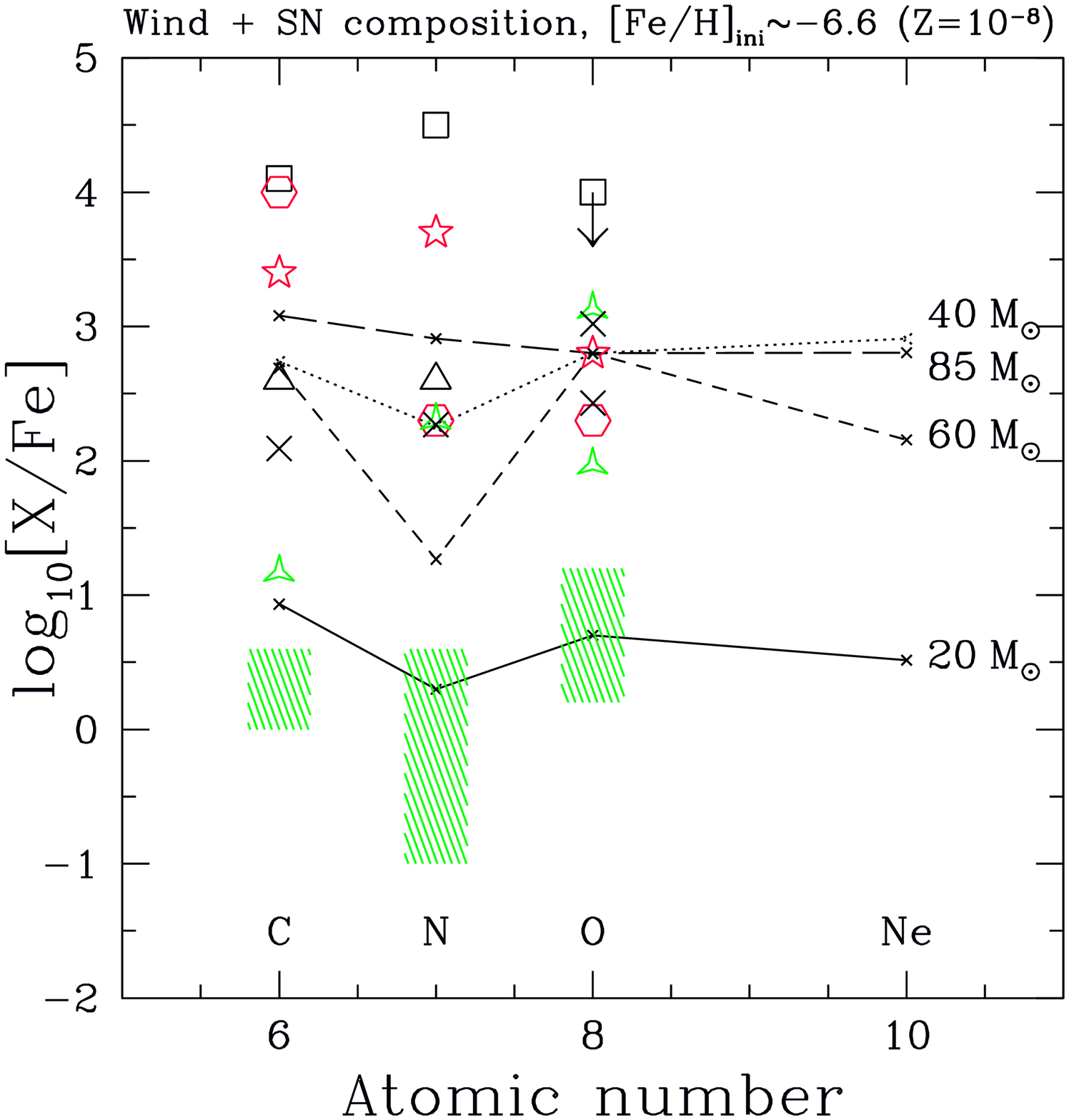}
\caption{Composition in [X/Fe] of the stellar wind ({\it left}) and the
mixture of the wind and SN ejecta ({\it right}) for the $Z=10^{-8}$
models.
The lines represent predictions from the models.
The following symbols correspond to the 
values obtained from measurements at the surface of giant CEMP stars: 
{\it red hexagons} for HE 0107-5240, [Fe/H]$\simeq$~-5.3 \citep{Ch04};
{\it green thin triangles} CS 22949-037, [Fe/H]$\simeq$~-4.0 \citep{NRBA02,FS02,I04};
{\it crosses} CS 29498-043, [Fe/H]$\simeq$~-3.5 \citep{AN04}.
Two values are given for oxygen for the last two stars. They come from
the determinations using O triplet or [OI] forbidden lines
\citep{AN04,I04}.
The {\it empty triangles} \citep{PC05}, [Fe/H]$\simeq -4.0$
and {\it squares} \citep{Fr05}, [Fe/H]$\simeq -5.4$
correspond to
non-evolved CEMP stars.
The new (3D/NLTE corrected) estimates for HE1327-2326 from \citet{Fr06} are
represented by the {\it red stars}. 
On the right, wind+SN ejecta are compared to the normal EMP
stars \citep[green hatched area with
lines going down to the right][]{FS5,FS6} for the 20 $M_\odot$ model 
and again to the CEMP stars for the more
massive models. For this purpose, the value [O/Fe] is chosen
to fall in the middle of the observed range (20 $M_\odot$: [O/Fe]=0.7 and
$M> 20\,M_\odot$: [O/Fe]=2.8).
For HE1327-2326, the best fit for the
CNO elements is
obtained by diluting the composition of the wind of the 40 $M_\odot$
model by a factor 600. }
\label{cempwt}
\end{figure}
HE1327-2326 could therefore have formed
from gas, which was mainly enriched by stellar winds of rotating very low
metallicity stars. In this scenario, a first generation of stars 
(PopIII) 
pollutes the interstellar medium to very low metallicities
([Fe/H]$\sim$-6). Then a PopII.5 star \citep{Br05,paris05,Ka06} like the 
40 $M_\odot$ model calculated here
pollutes (mainly through its wind) the interstellar medium out of
 which HE1327-2326 forms.
This would mean that HE1327-2326 is a third generation star.
In this scenario, 
the CNO abundances are well reproduced, in particular that of
nitrogen, which according the new values for a subgiant from \citet{Fr06}
is 0.9 dex higher in [X/Fe] than oxygen. 
This is shown in Fig. \ref{cempwt} where the new abundances are
represented by the red stars and the best fit is 
obtained by diluting the composition of the wind of the 40 $M_\odot$
model by a factor 600. On the right side of Fig. \ref{cempwt}, one sees that when the SN
contribution is added, the [X/Fe] ratio is usually lower for nitrogen
than for oxygen. The lithium depletion cannot be explained by rotational and convective
mixing in the massive star if the wind material is diluted in the ISM by 
a factor 600 (600 parts of ISM for 1 part of wind material) as is 
suggested above. 
However, if the wind material is less enriched in CNO elements, a lower
dilution factor would be necessary to reproduce the observations. 
Also if the massive star is born with a higher iron content, a lower dilution
factor is necessary. If this
dilution factor is of the order of unity, it becomes possible to explain
the lithium depletion by internal mixing in the massive star.
To investigate this possibility, more models have to be calculated with 
different initial metallicities. 
Although the existence of a lower limit for the minimum metallicity $Z$ for low mass stars
to form is still under debate,
It is interesting to note that the very high CNO yields of the 
40 $M_\odot$ stars brings the total
metallicity $Z$ above the limit for low mass star formation
obtained in \citet{BL03}.

For HE 0107-5240 (red hexagons in Fig. \ref{cempwt}), 
rotation does not help since neither the wind contribution nor the
total contribution produce such large overproduction of carbon compared
to nitrogen and oxygen. Possible origins for this star are presented
in \citet[][]{IUTNM05} and \citep{SAMFI04}.
For the other carbon rich stars  presented in Fig. \ref{cempwt}, the
oxygen abundances are either not determined or still quite uncertain
\citep{IT04}. The C and N surface abundances of G77-61 \citep{PC05} 
could originate from material similar to the wind of the 85 $M_\odot$. 
The C and N surface abundances of CS 22949-037 \citep{NRBA02,FS02,I04}
resemble the wind composition of the 60 $M_\odot$ model although some
oxygen needs be ejected from the supernova. 
The enrichments in C, N and O are very similar for CS 29498-043
\citep[around +2, see][]{AN04} and a partial ejection due to the supernova is
necessary to explain the oxygen enrichment since in the winds the oxygen
is usually under produced compared to C and N.
It will be interesting to follow the evolution of Na, Mg and Al 
since the high yields of $^{22}$Ne seem to indicate that there could an 
overproduction of
these elements in the wind \citep[see Table \ref{ytot} and the 
$Z=10^{-5}$ model
presented in][]{MEM06}. Since $^{22}$Ne is also a neutron source,
s--process calculations are also planned.

\section{Conclusion}

Two series of models were computed.
The  first series consists of 20 $M_\odot$ models with varying initial
metallicity (solar down to $Z=10^{-8}$) and rotation
($\rm \upsilon_{ini}=0-600$\,km\,s$^{-1}$). The second one consists of
models with an initial metallicity of $Z=10^{-8}$, masses between 9 and
85 $M_\odot$ and fast initial rotation velocities.
The results presented confirm the crucial role of rotation in stellar
evolution and its impact in very low metallicity stars \citep{MEM06}.
The evolution of the models with $Z=10^{-8}$ ([Fe/H]$\sim-6.6$) 
is very interesting.
In the course of helium burning, carbon and oxygen are mixed into the
hydrogen burning shell. This boosts the importance of the shell and
causes a reduction of the CO core mass. Later in the evolution,
the hydrogen shell deepens and produces large amount of primary
nitrogen. For the most massive models ($M\gtrsim 60$\,$M_\odot$),
significant mass
loss occurs during the red supergiant stage assuming that CNO elements
are important contributors to mass loss. This mass loss is due to
the surface enrichment in CNO elements via rotational and convective
mixing. 
The models predict the production of WR stars for an initial mass higher
than 60 $M_\odot$ at $Z=10^{-8}$ and the 85 $M_\odot$ model becomes a WO.
Therefore SNe of type Ib and Ic are predicted from single massive stars
at these low metallicities. The 85
$M_\odot$ model retains enough angular momentum to produce a GRB but
the calculations did not include the effects of magnetic fields.

The stellar yields are presented for light elements. These yields 
were used in a galactic 
chemical evolution model and successfully reproduce the early 
evolution of CNO elements \citep{CH06}. A scenario is also proposed to
explain the abundances of the most metal poor star known to date 
HE1327-2326 \citep{Fr05}. 
In this scenario, a first generation of stars 
(PopIII) 
pollutes the interstellar medium to very low metallicities
([Fe/H]$\sim$-6). Then a PopII.5 star \citep{Br05} like the 
40 $M_\odot$ model calculated in this study
pollutes only with its wind the interstellar medium out of which HE1327-2326 forms.

There are still many questions or issues that could not be treated in
this work. It is necessary to
determine over which metallicity range the large primary production
and the other specific features of the $Z=10^{-8}$  models occur. 
The impact of the yields on the initial composition and therefore on the
evolution of the next stellar 
generations also needs to be studied. 
It will also be interesting to follow the evolution of Na, Mg and Al 
since the high yields of $^{22}$Ne seem to indicate that there could an 
overproduction of
these elements in the wind \citep[see Table \ref{ytot} and the 
$Z=10^{-5}$ model
presented in][]{MEM06}. Since $^{22}$Ne is also a neutron source,
s--process calculations are also planned.
The effects of magnetic fields \citep{YL05,WH06,ROTBIII} 
on the results will be studied in the near future. 
The dependence of the mass loss rates on the metallicity,
especially in the RSG stage need to be further studied to see how
the results of \citet[mass loss in the RSG phase independent of
metallicity]{VL05} can be extrapolated to very low
metallicities.

\begin{acknowledgements} R. Hirschi is supported by SNF grant
200020-105328.
\end{acknowledgements}


\bibliographystyle{aa}

\begin{figure*}[!tbp]
\centering
\includegraphics[width=6cm]{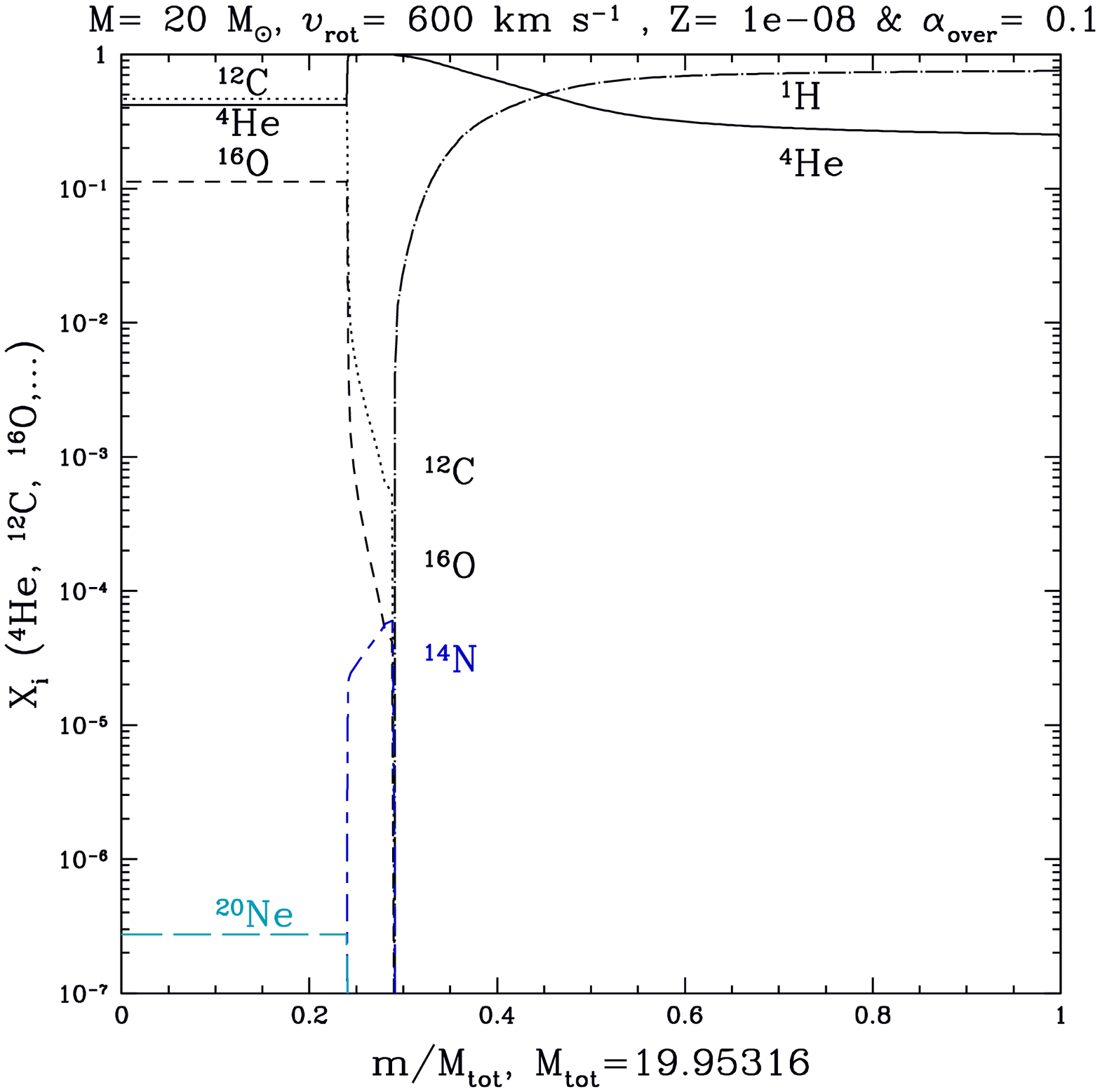}\includegraphics[width=6cm]{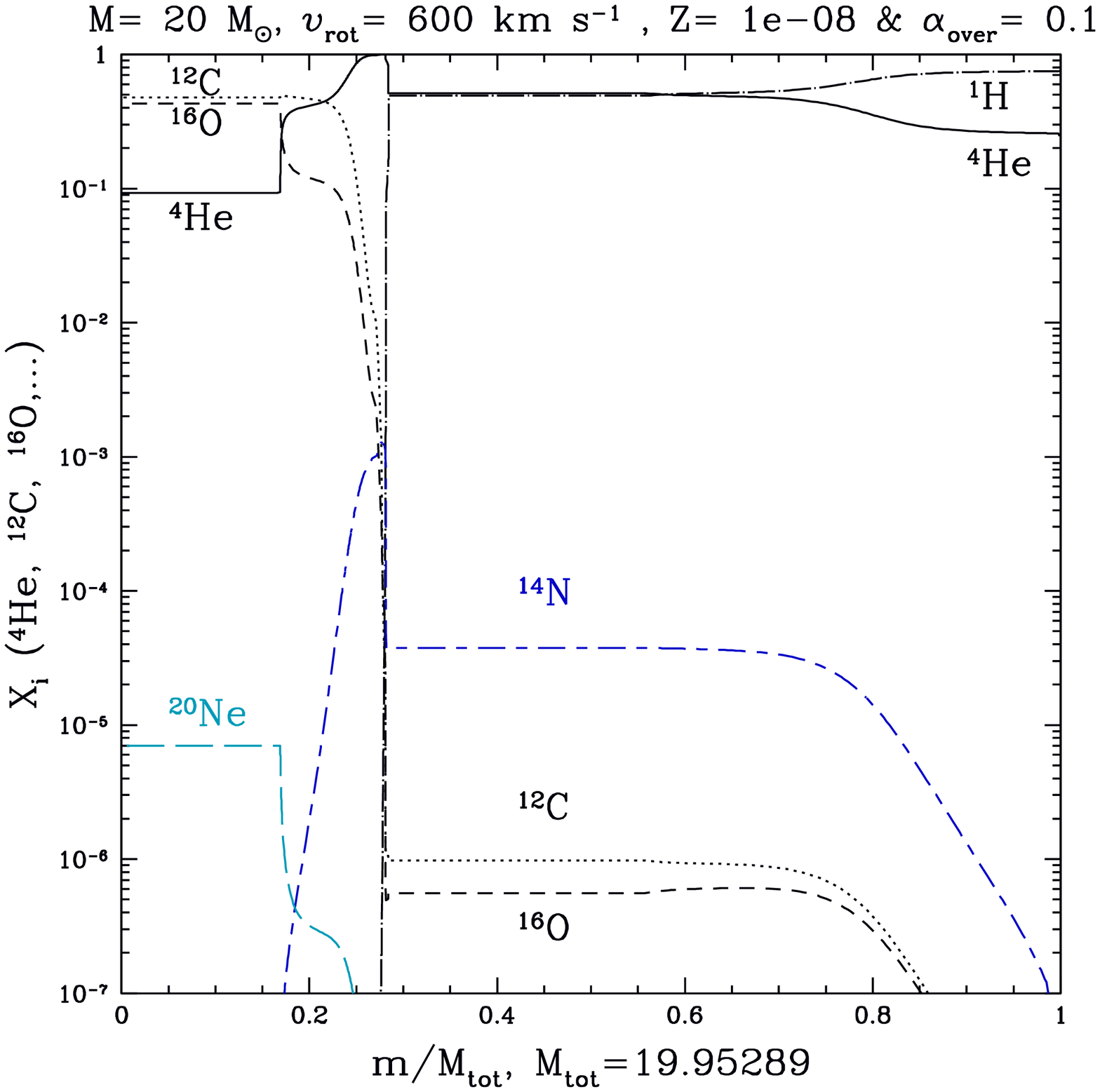}\includegraphics[width=6cm]{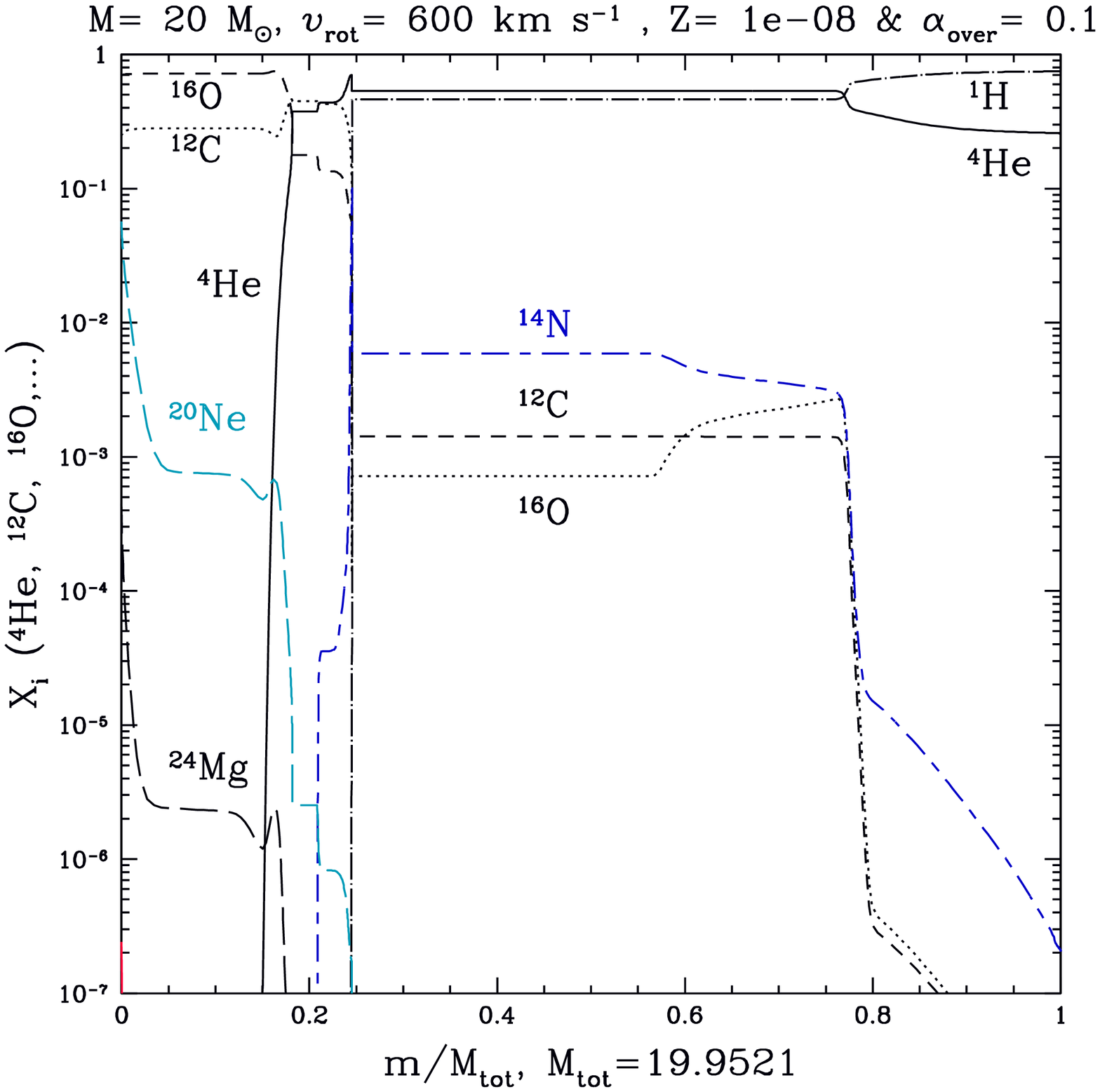}
\caption{Abundance profiles for the 20 $M_\odot$ with 
$\upsilon_{\rm ini}=$600\,km\,s$^{-1}$ at $Z=10^{-8}$: 
({\it left}) before shell H--burning boost (induced by mixing of carbon and oxygen
from the core), 
({\it middle}) just after the boost 
and ({\it right}) after shell H--burning has deepened during shell
He--burning.}
\label{ab20}
\end{figure*}
\begin{figure*}[!tbp]
\centering
\includegraphics[width=6cm]{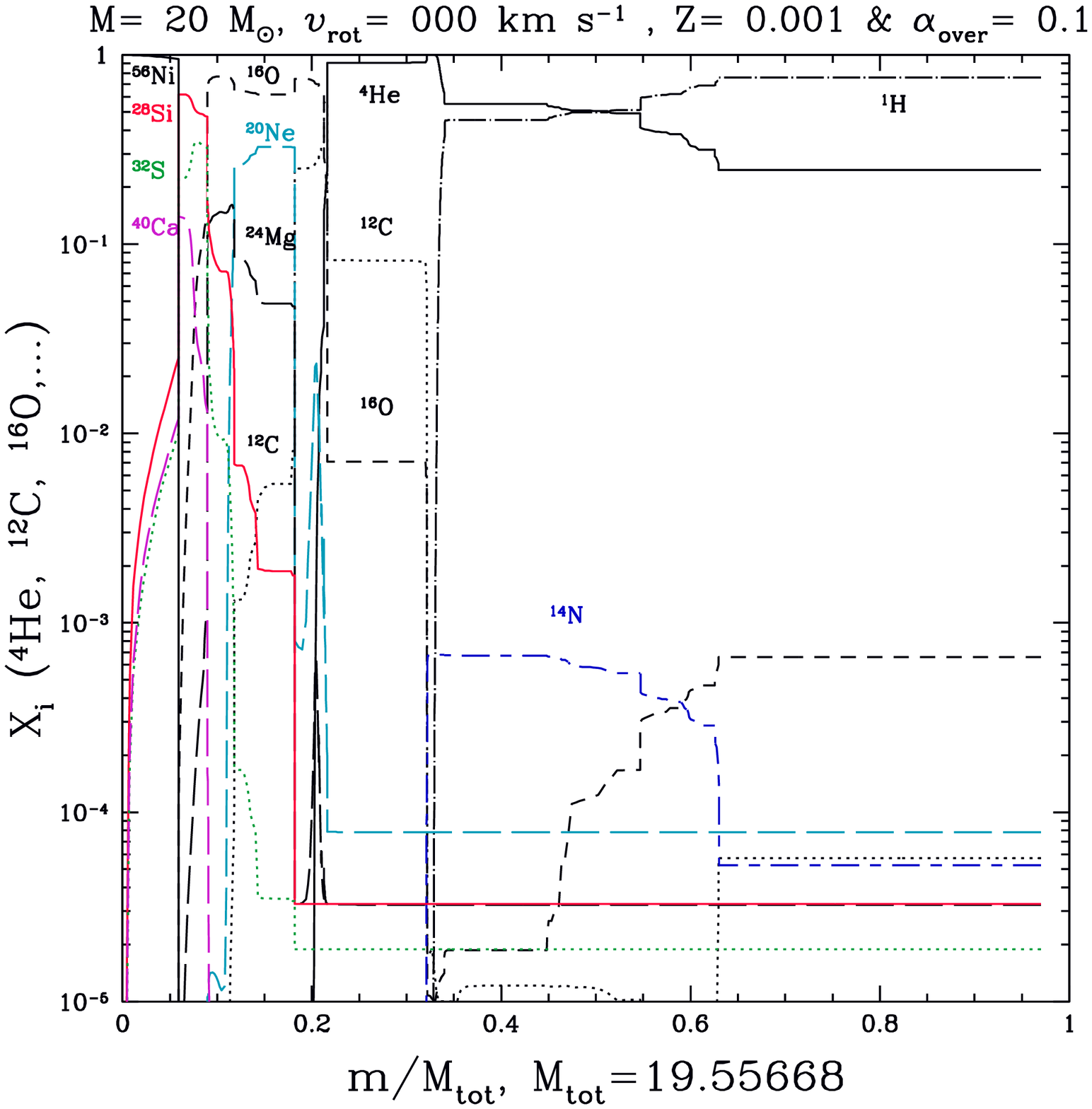}\includegraphics[width=6cm]{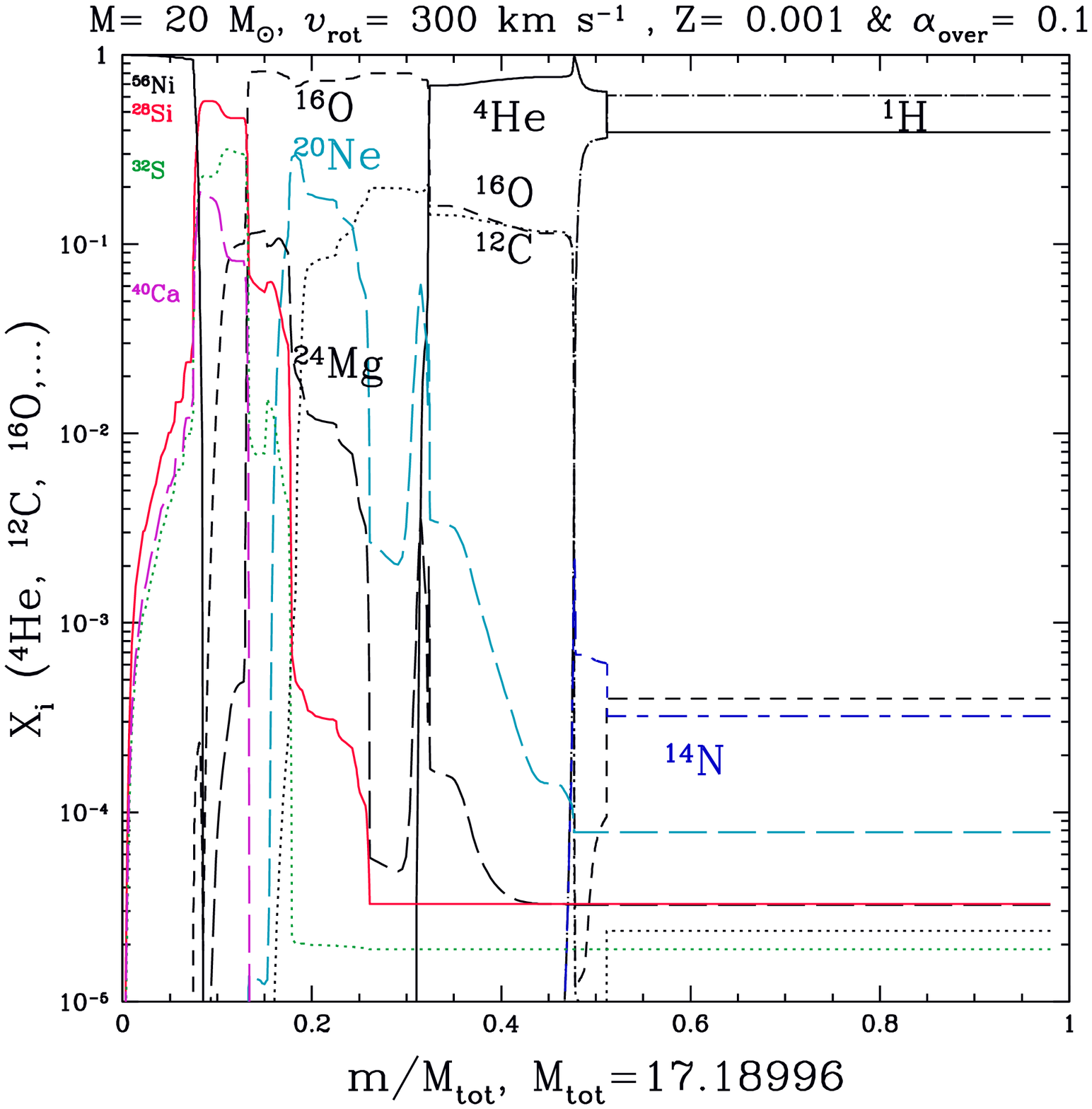}\includegraphics[width=6cm]{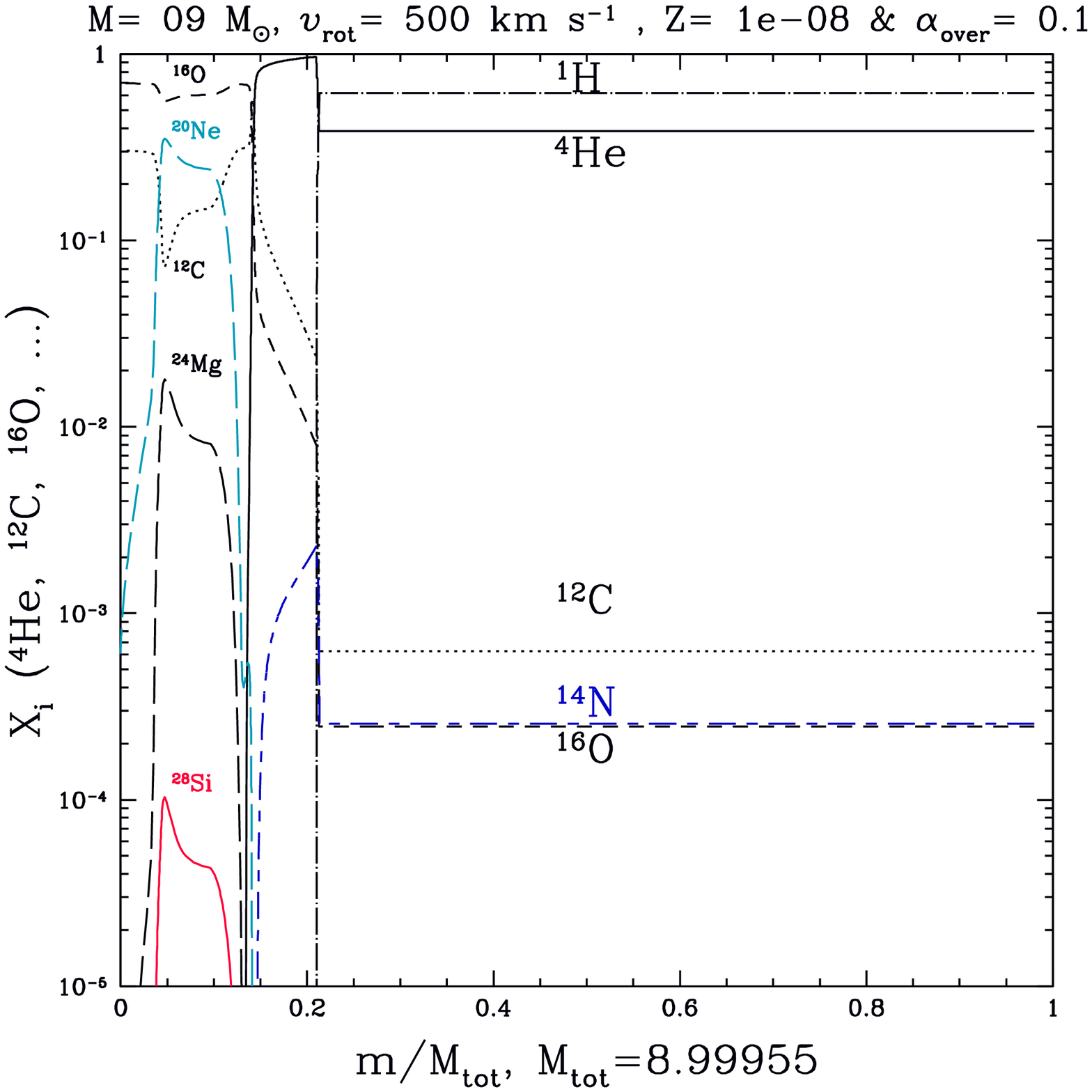}
\includegraphics[width=6cm]{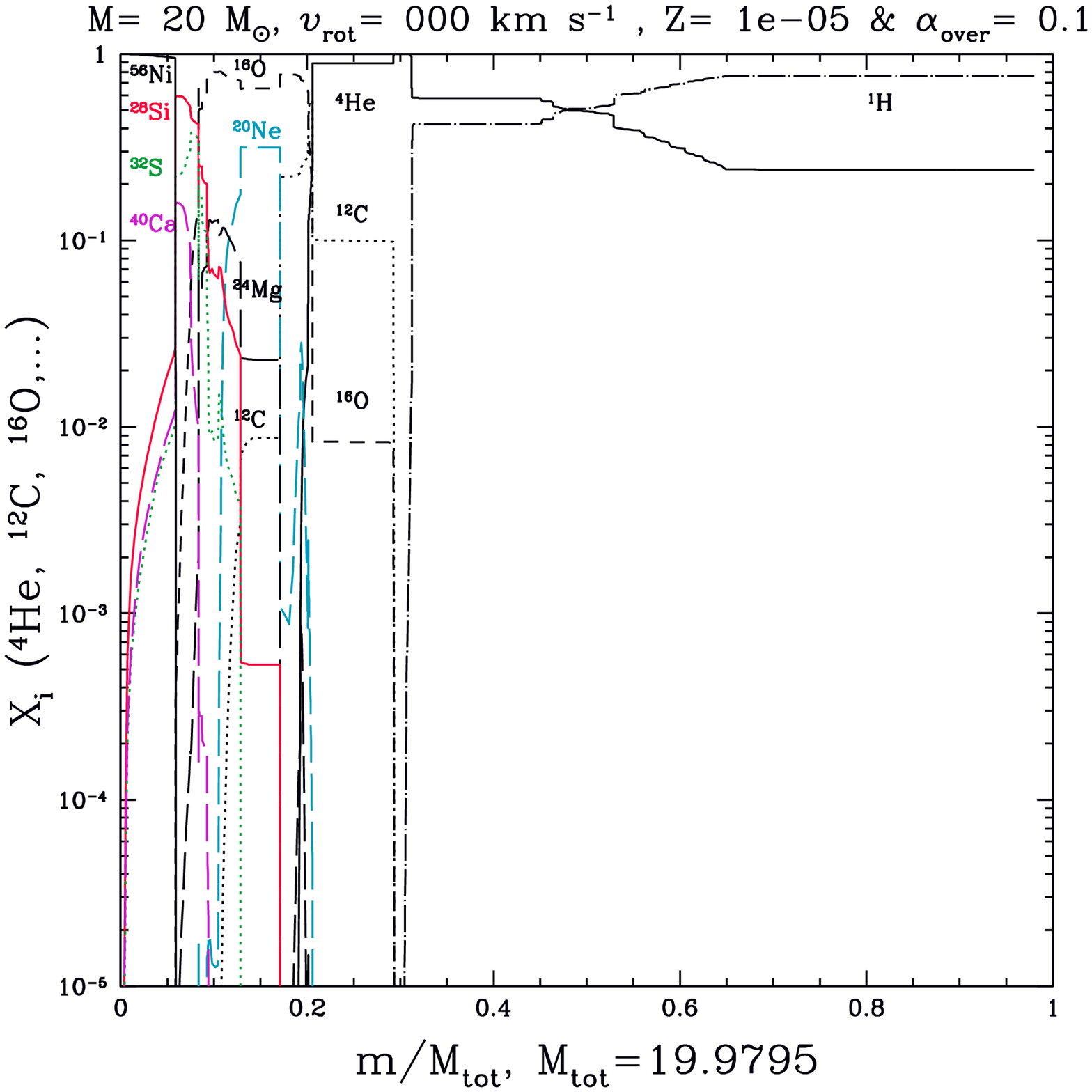}\includegraphics[width=6cm]{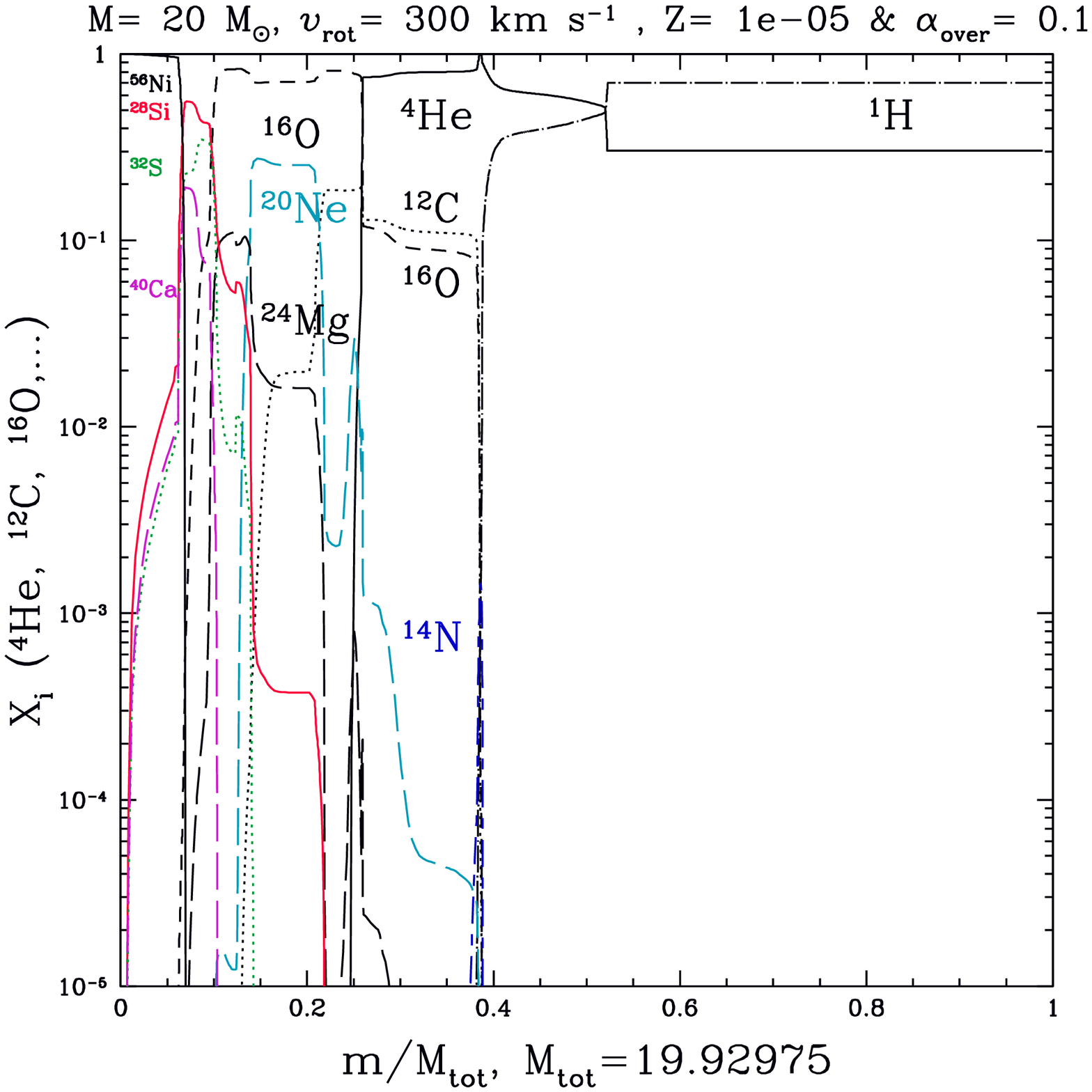}\includegraphics[width=6cm]{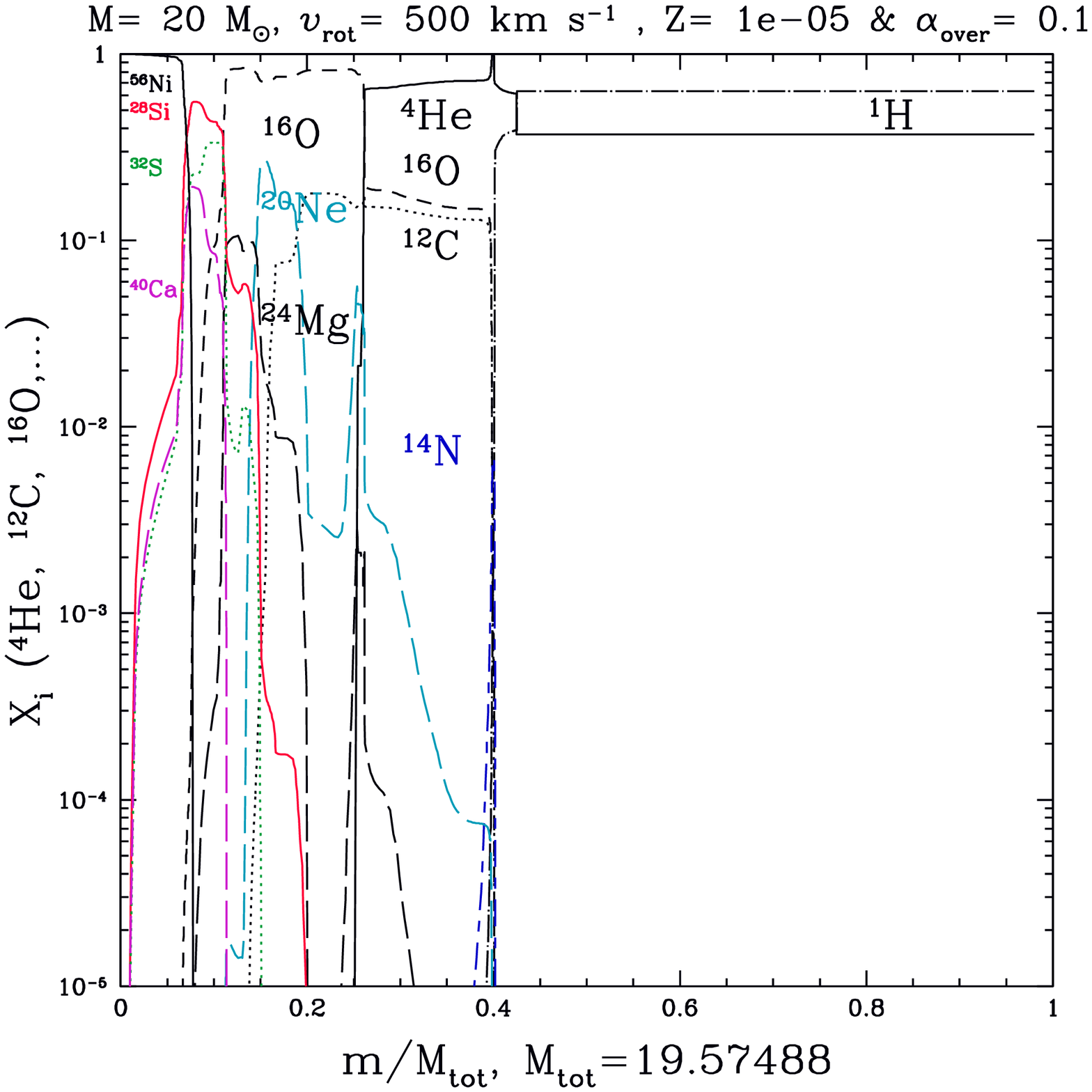}
\includegraphics[width=6cm]{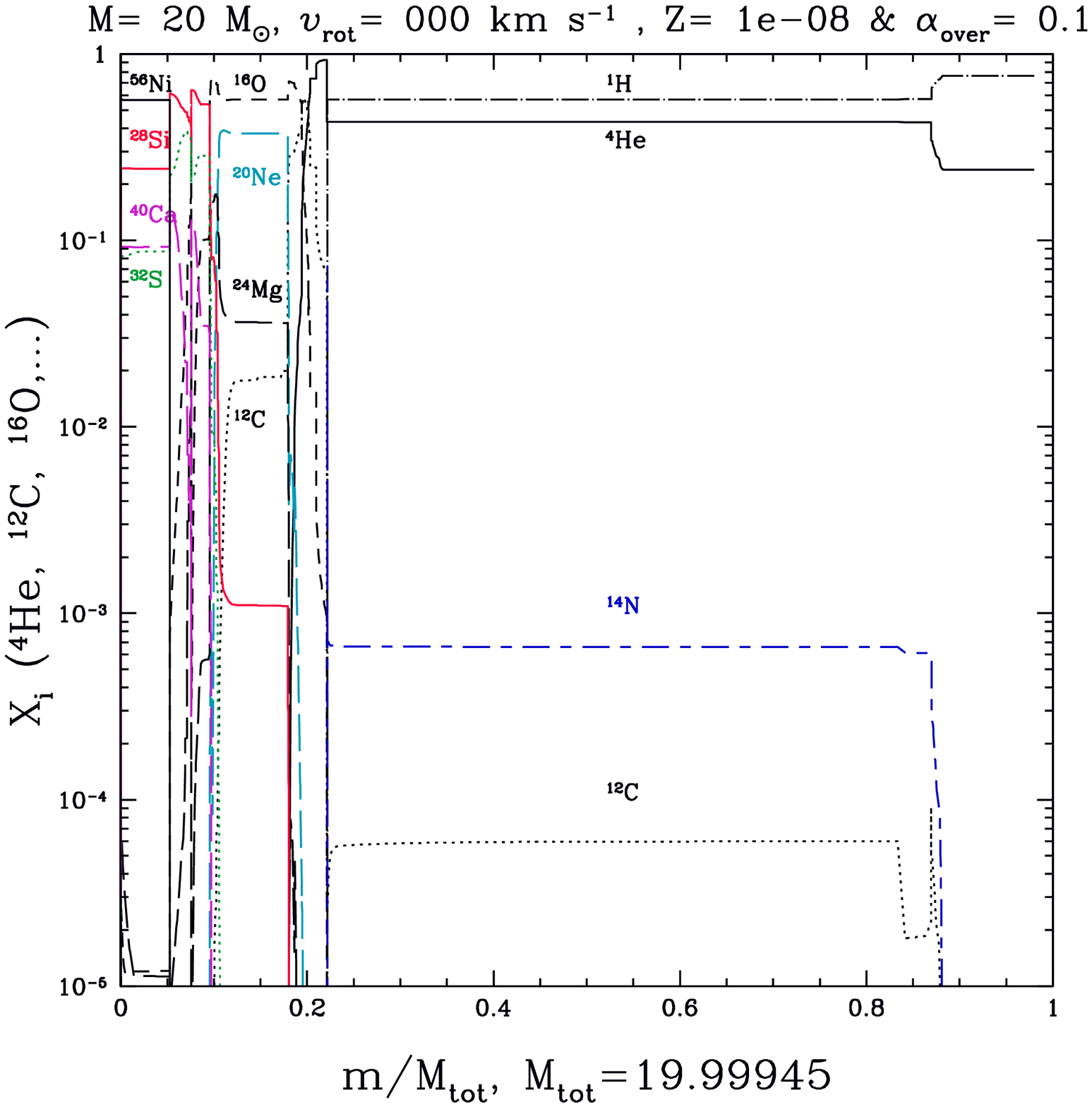}\includegraphics[width=6cm]{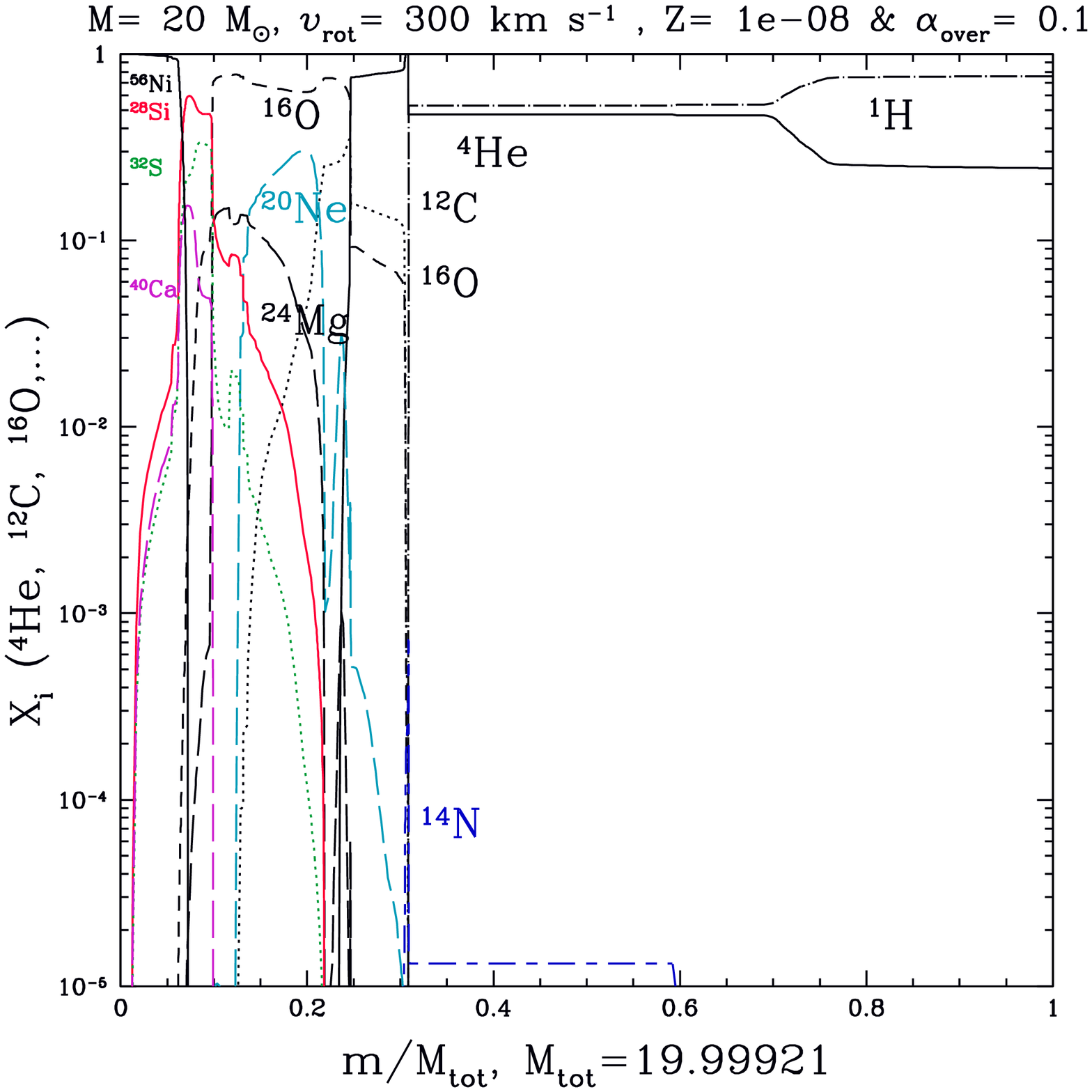}\includegraphics[width=6cm]{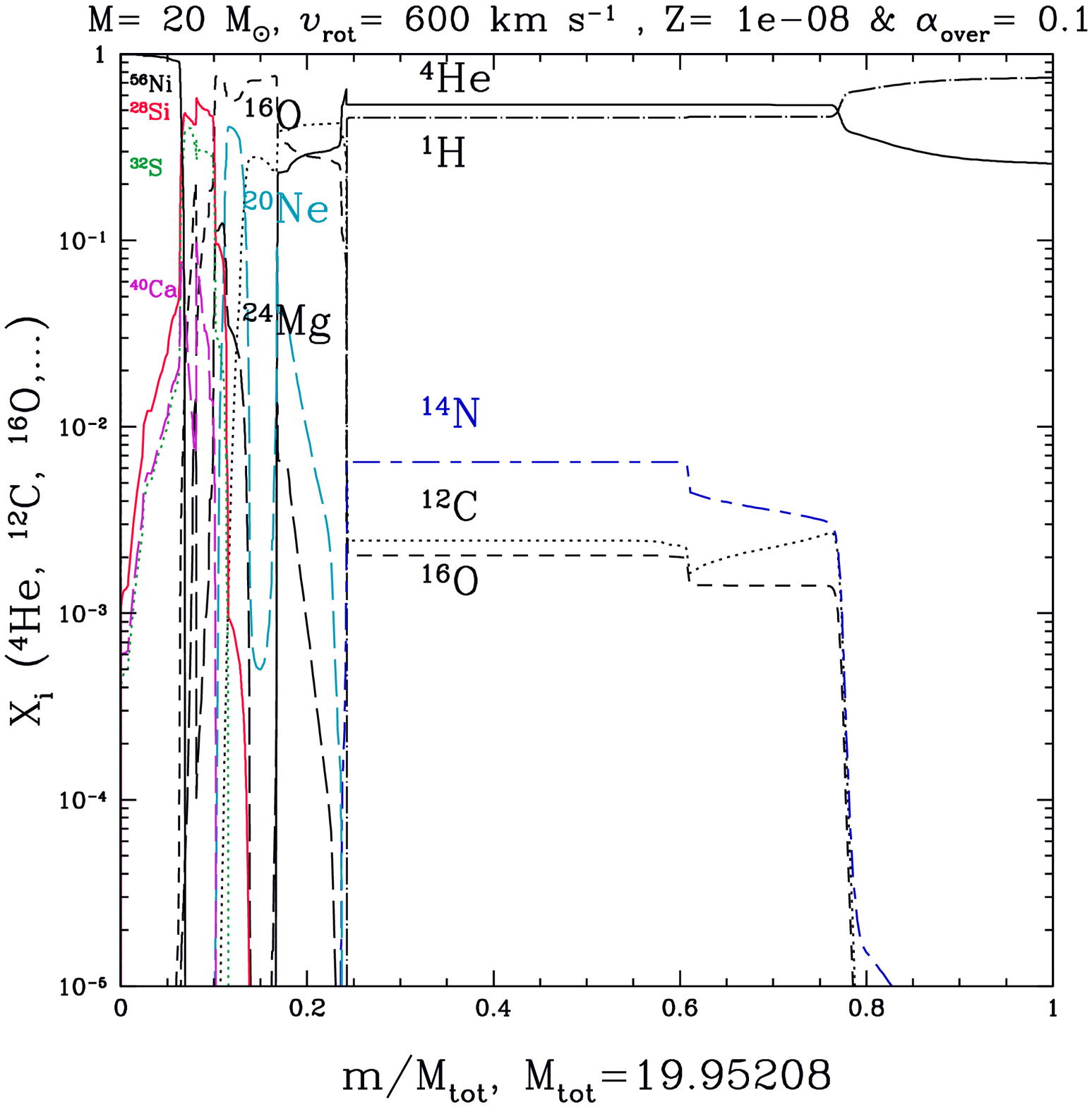}
\includegraphics[width=6cm]{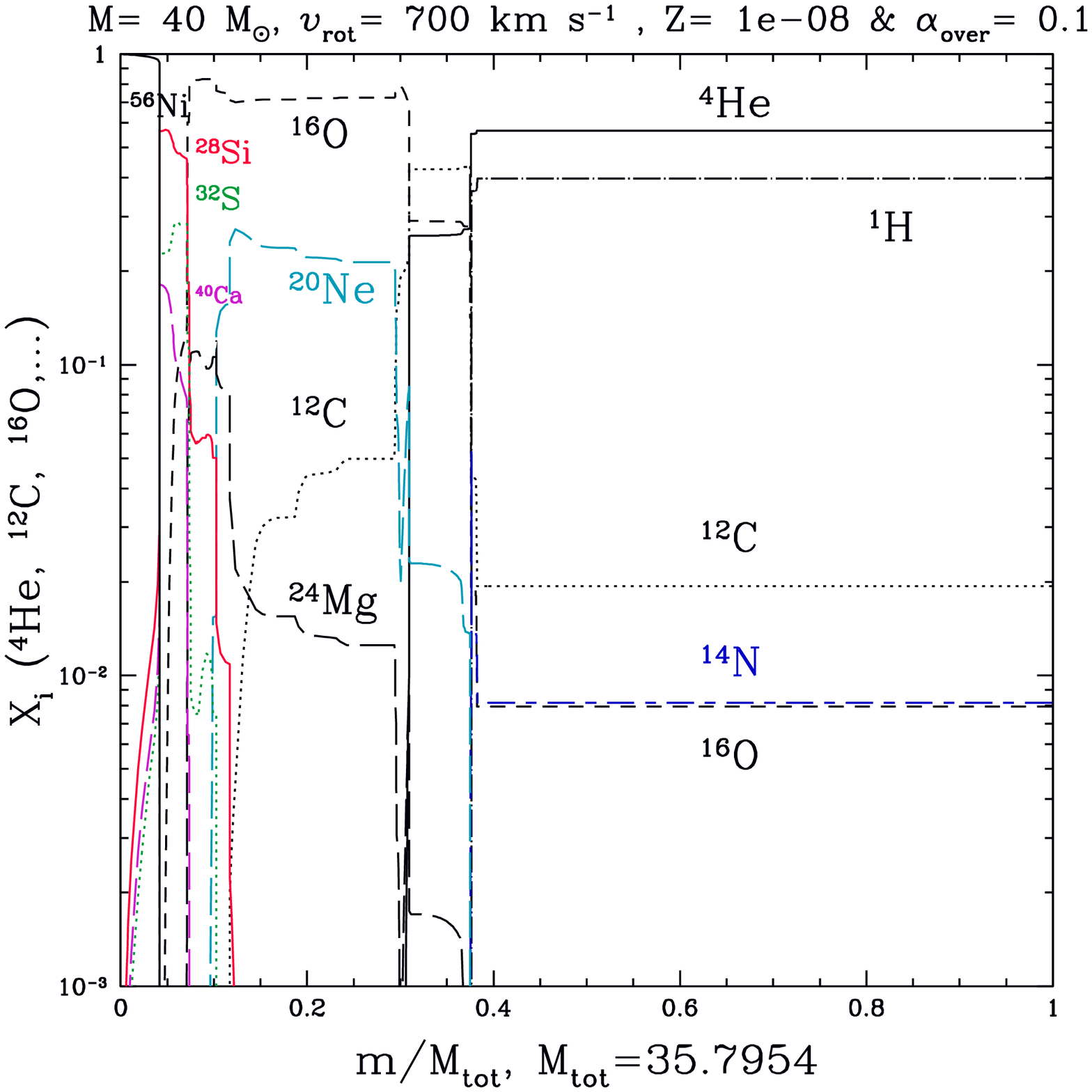}\includegraphics[width=6cm]{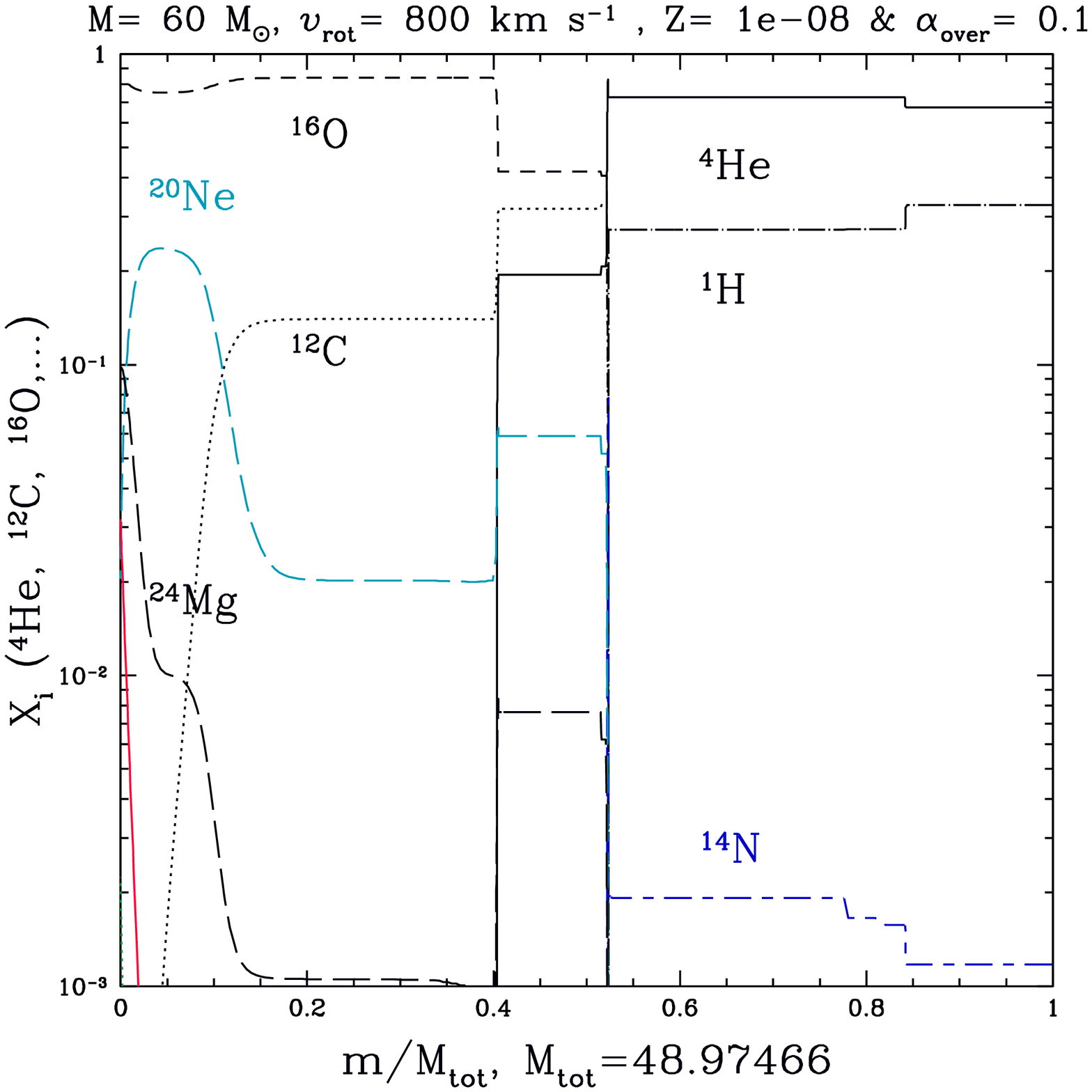}\includegraphics[width=6cm]{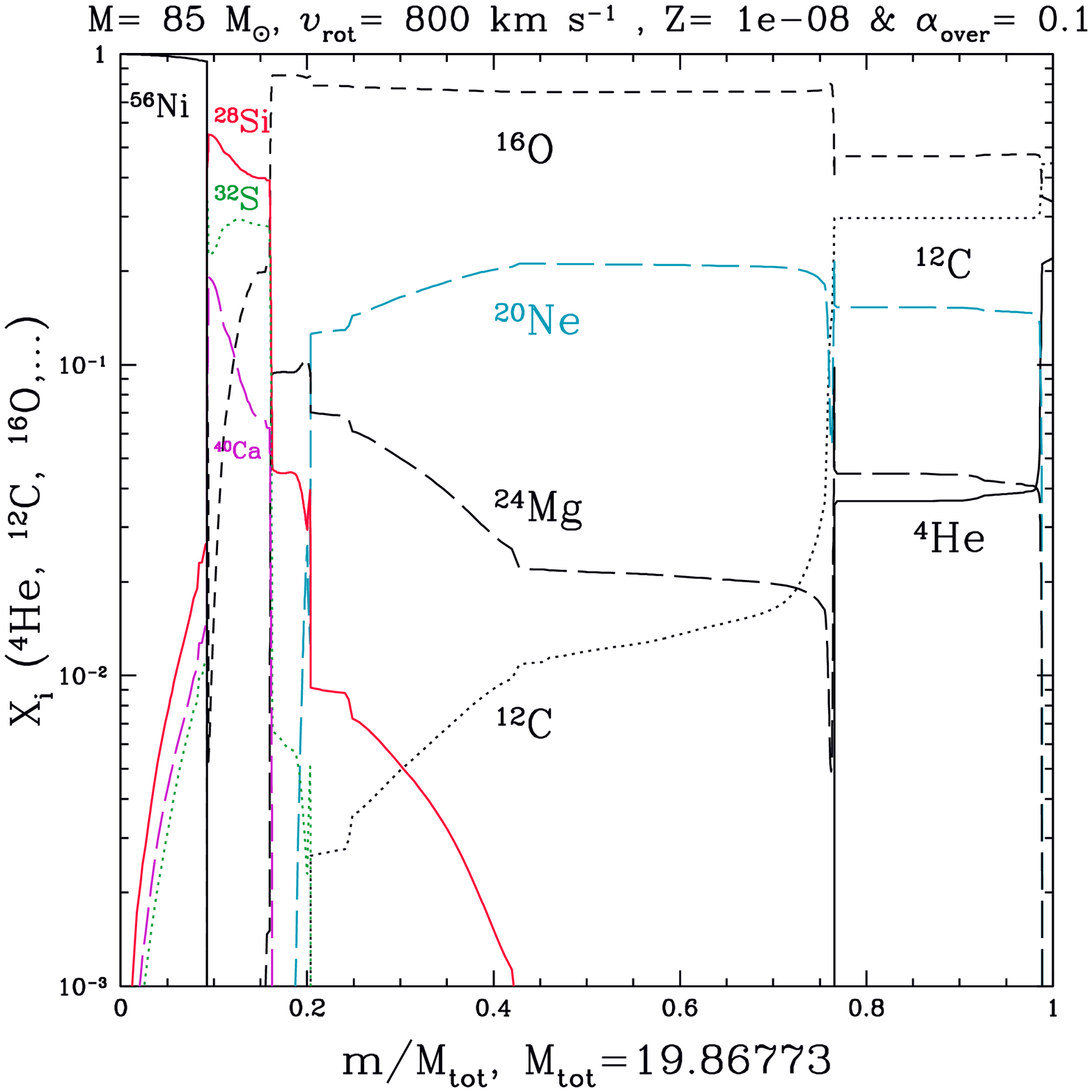}
\caption{Final abundance profiles. 
The initial parameters of the models are given on top of each plot.}
\label{ab1}
\end{figure*}

\end{document}